\RequirePackage{fix-cm}
\documentclass[smallextended]{svjour3}       
\smartqed  
\usepackage{graphicx}
\usepackage{rotating}
\usepackage{amssymb}
\usepackage{ascmac}
\usepackage{amsmath}
\usepackage{latexsym}
\usepackage[numbers]{natbib}
\usepackage{bm}
\usepackage[colorlinks=true, linkcolor=blue, citecolor=blue, urlcolor=blue]{hyperref}

\begin{document}

\title{Numerical Studies of Quantum Turbulence 
}


\author{Makoto Tsubota         \and
        Kazuya Fujimoto  \and
        Satoshi Yui
}


\institute{M. Tsubota, S. Yui \at
              Department of Physics, Osaka City University, Osaka 558-8585, Japan \\
              \email{tsubota@sci.osaka-cu.ac.jp, yui@sci.osaka-cu.ac.jp}           
           \and
           K. Fujimoto \at
              Department of Physics, The University of Tokyo, Tokyo 113-0033 , Japan \\
                            \email{fujimoto@cat.phys.s.u-tokyo.ac.jp}           
}

\date{Received: date / Accepted: date}

\maketitle

\begin{abstract}
We review numerical studies of quantum turbulence.
Quantum turbulence is currently one of the most important problems in low temperature physics and is actively studied for superfluid helium and atomic Bose--Einstein condensates.
A key aspect of quantum turbulence is the dynamics of condensates and quantized vortices.
The dynamics of quantized vortices in superfluid helium are described by the vortex filament model, while the dynamics of condensates are described by the Gross--Pitaevskii model.
Both of these models are nonlinear, and the quantum turbulent states of interest are far from equilibrium.
Hence, numerical studies have been indispensable for studying quantum turbulence.
In fact, numerical studies have contributed in revealing the various problems of quantum turbulence.
This article reviews the recent developments in numerical studies of quantum turbulence.
We start with the motivation and the basics of quantum turbulence and invite readers to the frontier of this research.
Though there are many important topics in the quantum turbulence of superfluid helium, this article focuses on inhomogeneous quantum turbulence in a channel, which has been motivated by recent visualization experiments.
Atomic Bose--Einstein condensates are a modern issue in quantum turbulence, and this article reviews a variety of topics in the quantum turbulence of condensates {\it e.g.} two-dimensional quantum turbulence, weak wave turbulence, 
turbulence in a spinor condensate, {\it etc.}, some of which has not been addressed in superfluid helium and paves the novel way for quantum turbulence researches. Finally we discuss open problems. 

\end{abstract}

\section{Introduction}
\label{intro}
Quantum turbulence (QT) describes turbulence in the hydrodynamics of quantum condensed systems.
The hydrodynamics in these systems are subject to severe quantum restrictions, and quantized vortices play a pivotal role in rotational motion of superfluid.
A quantized vortex is a stable topological defect with quantized circulation, which distinguishes quantum hydrodynamics (QHD) from classical hydrodynamics. 

QT was first discovered in superfluid $^4$He and was recently discovered in atomic Bose--Einstein condensates (BECs). As a result, QT has become one of the most important topics in low temperature physics.   
Numerical studies are indispensible for understanding QT because it is highly nonlinear and out of equilibrium.
This article reviews numerical studies of QT both in superfluid $^4$He and atomic BECs.  

Two models are usually available for studying QT, namely the vortex filament (VF) model and the Gross--Pitaevskii (GP) model \footnote{Another important model is the Hall--Vinen--Bekharevich--Khalatnikov (HVBK) model. It is briefly introduced in Sec. 2.1.1. }. The VF model represents a quantized vortex as a filament and follows the dynamics of the system. Because most characteristic lengths in superfluid $^4$He are much larger than the vortex core size, the VF model is very useful in studying superfluid $^4$He. On the other hand, the GP model is the mean-field approximation of a weakly interacting Bose system, which is useful in studying atomic BECs at very low temperatures.    
Here, the two models are compared from the viewpoint of numerical studies. The VF model supposes that the fluid is incompressible, and the vortex is represented as a string of points. Since the vortex core is assumed to be infinitely thin, the VF model cannot describe phenomena related to the cores such as vortex reconnection, nucleation, and annihilation. On the other hand, the GP model can describe not only the vortex dynamics but also the dynamics of the condensate, including the core-related phenomena. However, it is not so easy to consider finite temperature effects in the GP model.     
 
Though there are several existing review articles on QT \cite{vinenniemela,Halperin,Skrbekreview,TKT,SergeyPR,BarenghiPNAS}, this articles focuses on the numerical studies of QT. The contents of this article are as follows. Chapter 2 describes the numerical studies of QT in superfluid $^4$He. Starting with the two-fluid model, we briefly review the research history. Although there are many aspects on this issue, we focus on the topics of thermal counterflow. The readers should refer to other review articles on topics such as the energy spectrum of QT and QT created by an oscillating object \cite{Halperin,TKT}. Simulation of the homogeneous system pioneered by Schwarz has been successful, but recent visualization experiments have paved the way to the study of inhomogeneous systems in a channel. Chapter 3 reviews QT in atomic BECs. First, recent experiments of turbulence in the atomic BEC are introduced, where we describe 2D and 3D QT in single-component BECs and QT in spinor BECs. Such dependence on spatial dimension and multi-component system in QT are unique for the atomic BECs. Subsequently, we address theoretical studies of the GP model that investigate vortex turbulence, weak wave turbulence, 2D nature of QT, and spin turbulence. 
Chapter 4 is devoted to survey of the open problems. 

\section{QT in superfluid helium}
\label{sec:1}
This section describes QT in superfluid $^4$He by the VF model.

\subsection{Background and history}
\label{sec:2}
The main focus of this section is on numerical studies of thermal counterflow. This subsection briefly reviews the history of the modern problems of thermal counterflow and such other topics as energy spectra and QT caused by vibrating structures.
\subsubsection{Two-fluid model and thermal counterflow}
The hydrodynamics of superfluid helium is well described by the two-fluid model \cite{Tisza,Landau}
in which the system consists of an inviscid superfluid (density $\rho_{\rm s}$) and 
a viscous normal fluid (density $\rho_{\rm n}$) with two velocity fields ${\bm v}_{\rm s}$ and ${\bm v}_{\rm n}$. 
Only the normal fluid can carry entropy.
The mixing ratio of the two fluids depends on temperature. 
If two fluids are independent, the superfluid obeys the Euler-like equation and the normal fluid obeys the Navier--Stokes equation \cite{landau}: 
\begin{eqnarray}
 \rho_{\rm s}\left[ \frac{\partial {\bm v}_{\rm s}}{\partial t} + ({\bm v}_{\rm s} \cdot \nabla){\bm v}_{\rm s} \right] = -\frac{\rho_{\rm s}}{\rho} \nabla p + \rho_{\rm s} s \nabla T, \label{eq:super}\\
  \rho_{\rm n}\left[ \frac{\partial {\bm v}_{\rm n}}{\partial t} + ({\bm v}_{\rm n} \cdot \nabla){\bm v}_{\rm n} \right]= -\frac{\rho_{\rm n}}{\rho} \nabla p -\rho_{\rm s} s \nabla T+ \eta_{\rm n} \nabla^2 {\bm v}_{\rm n}
  \label{eq:normal}
\end{eqnarray}
with pressure $p$ and temperature $T$.
Here, $\rho=\rho_{\rm s}+\rho_{\rm n}$ is the total density of two fluids, $s$ is the entropy per unit mass, and $\eta_{\rm n}$ is the viscosity of the normal fluid.
The total momentum flux is ${\bm j}=\rho_{\rm s}{\bm v}_{\rm s}+\rho_{\rm n}{\bm v}_{\rm n}$.
This equation shows the characteristic behavior of two-fluid hydrodynamics. 
If a system is subject to a temperature gradient $\nabla T$, the superfluid is driven along it and the normal fluid is driven oppositely.
Thermal counterflow makes use of this behavior, and most of the early experimental studies of superfluid hydrodynamics focused on thermal counterflow \cite{Gorter}.

\begin{figure}
  \centering
  \includegraphics[width=0.5\textwidth]{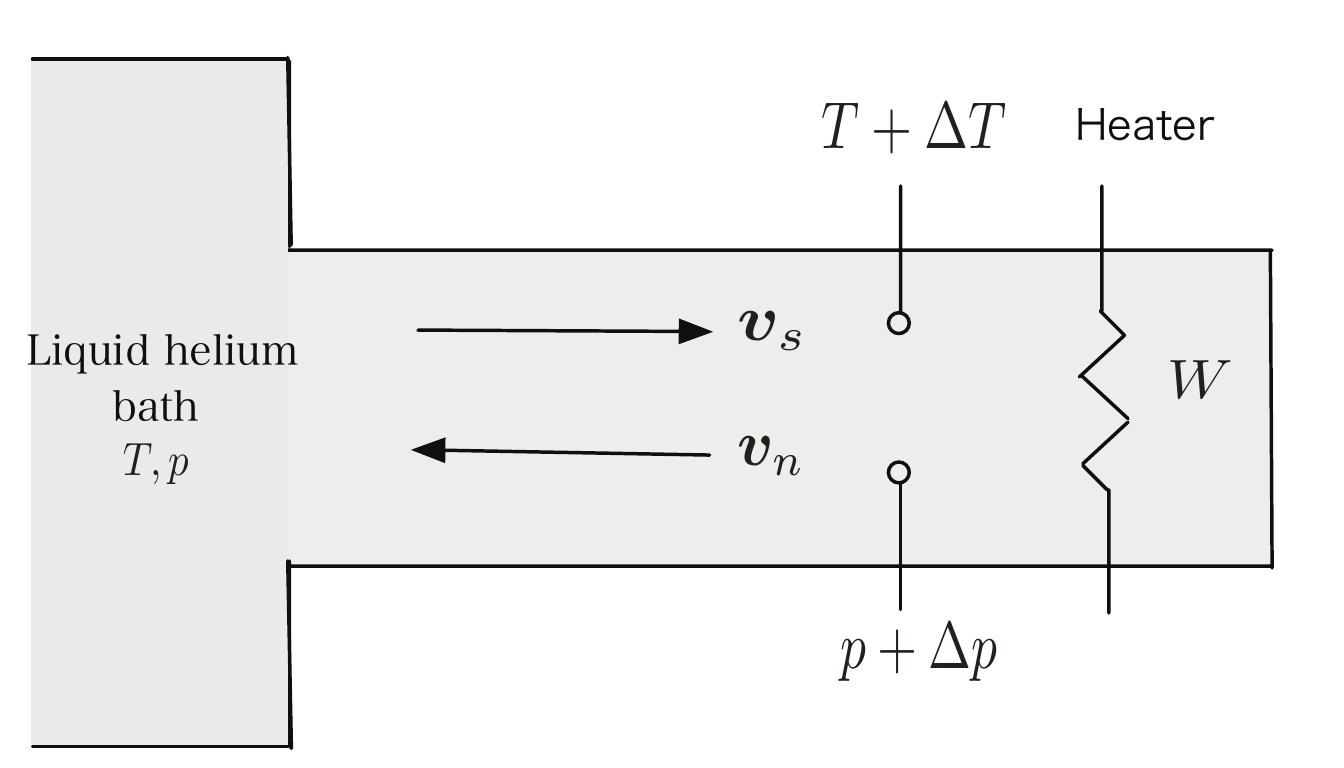}
  \caption
  {
    Schematic representation of thermal counterflow. The counterflow ${\bm v}_{\rm s}$ and ${\bm v}_{\rm n}$ are driven by the injected heat $W$. Compared with the temperature $T$ and the pressure $p$ of the liquid helium bath, the temperature difference $\Delta T$ and the pressure difference $\Delta P$ are observed near the heater. 
  }
  \label{counterflow}
\end{figure}

\begin{figure}
  \centering
  \includegraphics[width=0.5\textwidth]{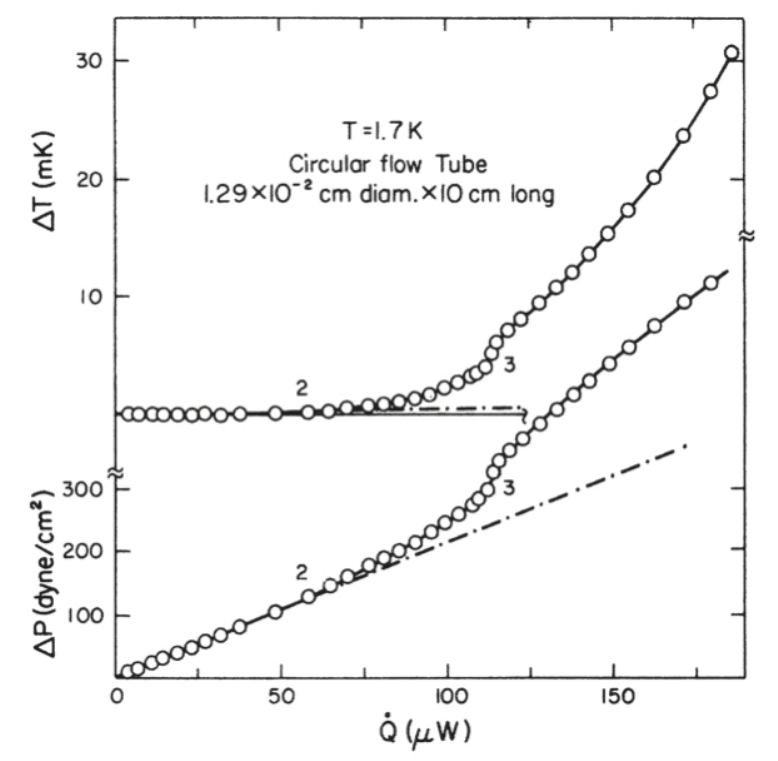}
  \caption
  {
 Typical temperature and pressure difference $\Delta T$ and $\Delta p$ as a
function of heat current $\dot{Q}$ for a circular flow tube at 1.7 K with thermal counterflow \cite{Ladner78}. The broken line is an extrapolation
of the linear region, consistent with Eq. (\ref{laminar}).
The region between points 2 and 3 is called T1, and the region above point 3 is called T2.
 [Reprinted figure with permission from 
 \href{https://doi.org/10.1103/PhysRevB.17.1455}{D. R. Ladner and J. T. Tough, Phys. Rev. B {\bf 17}, 1455 (1978)}.
 Copyright (1978) by the American Physical Society.]
  }
  \label{Tough}
\end{figure}

Figure \ref{counterflow} shows a schematic representation of thermal counterflow. Superfluid $^4$He is assumed to be confined in a channel of circular cross-section of radius $A$, with one closed end and the other end connected to a helium bath.
We suppose that heat $W$ per unit time is injected into the system near the closed end. 
Then, the normal fluid flows from the closed end towards the helium bath and the superfluid flows oppositely. 
Since the net mass flux vanishes, the counterflow ${\bm j}=\rho_{\rm s}{\bm v}_{\rm s}+\rho_{\rm n}{\bm v}_{\rm n}=0$ is driven.
The injected heat is carried away by the normal fluid with an energy flux of $W/\pi A^2=s \rho T v_{\rm n}$. 
Then, the relative velocity of the two fluids is: 
\begin{equation}
v_{\rm ns}=|v_{\rm n}-v_{\rm s}|=\frac{\rho}{\rho_{\rm s}}v_{\rm n}=\frac{W}{\pi A^2\rho_{\rm s} s T}.
\end{equation}
By assuming the counterflow is stationary and uniform along the channel, we can estimate what happens to the temperature and pressure gradients. 
When $W$ is small, both flows should be laminar with only axial components.
Then, Eq. (\ref{eq:super}) gives $\nabla p=\rho s \nabla T$ and Eq. (\ref{eq:normal}) gives $\eta_{\rm n}\nabla^2 {\bm v}_{\rm n}=\nabla p$, such that the superfluid flow becomes uniformly laminar and the normal fluid flow takes a laminar Poiseuille profile $v_{\rm n}(r)=-(\nabla p/4\eta_{\rm n})(A^2-r^2) $ with the radial coordinate $r$.  
Then, the gradients of pressure and temperature of the laminar cases are given by:
\begin{equation}
\nabla p_{\rm L}=-\frac{8\eta_{\rm n}}{A^2} v_{\rm n}, \quad \nabla T_{\rm L}=\frac{\nabla p_{\rm L}}{\rho s}= -\frac{8\eta_{\rm n}}{A^2\rho s} v_{\rm n}.  \label{laminar}
\end{equation}
Here $v_{\rm n}$ refers to the averaged $v_{\rm n}(r)$ over the cross-section, and the subscript L means that the flow is laminar.
When $W$ is small, these gradients are observed experimentally as shown in Fig. \ref{Tough}.

When the heat or the relative velocity exceeds some critical value, however, an additional temperature gradient to Eq. (\ref{laminar}) was observed \cite{Gorter}.  
Gorter and Mellink added the mutual friction term ${\bm F}_{\rm ns}$ into Eqs. (\ref{eq:super}) and (\ref{eq:normal}):
\begin{eqnarray}
 \rho_{\rm s}\left[ \frac{\partial {\bm v}_{\rm s}}{\partial t} + ({\bm v}_{\rm s} \cdot \nabla){\bm v}_{\rm s} \right] = -\frac{\rho_{\rm s}}{\rho} \nabla p + \rho_{\rm s} s \nabla T-{\bm F}_{\rm ns}, \label{mfeq:super}\\
  \rho_{\rm n}\left[ \frac{\partial {\bm v}_{\rm n}}{\partial t} + ({\bm v}_{\rm n} \cdot \nabla){\bm v}_{\rm n} \right]= -\frac{\rho_{\rm n}}{\rho} \nabla p -\rho_{\rm s} s \nabla T+ \eta_{\rm n} \nabla^2 {\bm v}_{\rm n}+{\bm F}_{\rm ns}.
  \label{mfeq:normal}
\end{eqnarray}
Then, the gradients of pressure and temperature become\footnote{The mutual friction adds nothing to the pressure gradient. The observed pressure gradient shown in Fig. \ref{Tough} is attributable to eddy viscosity of superfluid. We do not mention this issue.}:
\begin{equation}
\nabla p=\nabla p_{\rm L}, \quad \nabla T=\nabla T_{\rm L}+\frac{F_{\rm ns}}{\rho_{\rm s} s }. \label{delta T}
\end{equation}
By comparing their observations of thermal counterflow and the temperature gradient of Eq. (\ref{delta T}), Gorter and Mellink found $F_{\rm ns} \simeq A'\rho_{\rm s}\rho_{\rm n}v_{\rm ns}^3$ with a temperature-dependent coefficient $A'$. 
  

The mutual friction was later found to come from the interaction between quantized vortices and normal fluid \cite{HallVinen56a,HallVinen56b}.
The original idea of quantized circulation had already been suggested by Onsager \cite{Onsager}, but Feynman proposed the concept of a quantized vortex filament and a turbulent superfluid state consisting of a tangle of quantized vortices \cite{Feynman}.  
Feynman's idea was experimentally confirmed by Vinen \cite{Vinen57a,Vinen57b,Vinen57c,Vinen57d,Vinen61}. 
Subsequently, many experimental studies have examined superfluid turbulence (ST) in thermal counterflow systems and have revealed a variety of physical phenomena \cite{tough}. 

The idea of the mutual friction is developed to another macroscopic model of two fluids, namely the Hall--Vinen--Bekharevich--Khalatnikov (HVBK) model \cite{HallVinen56b,Hills}.  The HVBK model is a continuous coarse-grained model which assumes that superfluid contains a large number of parallel quantized vortices in each fluid parcel and ignores the details of individual vortices and their fast dynamics. The HVBK model is originally valid for the case where vortices were locally aligned like rotating superfluid helium \cite{HallVinen56a,HallVinen56b}, and succeeds in studying, for example Taylor--Couette flow \cite{Barenghi92,Henderson95}.  However, this model is not applicable to ST including randomly oriented vortices. 
If we confine ourselves to the scales much larger than the intervortex spacing, this model works well and the numerical simulation leads to the Kolmogorov's $-5/3$ and $4/5$ laws\footnote{ The Kolmogorov's $-$5/3 law is described in Sec. 2.1.1. The Kolmogorov's 4/5 law refers to a statistical law for the third-order longitudinal structure function\cite{Frisch}. This is the only exact relation derived from the Navier--Stokes equation.} in superfluid $^4$He \cite{Salort12}. This article does not address the HVBK model, because it is just beyond the scope. 

The observations of thermal counterflow depend strongly on the aspect ratio of the cross-section of the channel \cite{tough}.
Figure \ref{Tough} shows the typical observation for a circular channel. 
At low heat currents, both the temperature difference $\Delta T$ and the pressure difference $\Delta P$ are proportional to the heat current, which is consistent with the laminar solution of Eq. (\ref{laminar}).
As the heat current increases, the first nonlinear region appears at point 2 because of the mutual friction.
In the much larger heat current, the temperature and pressure differences increase greatly above point 3. 
The region between points 2 and 3 is called T1, and the region above point 3 is called T2. 
When the aspect ratio of the cross-section of the channel is low, the system has two such turbulent states.
Melotte and Barenghi suggested that the transition from T1 to T2 is caused by the transition of the normal fluid from a laminar state to a turbulent state \cite{melotte98}.
If the aspect ratio of the channel is high, the counterflow exhibits only a single turbulent state T3, but little information is available on the T3 state.

Numerical studies of the dynamics of quantized vortices have greatly contributed to revealing the above physics.
The equations of motion of the quantized vortex filament (described in Sec. 2.2) are well known. However, these equations are highly nonlinear and nonlocal. 
We are interested in QT that is far from the equilibrium state. Thus, numerical simulations of the VF model are necessary to understand QT.
The pioneering numerical works were performed by Schwarz \cite{schwarz78,schwarz85,schwarz88}. By performing direct simulation of the VF model, he showed that a vortex tangle is self-sustained in thermal counterflow by competition between excitation due to the applied flow and dissipation due to mutual friction. The observable quantities obtained by his
calculations agreed well with the experimental results for the steady state of vortex tangles.
Moreover, Schwarz’s work numerically confirmed the original Feynman picture for QT.  
Most of the Schwarz's simulations were performed under the localized-induction approximation (LIA) by simplifying the nonlocal integral of the Biot--Savart law; in LIA, the interaction between vortices is neglected. 
Thus, when Schwarz addressed a homogeneous system under periodic boundary conditions in all three directions, he had to introduce an artificial mixing procedure.
Adachi {\it et al.} performed the full Biot--Savart simulation and succeeded in obtaining a steady state without the mixing procedure \cite{adachi}.   
After the effort expended along this line, the problems of a homogeneous system were settled \cite{baggaley121,kondaurova14}. 

\begin{figure}
  \centering
  \includegraphics[width=0.9\textwidth]{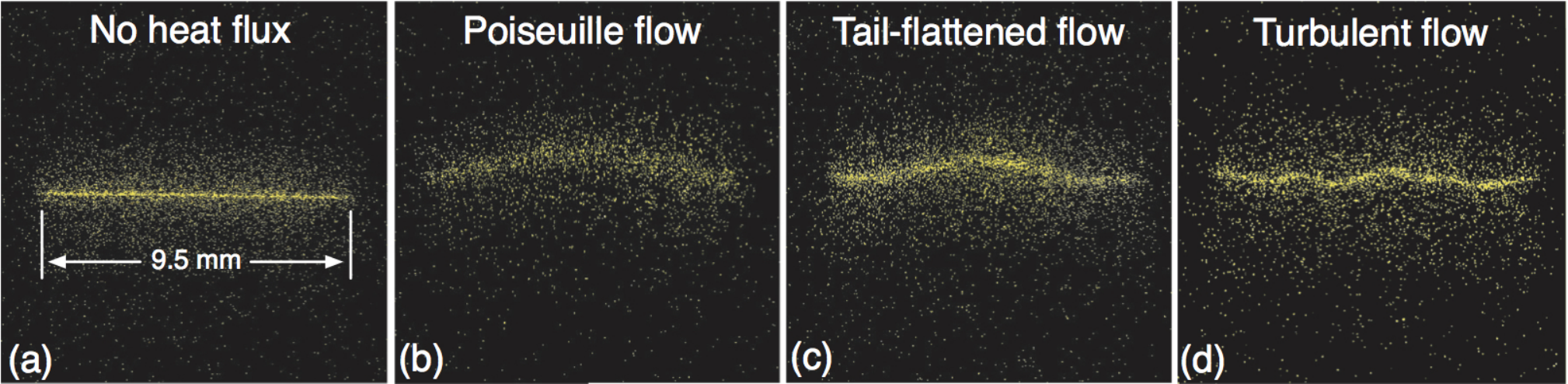}
  \caption
  {
Typical fluorescent images showing the motion of He$^*_2$ tracer lines in thermal counterflow of superfluid $^4$He at 1.83K \cite{marakov}.　
The square cross section of the channel is 9.5 mm$^2$. Superfluid flows downside, while normal fluid flows upside, whose profile is visualized by this technique. (a) shows the case of no heat flux. As the heat flux is increased, the profile changes from (b) a laminar Poiseuille: 10 mW/cm$^{2}$, via (c) a laminar tail-flattened: 62 mW/cm$^{2}$ to (d) a turbulent: 200 mW/cm$^{2}$.  The images shown in the laminar flow regime are averaged over nine
single shots.  By courtesy of Wei Guo.
[Figure adapted with permission from Ref. \cite{marakov}. Copyrighted by the American Physical Society.]
(Color figure online)
  }
  \label{Guo}
\end{figure}

However, recent visualization experiments changed the situation drastically \cite{GuoPNAS}. 
For example, Marakov {\it et al.} have observed the flow profiles of normal fluid in thermal counterflow in a square channel by following the motion
of seeded metastable He$_2^*$ molecules using a laser-induced fluorescence technique \cite{marakov}.
Figure \ref{Guo} shows how the normal fluid profile changes as the heat flux is increased. 
These observations have strongly triggered the studies of inhomogeneous QT in a channel, which is a main theme of this section.

\subsubsection{Energy spectra}
In turbulence, the velocity fields and the configuration of vortices are disordered and fluctuate both temporally and spatially.
Thus, it is important to focus on statistical quantities, which are universal and reproducible. 
In the field of classical turbulence, the typical statistical quantities would be energy spectra and the logarithmic velocity profile \cite{Frisch,davidson}.
Since the logarithmic velocity profile is described in Sec. \ref{loglaw}, this subsection addresses the topic of energy spectra. 

The important motivation comes from understanding the analogy between classical turbulence (CT) and QT.
A significant touchstone would be whether QT satisfies the statistical law of the energy spectra of CT.
The standard statistical law of CT is the Kolmogorov $-5/3$ law. The picture sustaining the Kolmogorov law is the following \cite{Frisch,davidson}. 
We assume a steady state of fully developed turbulence of an incompressible classical fluid.
The energy is injected into the fluid with a rate $\epsilon$ and at a scale comparable to the system size in the energy-containing range.
In the inertial range, this energy is transferred to smaller scales without being dissipated. 
In this range, the system is locally homogeneous and isotropic, which leads to the energy spectrum known as the Kolmogorov $-5/3$ law:
\begin{equation}
E(k)=C\epsilon^{2/3}k^{-5/3}.
\end{equation}
Here, the energy spectrum $E(k)$ is defined as $E=\int_0^{\infty} dk E(k)$, where $E$ is the kinetic energy per unit mass and $k$ is the wavenumber from the Fourier transformation of the velocity field. The Kolmogorov constant $C$ is a dimensionless parameter of order unity. The energy in this inertial range is believed to be transferred by the Richardson cascade\footnote{A picture of the energy cascade of instabilities of large eddies or vortices, whereby their breakdown and fragmentation bring energy from large to small scales. It comes from the Richardson's poem of 1922, "Big whirls have little whirls that feed on their velocity, and little whirls have lesser whirls, and so on to viscosity". }.  The energy transferred to much smaller scales in the energy-dissipative range is dissipated through the viscosity of the fluid. The Kolmogorov law has been confirmed experimentally and numerically in fully developed turbulence.

Our current interests are in what happens to the QT energy spectrum and what the mechanism of the energy cascade is \cite{BarenghiPNAS2}.
Thermal counterflow has no classical analogue, and thus novel experiments have appeared to study the energy spectrum in superfluid helium. 
The first important contribution was made by Maurer and Tabeling \cite{Maurer98}.
A turbulent flow was produced in a cylinder by driving two counter-rotating disks.
The authors observed the local pressure fluctuations to obtain the energy spectrum.  
The Kolmogorov spectrum was confirmed at three different temperatures 2.3 K, 2.08 K, and 1.4 K.
The next step was a series of experiments on grid turbulence performed for superfluid $^4$He above 1 K by the Oregon group \cite{Smith93, Stalp99, Skrbek00a,Skrbek00b,Stalp02}.  
The flow through a grid is typically used to generate turbulence in classical fluid dynamics. At a sufficient distance behind the grid, the flow displays a form of homogeneous isotropic turbulence. 
In the experiments by the Oregon group, the helium was contained in a channel with a square cross-section, through which a grid was pulled at a constant velocity. 
A pair of second-sound transducers was set into the walls of the channel.
When a vortex tangle appeared in a channel, it was detected by second-sound attenuation. 
This case is decaying turbulence.
The authors considered how turbulence obeying the Kolmogorov law decays and connected it with the decay of the vortex line density (VLD) $L$, which is the total length of the vortex lines per unit volume, to find $L \sim t^{-3/2}$. This behavior was consistent with the observations. 
The third contribution was made by the French groups who performed two independent experiments using wind tunnels \cite{Salort12,Salort10}.   
The authors confirmed Kolmogorov's $-5/3$ and 4/5 laws in superfluid $^4$He. 

These experiments have encouraged numerical studies of energy spectra. The above experiments were done at finite temperatures, where the normal fluid component was not negligible.
The simultaneous contribution of the two fluids complicates the problem. 
Thus, most efforts have been devoted to the case in the zero-temperature limit.
 Here, we should note that there are two extreme types of QT at 0 K, namely {\it quasiclassical turbulence} and {\it ultraquantum turbulence} \cite{GolovPNAS}.
This argument is closely related to how QT mimics CT or not.
A vortex tangle has a characteristic length, namely the intervortex spacing $\ell$, which is of the order of $L^{-1/2}$.
The scales larger and smaller than $\ell$ may be called the classical and quantum regions, respectively.
{\it Quasiclassical turbulence} is excited by forcing at length scales much larger than $\ell$, and thus there is a substantial amount of energy in the classical region. 
Then, the course-grained vorticity consisting of quantized vortices becomes a continuous function, which makes a classical-like description possible.
The energy should be transferred from large to small scales through the Kolmogorov-like cascade.
On the other hand, {\it ultraquantum turbulence} has forcing at length scales smaller than $\ell$ with a negligible amount of energy in the classical region.
 {\it Ultraquantum turbulence} consists of a nearly random tangle with no large-scale structures.  
 At these short length scales, the quantum nature of the vorticity is dominant, and the energy should be transferred by cascade along Kelvin waves (Kelvin-wave cascade) until some short wavelength where dissipation due to phonon emission is expected to be effective. 
It is important to know the condition that which of the two types of turbulence appears actually; the mechanism of preventing the formation of the quasiclassical Kolmogorov spectrum is recently discussed \cite{Barenghisrep}.   
 
 A simulation of the VF model in a three-dimensional (3D) periodic box confirms the Kolmogorov $-5/3$ spectrum \cite{Araki02,Baggaley122}. 
Numerical works based on the GP model are discussed in Sec. 3.3.2.  

The Kelvin-wave cascade is also an important problem of QT and a target of numerical simulation. The main interest  is the energy spectrum of the Kelvin-wave cascade. The standard theoretical procedure is to address small deformation around a straight vortex line, describe the deformation by the Fourier transformation and utilize the wave turbulence theory \cite{wt1,wt2}. The key issue is how to treat the interaction between these waves.  There were two theoretical proposals for the energy spectrum $E_k$. 
One is  the Kozik-Svistunov (KS) spectrum $E_K \propto k^{-17/5}$ based on the local six-wave interactions \cite{Kozik}.
The other is the L'vov-Nazarenko (LN) spectrum $E_k \propto k^{-5/3}$ \cite{L'vov10}, which is based on nonlocal six-wave interactions leading to local four-wave processes. 
 However, it was shown \cite{Laurie10} that the assumption of the locality of interaction used to derive the KS spectrum was incorrect, making the KS spectrum invalid.
 The recent numerical simulation of the VF model using the full Biot-Savart law supports the LN spectrum \cite{Baggaley14}.
 

\subsubsection{QT created by vibrating structures}
An important branch of current experimental studies relates to QT generated by vibrating structures. 
Various structures, such as spheres, thin wires, grids, and tuning forks have been used investigated in these studies.
In spite of the difference between the employed structures and geometry, the experiments show certain common phenomena. 
This subsection briefly describes what has been done experimentally and the contribution of numerical studies.

Referring to the pioneering experiments by J\"ager {\it et al.}\cite{Jager95}, we will show the typical observations. 
J\"ager {\it et al.} used a sphere with a radius of approximately 100 $\mu$m.
The sphere was made from a ferromagnetic material and its surface was rough.
The sphere was magnetically levitated in superfluid $^4$He, and its response with respect to the alternating drive was observed. 
When the driving force $F_{\rm D}$ was low, the velocity $v$ of the sphere was proportional to $F_{\rm D}$, taking the "laminar” form $F_{\rm D}=\lambda (T) v$ with the temperature-dependent coefficient $\lambda(T)$. When the driving force was increased, the response changed to the "turbulent" form $F_{\rm D}=\gamma(T) (v^2-v_0^2)$ above some critical velocity $v_0$ with another coefficient $\gamma(T)$. The transition from laminar to turbulent response was accompanied by significant hysteresis at relatively low temperatures. Subsequently, several groups have experimentally investigated the similar transition to turbulence in superfluids $^4$He and $^3$He-B by using various vibrating structures. The details are described in review articles \cite{VinenPLTP,VinenPNAS}.

\begin{figure}
  \centering
  \includegraphics[width=0.9\textwidth]{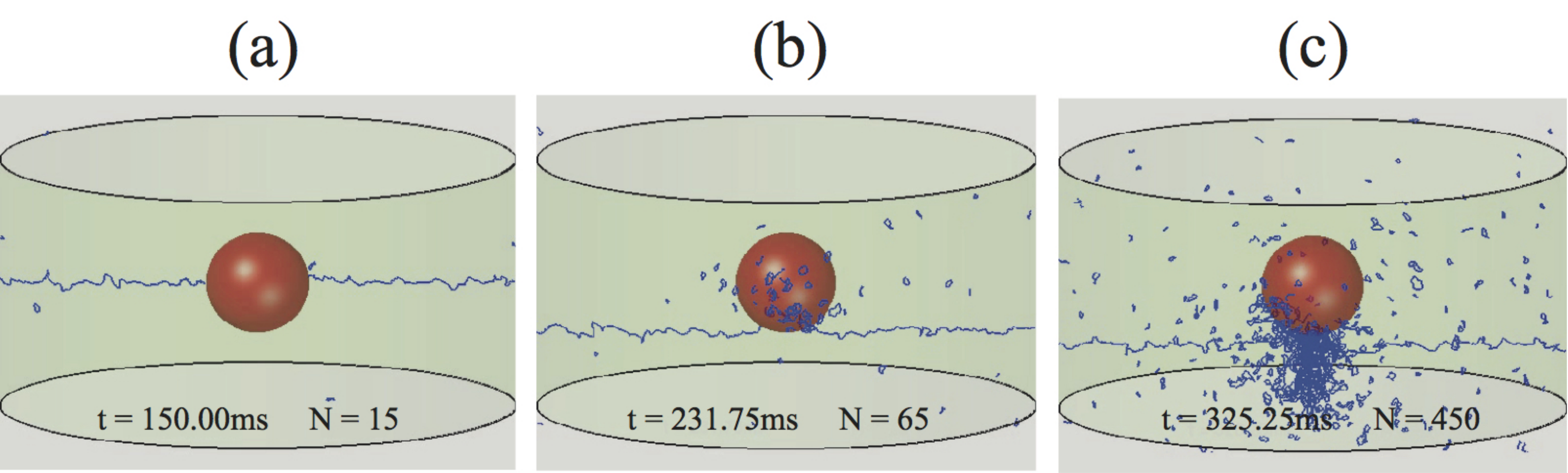}
  \caption
  {
Evolution of the vortex line near a sphere of radius
100~$\mu$m in an oscillating superflow of 150 mm$^{-1}$ at 200~Hz.
 [Reprinted figure with permission from 
 \href{https://doi.org/10.1103/PhysRevB.75.064502}{R. H\"anninen, M. Tsubota, and W. F. Vinen, Phys. Rev. B {\bf 75}, 064502 (2007)}.
 Copyright (2007) by the American Physical Society.]
 \label{Hanninen}
 }
\end{figure}

The observed critical velocities in superfluid $^4$He were in the range from 1 mm/s to approximately 200 mm/s and much lower than the Landau critical velocity of approximately 50 m/s. Hence, the transition to turbulence should come not from the intrinsic nucleation of vortices, but from the amplification of remnant vortices \cite{Awschalom}\footnote{This scenario is applicable only to $^4$He. The transition to turbulence observed in $^3$He-B  \cite{Bradley05b,Bradley06} can be related to the intrinsic nucleation of vortices, which is known to occur near a solid surface at very small velocities, on the order 4 mm/s, depending on the surface roughness \cite{Ruutu97}.}. The numerical simulation by the VF model shows such behavior \cite{Hanninen07}. Figure \ref{Hanninen} shows how the remnant vortices that are initially attached to a sphere develop into turbulence under an oscillating flow. A smooth solid sphere of radius 100~$\mu$m was placed in a cylindrical vessel filled with superfluid $^4$He. Generally, we do not know the geometry of remnant vortices, but they were assumed to initially extend between the sphere and the vessel wall. When an oscillating superflow of 150 mm$^{-1}$ at 200~Hz was applied, Kelvin waves resonant with the flow were gradually excited along the remnant vortices. The amplitude of the Kelvin waves grew large enough to lead to self-reconnection and the emission of small vortex rings. These vortices gathered around the stagnation points, repeatedly reconnected, and were amplified by the flow, eventually developing into a vortex tangle. This simulation likely captured the essence of what occurred in the experiments. Since the surface of the sphere is assumed to be smooth, however, this simulation does not explain the observed critical velocity, hysteresis, or lifetime of turbulence \cite{Yano10}. We discuss the boundary conditions of superfluid in Sec. 4.1. 

\subsection{Formulation and numerical scheme}
If we address the dynamics of a fluid at a scale much larger than the vortex core size, vortices can be regarded as filaments passing through a fluid.
This is known as the vortex filament (VF) model.
The core size of the quantized vortex in superfluid $^4$He is $0.1 ~\rm{nm}$, and the core is stable. Thus, the VF model is suitable for QT consisting of quantized vortices.
Indeed, the VF model has played an important role in study of QT \cite{schwarz85,schwarz88,barenghi01}.
In this subsection, we introduce the VF model for quantized vortices.

\subsubsection{Motion of vortex filament at 0 K} 
First, we find a velocity field produced by vortex filaments.
We regard vortex filaments as lines in three-dimensional  space and express the topological objects using differential geometry.
We introduce a parameter $\xi$ as a one-dimensional coordinate along those lines and specify a point on the vortex lines at time $t$ using vector ${\bm s}(\xi,t)$.
The vector ${\bm s}' \equiv \partial {\bm s}/\partial \xi$ is a unit vector along the vortex lines.
Moreover, ${\bm s}'' \equiv \partial^2 {\bm s}/\partial \xi^2$ is a normal vector to the vector ${\bm s}'$, and the magnitude $|{\bm s}''|$ is a curvature $R^{-1}$ with the curvature radius $R$.
Figure \ref{filament.pdf} is a schematic of the vortex filament.
We suppose that the superfluid component is incompressible:
\begin{equation}
  \nabla \cdot {\bm v}_{\rm s} = 0.
  \label{eq:comp_vs}
\end{equation}
In the VF model, the superfluid vorticity ${\bm \omega}_{\rm s}$ is completely localized at positions of vortex filaments, so that
\begin{equation}
  {\bm \omega}_{\rm s}({\bm r},t) \equiv \nabla \times {\bm v}_{\rm s} = \kappa \int_{\mathcal L} d\xi {\bm s}'(\xi,t) \delta({\bm r}-{\bm s}(\xi,t)),
  \label{eq:vor_vs}
\end{equation}
where the integral path ${\mathcal L}$ represents curves along all vortex filaments.
From these two equations, we obtain the Biot--Savart law for the superfluid velocity:
\begin{equation}
  {\bm v}_{\rm s}({\bm r},t) 
  = \frac{1}{4\pi} \int {\bm \omega}_{\rm s}({\bm x},t) \times \frac{{\bm x} -{\bm r}}{|{\bm x} -{\bm r}|^3} d{\bm x}
  = \frac{\kappa}{4\pi} \int_{\mathcal L} \frac{({\bm s}_1 - {\bm r}) \times d{\bm s}_1}{|{\bm s}_1 - {\bm r}|^3}.
\end{equation}
This law determines the superfluid velocity filed via the arrangement of the vortex filaments.

\begin{figure}
  \centering
  \includegraphics[width=0.4\textwidth]{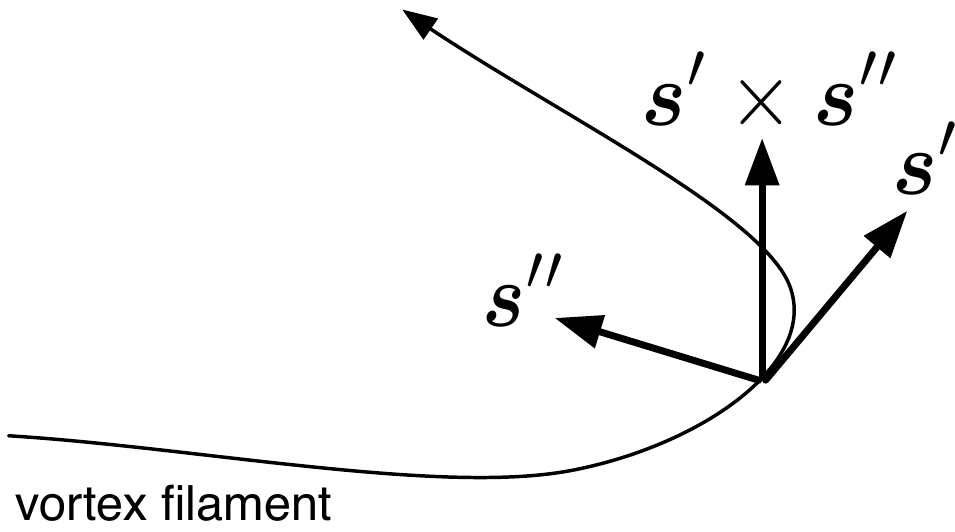}
  \caption
  {
  Schematic of the vortex filament and the triad vectors characteristic of its structure.
  }
  \label{filament.eps}
\end{figure}

If the normal fluid component is negligible and the system is free from boundaries, the vortex filaments move with the superfluid component because of Helmholtz's vortex theorem.
Thus, by calculating the superfluid velocity on the vortex filament, we can obtain the dynamics of a vortex filament.
The velocity ${\bm v}_{\rm s,\omega} ({\bm s},t)$ of the point ${\bm r} = {\bm s}(\xi,t)$ on the filaments is written by
\begin{equation}
  {\bm v}_{\rm s, \omega} ({\bm s},t) = \frac{\kappa}{4\pi} \int_{\mathcal L} \frac{({\bm s}_1 - {\bm s}) \times d{\bm s}_1}{|{\bm s}_1 - {\bm s}|^3}
  = \frac{\kappa}{4\pi} \int_{\mathcal L} \frac{[{\bm s}_1 (\xi,t)-{\bm s}(\xi_0,t)] \times {\bm s}'_1 (\xi,t)}{|{\bm s}_1 (\xi,t) - {\bm s}(\xi_0,t)|^3} d\xi,
\end{equation}
where $\partial {\bm s}_1 / \partial \xi \equiv {\bm s}_1 ' (\xi,t)$.
This integral diverges as ${\bm s}_1 \to {\bm s}$ because the core structure of the quantized vortices is neglected in the VF model.
We avoid this divergence by splitting the integral path into two parts, namely the neighborhood of the point ${\bm s}$ and the other path ${\mathcal L}'$.
The Taylor expansions of ${\bm s}$ and ${\bm s}'$ are \cite{arms}
\begin{eqnarray*}
  {\bm s}(\xi) &\approx& {\bm s}(\xi_0) + (\xi - \xi_0){\bm s}'(\xi_0) + \frac{(\xi-\xi_0)^2}{2} {\bm s}''(\xi_0), \\ 
  {\bm s}'(\xi) &\approx& {\bm s}'(\xi_0) + (\xi - \xi_0){\bm s}''(\xi_0). \nonumber
\end{eqnarray*}
Replacing $\xi-\xi_0$ with $\xi$, we obtain
\begin{eqnarray}
  {\bm v}_{\rm s,\omega}({\bm s},t) &=& \frac{\kappa}{4\pi} \int_{a} ^{R_1} \frac{[{\bm s}(\xi,t)-{\bm s}(\xi_0,t)] \times {\bm s}'(\xi,t)} {|{\bm s}(\xi,t) - {\bm s}(\xi_0,t)|^3} d\xi + \frac{\kappa}{4\pi} \int_{\mathcal L'} \frac{({\bm s}_1 - {\bm s}) \times d{\bm s}_1}{|{\bm s}_1 - {\bm s}|^3} \nonumber \\ 
  &=& \beta {\bm s}' \times {\bm s}''+ \frac{\kappa}{4\pi} \int_{\mathcal L'} \frac{({\bm s}_1 - {\bm s}) \times d{\bm s}_1}{|{\bm s}_1 - {\bm s}|^3},
\end{eqnarray}
where $\beta \equiv (\kappa/4\pi) \ln (R_1/a)$.
The velocity ${\bm v}_{\rm s }$ is composed of two components, namely a local term produced by the neighborhood of ${\bm s}$
\begin{equation}
  {\bm v}_{\rm s,local} =  \beta {\bm s}' \times {\bm s}'',
\end{equation}
and a nonlocal term produced by distant vortices
\begin{equation}
  {\bm v}_{\rm s,nonlocal} = \frac{\kappa}{4\pi} \int_{\mathcal L'} \frac{({\bm s}_1 - {\bm s}) \times d{\bm s}_1}{|{\bm s}_1 - {\bm s}|^3}.
\end{equation}
The local term is proportional to the local curvature because $|{\bm s}''| = R^{-1}$, and this term is known as a self-induced velocity.
Schwarz assessed that the nonlocal term is not important for homogeneous quantum turbulence and performed numerical simulations of quantum turbulence by neglecting the nonlocal term, which is the LIA \cite{schwarz88}.
However, the numerical study of Adachi {\it et al.} showed that the nonlocal term plays an important role even for homogeneous quantum turbulence \cite{adachi}.

\subsubsection{Boundary conditions of quantized vortices}
It is very important to treat properly boundary conditions of quantized vortices. 
Since the vortex core of a quantized vortex in superfluid $^4$He is very thin of the order of atomic scale, any solid surface can be rough for it.

The general procedure of treating boundary conditions of superfluid and quantized vortices is the following.
The superfluid component can slip on solid surfaces due to its superfluidity.
There is no flow into solid surfaces:
\begin{equation}
  {\bm v}_{\rm s} \cdot {\bm {\hat n}} = 0 ~~~ (\mbox{on solid surfaces}),
\end{equation}
where ${\bm {\hat n}}$ is a unit vector perpendicular to the surface.
Thus, solid surfaces induce an additional velocity ${\bm v}_{\rm s,b}$ to satisfy the boundary conditions; then, the superfluid velocity becomes
\begin{equation}
  {\bm v}_{\rm s} = {\bm v}_{\rm s,\omega} + {\bm v}_{\rm s,b}.
\end{equation}
To determine the velocity ${\bm v}_{\rm s,b}$ induced by solid surface, we solve Eq. (\ref{eq:comp_vs}) and (\ref{eq:vor_vs}) under the boundary conditions.
If the solid surface is a smooth flat plane, it is easy and straightforward to obtain ${\bm v}_{\rm s,b}$ by using the method of image vortices.
The induced velocity ${\bm v}_{\rm s,b}$ is produced by image vortices, which are symmetrical with the surfaces, and has the opposite circulation.

However, realistic solid surface is not smooth. This case makes the procedure of obtaining ${\bm v}_{\rm s,b}$ so complicated. 
Schwarz obtained analytically ${\bm v}_{\rm s,b}$ when there was a hemispherical pinning site on a flat surface \cite{schwarz85} and applied this method to 
study the vortex dynamics in a channel with rough surface \cite{schwarz92}. We will come back to this problem in Chap. 4. 

Finally, we find that the equation of motion of vortex filaments at 0 K is given by
\begin{equation}
  \dot{\bm s}_0 = \beta {\bm s}' \times {\bm s}'' + \frac{\kappa}{4\pi} \int_{\mathcal L '} \frac{({\bm s}_1 - {\bm s})\times d{\bm s}_1}{|{\bm s}_1 - {\bm s}|^3} + {\bm v}_{\rm s,b} + {\bm v}_{\rm s,a}.
\end{equation}
The additional velocity ${\bm v}_{\rm s,a}$ is a external irrotational velocity field, which is not caused by quantized vortices;
for example, this velocity is a uniform flow in a thermal counterflow.

\subsubsection{Motion of the vortex filament at finite temperatures}
If quantized vortices move at finite temperatures, mutual friction, which is the interaction between the normal fluid and the vortex core, works for the vortex dynamics.
By splitting the drag force into tangent and vertical components with the filaments, the drag force per unit length can be written as
\begin{equation}
  {\bm f}_{\rm D} = \gamma_0 \kappa {\bm s}' \times [ {\bm s}' \times (\dot{\bm s} - {\bm v}_{\rm n})] - \gamma_0 ' \kappa {\bm s}' \times (\dot{\bm s} - {\bm v}_{\rm n}),
  \label{eq:drag}
\end{equation}
where $\dot{\bm s}$ is a velocity of the vortex filaments at a finite temperature.
Moreover, the vortex filaments are affected by Magnus forces:
\begin{equation}
  {\bm f}_{\rm M} = \rho_{\rm s} \kappa {\bm s}' \times (\dot{\bm s} - \dot{\bm s}_0).
  \label{eq:magnus}
\end{equation}
Since the inertia of quantized vortex is negligible, the sum of these forces is equal to zero:
\begin{equation}
  {\bm f}_{\rm M} + {\bm f}_{\rm D} = {\bm 0}.
\end{equation}
Hence, we obtain the equation of motion of quantized vortices (see Appendix A):
\begin{equation}
  \dot{\bm s} = \dot{\bm s}_0 + \alpha {\bm s}' \times ({\bm v}_{\rm n} - \dot{\bm s}_0) - \alpha' {\bm s}' \times \left[ {\bm s}' \times ({\bm v}_{\rm n} - \dot{\bm s}_0) \right],
  \label{eq:vor_fric}
\end{equation}
where 
\begin{equation}
  \alpha = \frac{\rho_{\rm s}\kappa\gamma_0}{\gamma_0^2 + (\rho_{\rm s} \kappa - \gamma_0')^2}, ~~~
  \alpha' = \frac{\gamma_0^2 - \gamma_0' (\rho_{\rm s}\kappa - \gamma_0')}{\gamma_0^2 + (\rho_{\rm s} \kappa - \gamma_0')^2}.
\label{mutual friction}
\end{equation}
Experimental values of $\alpha$ and $\alpha'$ are shown in Table \ref{tab:friction}.

\begin{table}
  \begin{center}
  \caption{Mutual friction coefficients as a function of temperature \cite{schwarz85,barenghi83}．}
  \label{tab:friction}
  \begin{tabular}{ccc}
    \hline
    \hline
        $T$ (K)   & $\alpha$    & $\alpha'$  \\ \hline
            $1.0$ & $0.006$       & $0.003$ \\
            $1.1$ & $0.012$       & $0.006$ \\
            $1.2$ & $0.023$       & $0.011$ \\
            $1.3$ & $0.036$       & $0.014$ \\
            $1.4$ & $0.052$       & $0.017$ \\
            $1.5$ & $0.073$       & $0.018$ \\
            $1.6$ & $0.098$       & $0.016$ \\
            $1.7$ & $0.127$       & $0.012$ \\
            $1.8$ & $0.161$       & $0.008$ \\
            $1.9$ & $0.21$         & $0.009$ \\
            $2.0$ & $0.29$         & $0.011$ \\
            $2.05$ & $0.36$       & $0.003$ \\
            $2.10$ & $0.50$       & $-0.030$ \\
            $2.15$ & $1.09$       & $-0.27$   \\
    \hline
    \hline
  \end{tabular}
  \end{center}
\end{table}

We consider the characteristic motion of vortex filaments with mutual friction in a thermal counterflow.
By using the LIA,  we consider the motion of a vortex ring with radius $R$.
The term of $\alpha$ in Eq. (\ref{eq:vor_fric}) becomes
\begin{equation}
  \alpha {\bm s}' \times ({\bm v}_{\rm n} - {\bm v}_{\rm s,a} - \beta {\bm s}' \times {\bm s}'').
\end{equation}
If we consider the case that the applied velocity ${\bm v}_{\rm n} - {\bm v}_{\rm s,a}$ is parallel to the self-induced velocity ${\bm v}_{\rm s,local} = \beta {\bm s}' \times {\bm s}''$, this vector inverts its direction at radius
\begin{equation}
  R_0 \approx \frac{\beta}{|{\bm v}_{\rm n} - {\bm v}_{\rm s,a}|},
\end{equation} 
since $|{\bm s}' \times {\bm s}''| = R^{-1}$.
Thus, the vortex ring expands outward if $R>R_0$ and decreases if $R<R_0$.
The term with $\alpha$ tends to grow with large structures and decrease with small structures.
When  the applied velocity is antiparallel to the self-induced velocity, any vortex ring shrinks.

\subsubsection{Reconnection of vortex filaments}
\begin{figure}
  \centering
  \includegraphics[width=0.5\textwidth]{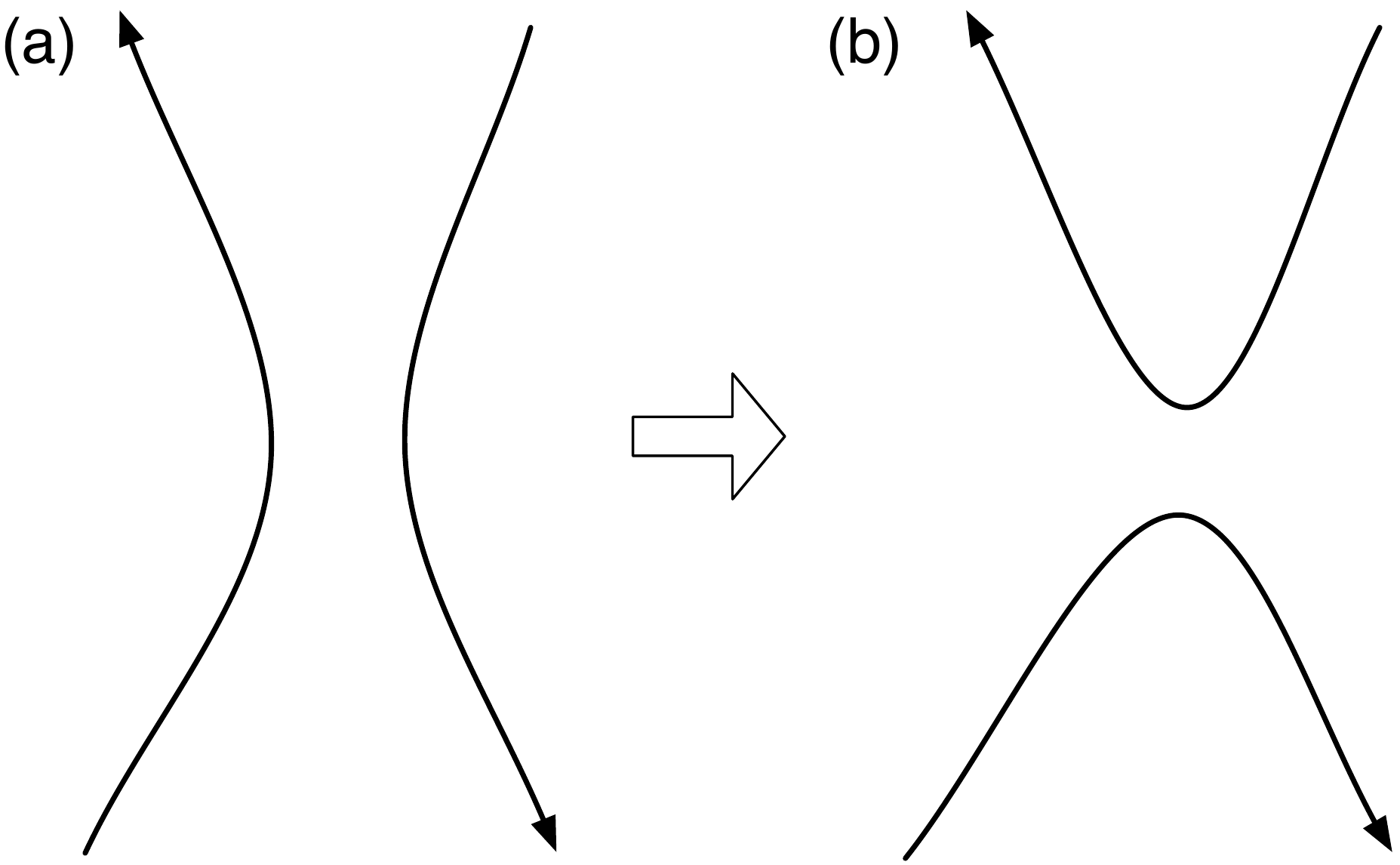}
  \caption
  {
    Schematic of the reconnection of quantized vortices.
    (a) Two vortices before reconnection.
    The vortices will approach and contact each other.
    (b) The vortices reconnect at the contact point.
  }
  \label{reconnect.eps}
\end{figure}

When two quantized vortices approach and contact each other, what happens?
From analogies with vortex dynamics in the Navier--Stokes equation, many researchers suppose that the vortices will reconnect, as shown in Fig. \ref{reconnect.eps}.
Indeed, numerical studies of the Gross--Pitaevskii equation showed that quantized vortices can reconnect \cite{koplik}.

The VF model cannot describe the reconnection process because this model neglects the vortex core structure.
Hence, we must introduce some artificial procedures to simulate vortex reconnection.
For instance, when two vortices approach within a critical distance, which is often the spatial resolution for a vortex filament, the vortices are artificially reconnected.
The sensitivity of the reconnection algorithm was analyzed by Baggaley \cite{baggaley123}.
They performed numerical simulation of QT with different reconnection models and showed that results of full Biot--Savart integral were not sensitive to those models.

\subsubsection{Coupled dynamics of the two-fluid model}
As explained above, the dynamics of vortex filaments are expressed by
\begin{equation}
  \dot{\bm s} = \dot{\bm s}_0 + \alpha {\bm s}' \times ({\bm v}_{\rm n} - \dot{\bm s}_0) - \alpha ' {\bm s}' \times [{\bm s}' \times ({\bm v}_{\rm n} - \dot{\bm s}_0)],
\end{equation}
where 
\begin{equation}
  \dot{\bm s}_0 = \beta {\bm s}' \times {\bm s}'' + \frac{\kappa}{4\pi} \int_{\mathcal L'} \frac{ ({\bm s}_1 - {\bm s})\times d{\bm s}_1}{|{\bm s}_1 - {\bm s}|^3} + {\bm v}_{\rm s,b} + {\bm v}_{\rm s,a}.
\end{equation}
The terms with $\alpha$ and $\alpha'$ arise from the mutual friction of the two fluids.
The normal fluid affects the superfluid through these terms.
On the other hand, the dynamics of the normal fluid component are described by the forced Navier--Stokes equation\cite{barenghi83,idowu00,kivotides00,kivotides07}
\begin{equation}
  \frac{\partial {\bm v}_{\rm n}}{\partial t} + ({\bm v}_{\rm n} \cdot \nabla){\bm v}_{\rm n} = \frac{1}{\rho_{\rm n}} \nabla P + \nu_{\rm n} \nabla^2 {\bm v}_{\rm n} + \frac{1}{\rho_{\rm n}} {\bm F}_{\rm ns},
  \label{eq:vn_navier}
\end{equation}
where
\begin{equation}
  {\bm F}_{\rm ns} ({\bm r})= \frac{\rho_{\rm s} \kappa}{\Omega' ({\bm r})} \int_{{\mathcal L}'({\bm r})} d\xi \lbrace \alpha' {\bm s}' \times ({\bm v}_{\rm n} - \dot{\bm s}_0) + \alpha {\bm s}' \times [{\bm s}' \times ({\bm v}_{\rm n} - \dot{\bm s}_0)] \rbrace.
\end{equation}
Here, the effective pressure gradient $\nabla P$ is defined by
\begin{equation}
  \nabla P = - \frac{\rho_{\rm n}}{\rho} \nabla p - \rho_{\rm s} s \nabla T,
\end{equation}
and $\nu_{\rm n}=\eta_{\rm n}/\rho_{\rm n}$ is the kinematic viscosity of the normal fluid, $\Omega ' ({\bm r})$ is a small volume at position ${\bm r}$, and ${\mathcal L}'({\bm r})$ represents vortex filaments in volume $\Omega ' ({\bm r})$.
These equations show that the dynamics of the two fluids are coupled through mutual friction.
The normal fluid is affected by the tangle of quantized vortices through ${\bm F}_{\rm ns}$, whereas the quantized vortices are also affected by normal flow through the mutual friction terms with $\alpha$ and $\alpha'$.
Most numerical simulations of the VF model have never solved Eq. (\ref{eq:vn_navier}) but use a prescribed profile for a normal flow.
The main reason is that the coordinates of the two fluids are different.
The VF model for a superfluid is described by the Lagrange coordinate along the vortex filament, whereas the normal fluid velocity is expressed by the Euler coordinate.
To investigate the two-fluid coupled dynamics, we must overcome these difficulties in solving both of the Navier--Stokes equation and the VF model.

\subsubsection{Vinen's equation}
When we study turbulence, it is important to consider statistical quantities. 
The important statistical quantity characterizing quantized vortices would be the vortex line density (VLD), defined by
$L=(1/\Omega) \int_{\cal L}d\xi$, where the integral is performed over all vortices in the sample volume $\Omega$. 
To understand the experimental results of counterflow of relative velocity ${\bm v}_{\rm ns}={\bm v}_{\rm n}-{\bm v}_{\rm s}$, Vinen proposed an equation for the evolution of $L(t)$, which we call Vinen's equation:
\begin{equation}
\frac{dL}{dt}=\frac{\chi_1 B \rho_n}{2\rho} |{\bm v}_{\rm ns}|L^{3/2}-\chi_2 \frac{\kappa}{2\pi}L^2,
\end{equation}
where $\chi_1$ is a constant, and $B$ and $\chi_2$ are temperature-dependent parameters \cite{Vinen57c}. 
The first term represents the energy injection from the normal fluid to the vortices. 
The second term denotes the energy dissipation of vortices.
In the steady state, the VLD is given by
\begin{equation}
L=\gamma^2v_{\rm ns}^2,
\label{Lvns}
\end{equation}
where $\gamma=\pi B \rho_{\rm n} \chi_1/ \kappa \rho \chi_2$ is a temperature-dependent parameter.
According to experiments, a small constant parameter $v_0$ is required: $L = \gamma^2 (v_{\rm ns} - v_0)^2$.
The behavior of Eq. ({\ref{Lvns}) is consistent with a large number of observations and has also used as a benchmark of numerical simulation. 

\subsubsection{Numerical method}
In the numerical simulation, we regard vortex filaments as a series of discrete points, where those points are connected by line segments.
Each point is specified by number $i=1, 2, 3, \cdots$.
A velocity $\delta {\bm v}^{j}_{\rm s,nonlocal}$ at a point ${\bm s}_{i}$ on the vortex lines, which is produced by a line segment ${\bm s}_{j+1} - {\bm s}_{j}$, is written by \cite{schwarz85,barenghi01}:
\begin{equation}
  \delta {\bm v}^{j}_{\rm s,nonlocal} ({\bm s}_{i}) = \frac{\kappa}{4 \pi} \frac{ (R_{j} + R_{j+1}) {\bm R}_{j} \times {\bm R}_{j+1}}{ R_{j} R_{j+1} (R_{j}  R_{j+1} + {\bm R}_{j} \cdot {\bm R}_{j+1})},
\end{equation}
where ${\bm R}_{j} = {\bm s}_{j} - {\bm s}_{i}$, ${\bm R}_{j+1} = {\bm s}_{j+1} - {\bm s}_{i}$.
A nonlocal term ${\bm v}_{\rm s,nonlocal}$ is $\sum_{j} \delta {\bm v}_{\rm s,nonlocal}$, except for the adjacent line segments ${\bm l}_{+} = {\bm s}_{i+1} - {\bm s}_{i}$ and ${\bm l}_{-} = {\bm s}_{i} - {\bm s}_{i-1}$.
A local term ${\bm v}_{\rm s,local}$ on ${\bm s}_{i}$ is a velocity produced by those adjacent segments ${\bm l}_{+}$ and ${\bm l}_{-}$, as shown in Fig. \ref{local.eps}.
By drawing a circle passing through those adjacent points ${\bm s}_{i+1}$, ${\bm s}_{i}$, ${\bm s}_{i-1}$, we obtain differential coefficients (see Appendix B) \cite{schwarz85}:
\begin{equation}
  {\bm s}'_{i} = d_i^{+} {\bm l}_{+} + d_i^{-} {\bm l}_{-}, ~{\bm s}''_{i} = c_i^{+} {\bm l}_{+} - c_i^{-} {\bm l}_{-},
\label{Numerics1}
\end{equation}
where
\begin{equation}
  d_i^{\pm} = \frac{l_{\mp} ^2}{ |{\bm l}_{+} l_{-} ^2 + {\bm l}_{-} l_{+} ^2| },~
  c_i^{\pm} = \frac{a_{\pm}}{ |a_{+} {\bm l}_{+} - a_{-} {\bm l}_{-} |^2 },
\end{equation}
with
\begin{equation}
  a_{\pm} = \frac{1}{2} \frac{l_+^2 l_-^2 + l_{\mp}^2 ({\bm l}_+ \cdot {\bm l}_-)}{l_+^2 l_-^2 - ({\bm l}_+ \cdot {\bm l}_- )^2}.
 \label{Numerics2}
\end{equation}
Hence, a local term is given by
\begin{equation}
  {\bm v}_{\rm s,local} ({\bm s}_i) = \beta_i {\bm s}_{i} ' \times {\bm s}_{i} '' = - \beta_i (d_i^+ c_i^-+ d_i^- c_i^+)({\bm l}_{+} \times {\bm l}_{-}),
\end{equation}
where
\begin{equation}
  \beta_i = \frac{\kappa}{4\pi} \ln \left[ \frac{2(l_+ l_-)^{1/2}}{e^{1/2} a} \right].
\end{equation}
Here, $a$ is the core size of quantized vortex.

\begin{figure}
  \centering
  \includegraphics[width=0.6\textwidth]{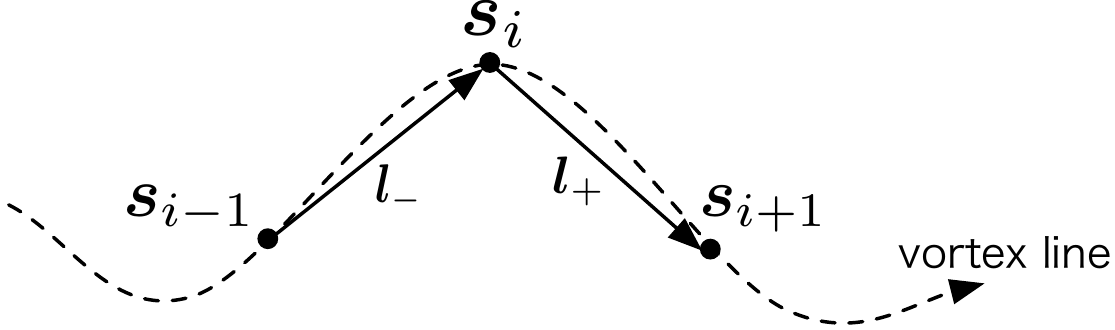}
  \caption
  {
  Adjacent points ${\bm s}_{i-1}$, ${\bm s}_{i}$, ${\bm s}_{i+1}$, and the vectors between those points.
  }
  \label{local.eps}
\end{figure}

Representative points of vortex lines are added or removed to adjust the spatial resolution.
Typically, when the spacing between ${\bm s}_{i}$ and ${\bm s}_{i+1}$ becomes larger than some upper limit $h_{\rm max}$, a new point is added between these two points.
Here, the new point has a principal normal vector $({\bm s}''_{i} + {\bm s}''_{i+1})/2$. 
Alternatively, when a spacing becomes smaller than some lower limit $h_{\rm min}$, that point is removed.
That is, $h_{\rm min}$ is a numerical space resolution $\delta \xi$.

In a typical simulation, a vortex loop composed of a few points is removed, because the number of points is too low to describe vortex motion accurately.
Similarly, small vortices attached to solid surfaces are removed.
These numerical procedures can be interpreted physically as energy dissipation at small scales, such as emission of sound waves.

\subsection{Homogeneous thermal counterflow}
\label{sec:3}
This subsection reviews simulations of homogeneous counterflow of the LIA by Schwarz \cite{schwarz88} and of the full Biot--Savart model by Adachi {\it et al.}\cite{adachi}.

Starting with several remnant vortices under thermal counterflow, Schwarz numerically studied how these vortices developed into a vortex tangle.
The tangle was self-sustained by competition between excitation due to the applied flow and dissipation through mutual friction. 
By combining the numerical results with dynamical scaling, Schwarz could calculate the VLD as a function of temperature and $|{\bm v}_{ns}|$, which agreed well with typical observations including the VLD of Eq. ({\ref{Lvns}). This was a great accomplishment in numerical research.
However, his calculation had serious difficulties when the system was subjected to periodic boundary conditions along all three directions.
As the vortices developed, they gradually began to form layered structures perpendicular to the counterflow and eventually degenerated. 
This behavior comes from the LIA, which is not a realistic approximation. 
In order to address this difficulty, an unphysical, artificial mixing procedure was employed, in which half the vortices are randomly selected to be rotated by 90$^\circ$ around the axis defined by the flow velocity. 
This method enables the steady state to be sustained under periodic boundary conditions.

 \begin{figure}
  \centering
  \includegraphics[width=1.0\textwidth]{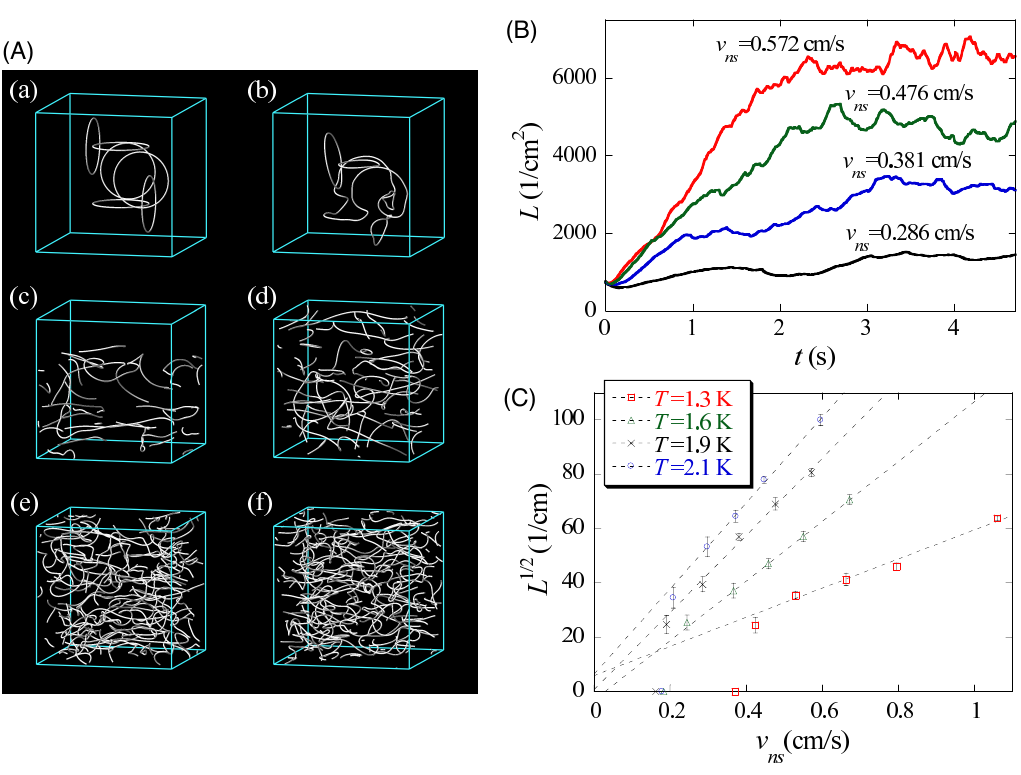}
  \caption
  {
  (A)
  Development of a vortex tangle by the full Biot--Savart calculation in a periodic box with a size of 0.1 cm. 
  Here, the temperature is $T=1.9\,{\rm K}$, and the counterflow velocity $v_{ns}=0.572\,{\rm cm/s}$ is along the vertical axis: 
  (a) $t=0\,{\rm s}$, (b) $t=0.05\,{\rm s}$, (c) $t=0.5\,{\rm s}$, (d) $t=1.0\,{\rm s}$, (e) $t=3.0\,{\rm s}$, and (f) $t=4.0\,{\rm s}$. 
  (B) Vortex line density as a function of time.
  (C) Mean vortex line density in statistically steady states as a function of the counterflow velocity $v_{\rm ns}$.
  The error bars are the standard derivation.
  [Reprinted figures with permission from 
  \href{https://doi.org/10.1103/PhysRevB.81.104511}
  {H. Adachi, S. Fujiyama, and M. Tsubota, Phys. Rev. B {\bf 81}, 104511 (2010)}.
  Copyright (2010) by the American Physical Society.]
  (Color figure online)
  }
  \label{Adachi} 
\end{figure}

 Adachi {\it et al.} performed numerical simulations using the full Biot--Savart law under periodic boundary conditions and succeeded in obtaining a statistically steady state without any unphysical procedures \cite{adachi}.
 Figure \ref{Adachi}(A) shows a typical result of the time evolution of the vortices, whose VLD grows as shown in Fig. \ref{Adachi}(B). 
 The obtained steady states almost satisfy the relation of Eq. (\ref{Lvns}) when $v_{ns}$ and $L$ are relatively large, as shown in Fig. \ref{Adachi}(C).
The results quantitatively agree with the typical experimental observations \cite{childers}.
Thus, the LIA does not work well for turbulence, and vortex interaction is essential for studying QT.  

\subsection{Inhomogeneous thermal counterflow in a realistic channel}
Modern numerical studies of counterflow move from homogeneous systems in bulk to inhomogeneous systems in a channel for two primary reasons. The first reason is old and traditional. Previous experiments \cite{tough} found that the turbulent state depended strongly on the aspect ratio of the cross section of a channel,  and different states such as the T1, T2 and T3 states were observed. These states must be attributable to the boundary conditions, so we have to consider the systems in a channel. The second reason is based on the recent visualization experiments.  Marakov {\it et al.} \cite{marakov} recently observed that the normal fluid profile showed different states such as Poisueille, tail-flattened, and turbulent states. Understanding these states also requires numerical studies of the system in a channel.  

In this subsection, we review numerical studies related to the T1 state of quantum turbulence.
Most previous studies were performed by using a prescribed nonuniform velocity profile for a normal fluid.
These studies observed that the characteristic dynamics of inhomogeneous vortices were much different from the homogeneous cases.
Later numerical studies using the VF model were performed with the Navier--Stokes equation for laminar flow and addressed two-fluid coupled dynamics for thermal counterflow.
Here, we review the two-dimensional (2D) simulation of quantized vortices for counterflow quantum turbulence, which gives us important knowledge even though it is essentially different from three-dimensional (3D)  quantum turbulence.

\subsubsection{Poiseuille normal flow between parallel two plates}
Baggaley {\it et al.} performed numerical simulations of thermal counterflow between two parallel plates, as shown in Fig. \ref{baggaley1.eps} \cite{baggaley13,baggaley15}.
The flow is along the $x$-direction.
The calculations are performed in a cuboid of size  ${\mathcal D}_x \times {\mathcal D}_y \times {\mathcal D}_z = 0.2 ~{\rm cm} \times 0.1 ~{\rm cm} \times 0.1 ~{\rm cm}$ with solid boundaries in the $y$-direction and periodic boundaries in the $x$- and $z$-directions.
They used a laminar Poiseuille profile for the normal fluid velocity
\begin{equation}
  {\bm v}_{\rm n} (y) = \left( 1-\frac{y^2}{h^2} \right) U_{\rm c} {\bm e}_x,
\end{equation}
where $h={\mathcal D}_y /2$ is half of the channel width, ${\bm e}_x$ is the unit vector along the $x$-direction, and $U_{\rm c}$ is the velocity at the center of the channel.
The assumption of the laminar profile for normal flow is valid if its Reynolds number is small and if the mutual friction is too small to change the velocity profile of the normal fluid.

\begin{figure}
  \centering
  \includegraphics[width=1.0\textwidth]{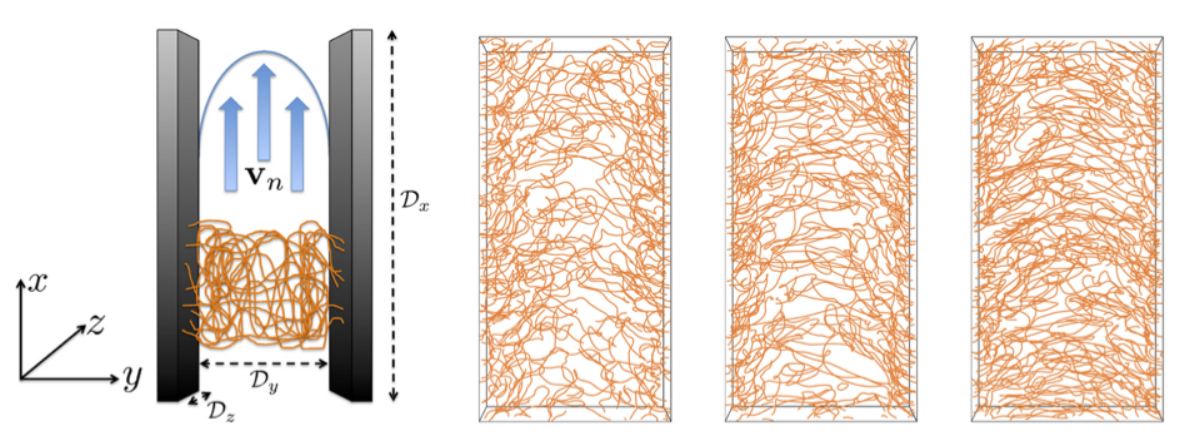}
  \caption
  {
  ({\it far left})
  Schematic of the numerical box used in the simulation of Baggaley and Laurie \cite{baggaley15}.
  They have solid boundaries in the $y$-direction, and periodic boundaries in the $x$- and $z$-directions.
  The orange lines denote vortex filaments.
  Snapshots of the vortex tangle in the statistically steady state with $T=1.3 ~\rm{K}$, $\overline{v}_{\rm ns} = 1.5 ~\rm{cm/s}$ ({\it second left}); $T=1.6 ~\rm{K}$, $\overline{v}_{\rm ns} = 1.13 ~\rm{cm/s}$ ({\it second right}); and $T=1.9 ~\rm{K}$, $\overline{v}_{\rm ns} = 0.90 ~\rm{cm/s}$ ({\it far right}).
  [Reprinted figure with permission of Springer from A. W. Baggaley and J. Laurie, Thermal Counterflow in a Periodic Channel with Solid Boundaries, J. Low Temp. Phys. {\bf 178}, Issue 1, 35 (2015). Copyright by the Springer Science+Business Media New York 2014.]
  (Color figure online)
  }
  \label{baggaley1.eps}
\end{figure}

Moreover, Baggaley {\it et al.} used a ``frozen'' turbulent profile for the normal fluid velocity in the thermal counterflow.
Here, the ``frozen'' turbulent profile means the velocity profile of a snapshot of the stationary turbulence obtained by simulation of the ordinary Navier--Stokes equation.
Simulations of the laminar and ``frozen'' turbulent normal flow were compared, and the T1-T2 transition was shown to be caused by the transition of the normal flow from laminar to turbulent.
Here, we do not mention the study of the ``frozen'' turbulent profile for the T2 state, and focus on the study of the laminar profile for the T1 state.

Figure \ref{baggaley1.eps} shows snapshots of the vortex tangle in the stationary thermal counterflow with the Poiseuille normal fluid velocity viewed along the $z$-axis.
The temperature in the snapshots is $1.3 ~\rm{K}$, $1.6 ~\rm{K}$, and $1.9 ~\rm{K}$ from the left to the right.
We can see higher concentrations of vortex line density, and these structures are never observed in homogeneous thermal counterflow.
By analyzing the coarse-grained vortex line density, Baggaley {\it et al.} argued that the spatial distribution of vortices depends not on the counterflow velocity but on the temperature.
Moreover, the position of the peak VLD was determined by the balance of two competing effects, namely the rate of turbulent diffusion of an isotropic tangle near the channel walls and the rate of quantized vorticity production at the central region.

How is the distribution of quantized vortices related to a boundary layer or a wall-bounded turbulent flow of an ordinary viscous fluid?
In ordinary turbulent flow in a channel, there is a turbulent core region in the center of the channel that has a large velocity fluctuation, while there is a viscous region near the channel walls in which turbulent fluctuations are suppressed by viscosity \cite{davidson,landau,marusic}.
On the other hand, in wall-bounded quantum turbulence, the fluctuations in turbulence become large near the channel wall, while the fluctuation is suppressed in the center of the channel.
Hence, in this article, we can call wall-bounded QT structures an {\it inverse distribution of quantized vortices} that may be characteristic of the T1 state.
The origins of the {\it inverse distribution} may be that the mutual friction term
\begin{equation}
  |\alpha {\bm s}' \times ({\bm v}_{\rm n} - \dot{\bm s}_0)| \sim \alpha \left[ \left( 1-\frac{y^2}{h^2} \right)U_{\rm c} + v_{\rm s,a} \right]
\end{equation}
in Eq. (\ref{eq:vor_fric}) becomes stronger in the central region, which dissipates the QT and pushes the vortices toward the walls.

\subsubsection{Poiseuille and tail-flattened normal flow in a duct}
Yui and Tsubota studied counterflow quantum turbulence in a duct by performing numerical simulations \cite{yui14,yui15}.
The visualization experiment of thermal counterflow by Marakov {\it et al.} showed that the normal fluid velocity distribution becomes a Poiseuille profile, a tail-flattened profile, and a turbulent profile \cite{marakov}, as explained above.
Recall that the Melotte--Barenghi assumption argued that the laminar normal flow corresponds to the T1 state of QT \cite{tough,melotte98}.
Does the T1 state correspond to the Poiseuille or the tail-flattened flow?
We could not answer this problem using simulations with the prescribed normal flow, but it is important to investigate what happens to vortex tangles for these two profiles of the normal fluid.

\begin{figure}
  \centering
  \includegraphics[width=1.0\textwidth]{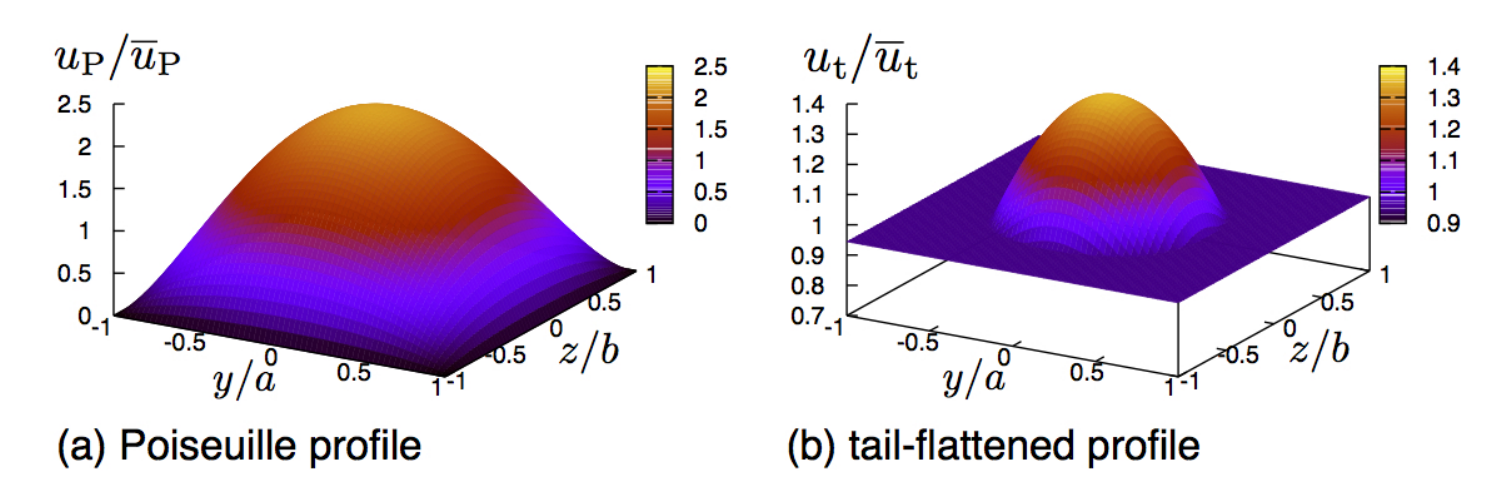}
  \caption
  {
  Laminar velocity profile of the normal fluid component in a duct $2a \times 2b$ used in the numerical study of Yui and Tsubota \cite{yui15}.
  (a) Poiseuille profile $u_{\rm P}$.
  The value is scaled by the mean value $\overline{u}_{\rm P}$.
  (b) Tail-flattened profile modeled by the study $u_{\rm t}$ with $h=0.7$.
  In the central region, the profile becomes Poiseuille.
  (Color figure online)
  }
  \label{yui1.pdf}
\end{figure}

Yui and Tsubota used prescribed profiles for the normal fluid velocity ${\bm v}_{\rm n}$, namely the Poiseuille profile $u_{\rm P}$ and the tail-flattened flow $u_{\rm t}$.
Here, the flow direction is the $x$-direction, and the channel walls are applied to the $y$- and $z$-directions. 
The laminar normal fluid velocity is written as ${\bm v}_{\rm n} = u(y,z) {\bm e}_x$, where the center of the channel is $(y,z)=(0,0)$.
The Poiseuille profile in a duct is
\begin{equation}
  u_{\rm P}(y,z) = u_0 \sum_{m=1,3,5, \cdots}^{\infty} (-1)^{(m-1)/2} \times \left[ 1-\frac{\cosh(m\pi z/2a)}{\cosh(m\pi b/2a)} \right] \frac{\cos(m\pi y/2a)}{m^3},
\end{equation}
where $u_0$ is a normalization factor and $a$ and $b$ are halves of the channel width along the $y$- and $z$-axes, respectively \cite{poiseuille}.
The tail-flattened profile observed by Marakov {\it et al.} \cite{marakov} is modeled as
\begin{equation}
  u_{\rm t} (y,z) =  u_0 \max[u_{\rm P}(y,z), hu_{\rm P}(0,0)],
\end{equation}
where $\max[A,B]$ refers to the larger value of the two arguments, and $0 < h < 1$ is a fitting parameter, which determines the flattened area.
If $h=1$, the profile of $u_{\rm t}$ becomes uniform, whereas if $h=0$, the profile becomes the Poiseuille profile. 
Figure \ref{yui1.pdf} shows the Poiseuille profile and the tail-flattened profile with $h=0.7$.

The simulations are performed in the computing cube $0.1 \times 0.1 \times 0.1 ~\rm{cm^3}$.
Periodic boundary conditions are used along the flow direction, whereas solid boundary conditions are applied for the channel at $y= \pm 0.05 ~\rm{cm}$ and $z= \pm 0.05 ~\rm{cm}$.
Temperatures are $T=1.9 ~\rm{K}$, $1.6 ~\rm{K}$, and $1.3 ~\rm{K}$ for the Poiseuille profile but only $T=1.9 ~\rm{K}$ for the tail-flattened profile.

\begin{figure}
  \centering
  \includegraphics[width=1.0\textwidth]{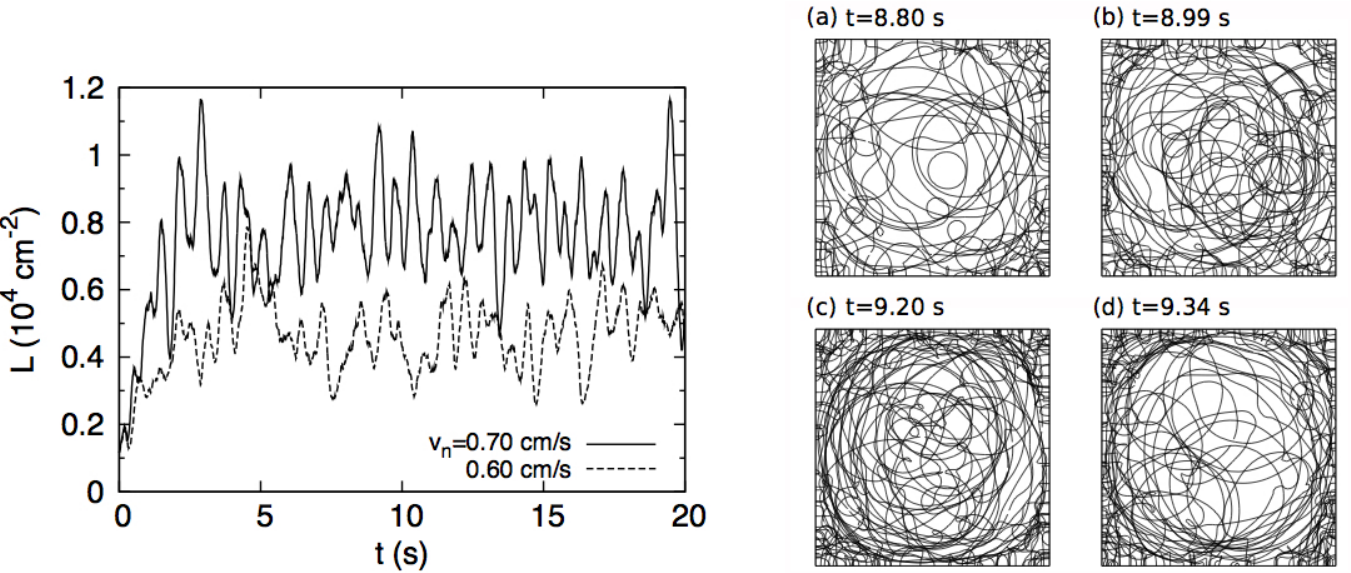}
  \caption
  {
  Results of simulation of Fig. \ref{yui1.pdf} for thermal counterflow with the Poiseuille normal flow in a duct at $T=1.9 ~\rm{K}$.
  ({\it left})
  Vortex line density as a function of time.
  The vortex tangle reaches a statistically steady state.
  ({\it right})
  Snapshots of the vortex tangle in the duct viewed along the flow direction in the stationary state at $T=1.9 ~\rm{K}$, $\overline{v}_{\rm n}=0.7 ~\rm{cm/s}$.
  The snapshots correspond to a local minimum of $L$ (a), the middle of the increase (b), a local maximum (c), and the middle of the decrease (d).
  [Reprinted figures with permission from 
  \href{https://doi.org/10.1103/PhysRevB.91.184504}
  {S. Yui and M. Tsubota, Phys. Rev. B {\bf 91}, 184504 (2015)}.
  Copyright (2015) by the American Physical Society.]
  }
  \label{yui2.pdf}
\end{figure}

Figure \ref{yui2.pdf}({\it left}) shows the vortex line density $L$ as a function of time.
The temperature is $1.9 ~\rm{K}$, and the mean velocity of the normal fluid component is $0.6 ~\rm{cm/s}$ and $0.7 ~\rm{cm/s}$.
The vortex tangle reaches the statistically steady state even if the counterflow is spatially inhomogeneous.
The fluctuations are much larger than those in uniform counterflow \cite{adachi} or those between two parallel plates \cite{baggaley13,baggaley15}.
The period of the oscillation is about $0.7 ~\rm{s}$, and the oscillation consists of four stages (a)--(d).
The vortex configurations in one cycle of the large fluctuation at $T=1.9 ~\rm{K}$, $\overline{v}_{\rm n} = 0.7 ~\rm{cm/s}$ are shown in Fig. \ref{yui2.pdf}({\it right}).
In Fig. \ref{yui2.pdf}(a), corresponding to a local minimum of $L$, the vortices are dilute, and the vortices remain only near the channel walls.
Then, the vortices invade the central region in Fig. \ref{yui2.pdf}(b).
These vortices make reconnections in the central region, and the vortex tangle becomes the local maximum state in Fig. \ref{yui2.pdf}(c).
Eventually, in Fig. \ref{yui2.pdf}(d), strong mutual friction in the central region pushes the vortices toward the walls.
These periodic dynamics produce the long-term oscillation of $L$.

We observe the {\it inverse distribution of quantized vortices} in this case too.
In the stationary state, the distribution temporally fluctuates, while $L$ shows large fluctuations.
This fluctuations of the distribution can be called a space-time oscillation of QT.

\begin{figure}
  \centering
  \includegraphics[width=1.0\textwidth]{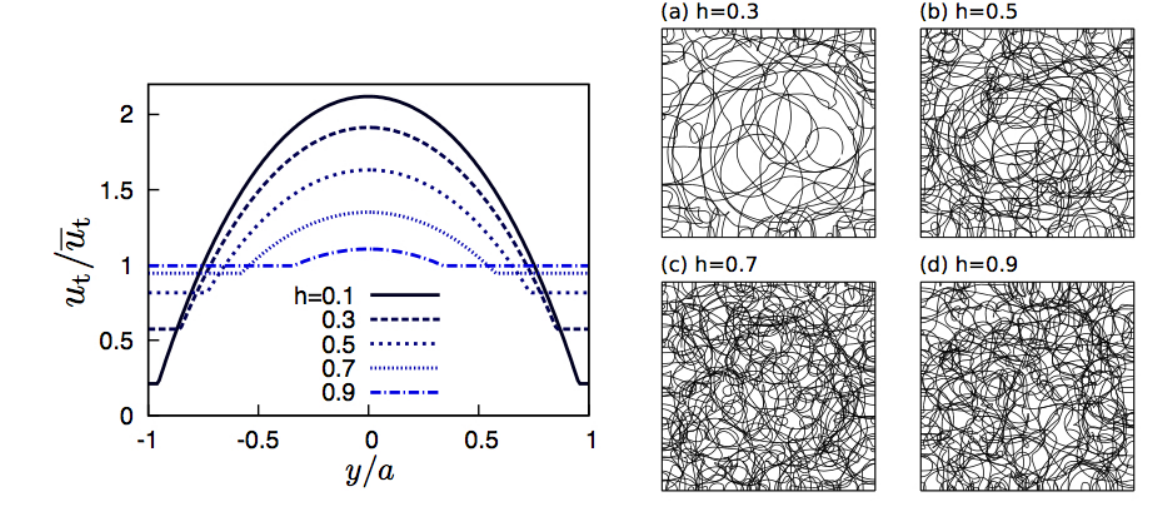}
  \caption
  {
  Results of the simulation of Fig. \ref{yui1.pdf} for thermal counterflow with the tail-flattened normal flow in a duct.
  ({\it left})
  Tail-flattened profile of the normal fluid velocity with different $h$.
  The values are scaled by its mean value.
  (Color figure online)
  ({\it right})
  Snapshots of the vortex tangle in tail-flattened flow viewed along the flow direction with various values of $h$ at $T=1.9 ~\rm{K}$, $\overline{v}_{\rm n} = 0.5 ~\rm{cm/s}$.
  [Reprinted figure with permission from 
  \href{https://doi.org/10.1103/PhysRevB.91.184504}
  {S. Yui and M. Tsubota, Phys. Rev. B {\bf 91}, 184504 (2015)}.
  Copyright (2015) by the American Physical Society.]
  }
  \label{yui3.pdf}
\end{figure}

Under thermal counterflow with the tail-flattened normal flow, QT also reaches a statistically steady state.
Figure \ref{yui3.pdf}({\it left}) shows the tail-flattened profiles with various $h$, and the flatness becomes stronger with increasing $h$.
The snapshots of the vortex tangle in the stationary state with different values of $h$ are shown in Fig. \ref{yui3.pdf}({\it right}), where the parameters are $T=1.9 ~\rm{K}$, $\overline{v}_{\rm n}=0.5 ~\rm{cm/s}$.
With increasing $h$, the vortex tangle becomes denser, and the curvature of the quantized vortices becomes higher.
In the simulation, the value of $L$ in the stationary state becomes larger with $h$ and saturates around $h\sim 0.7$.
This implies that the values of $L$ as a function of $v_{\rm ns}$ does not jump at the the T1-T2 transition, but $\gamma$ increases, which is consistent with experiments \cite{marakov}.

\subsubsection{Two-fluid coupled dynamics}
Khomenko {\it et al.} studied the coupled dynamics of the normal and superfluid components of superfluid $^4$He between two parallel plates, considering the counterflow turbulence with a laminar normal component \cite{khomenko}.
They numerically investigated the velocity profile of the normal fluid component by coupling the averaged Navier--Stokes equation with the VF model.
The Hall--Vinen--Bekarevich--Khalatnikov (HVBK) equation \cite{HallVinen56b,schwarz85,bekarevich} for the normal fluid component is
\begin{equation}
  \frac{\partial {\bm v}_{\rm n}}{\partial t} + ({\bm v}_{\rm n} \cdot \nabla){\bm v}_{\rm n} = \frac{\nabla P}{\rho_{\rm n}} + \frac{{\bm F}_{\rm ns}}{\rho_{\rm n}} + \nu_{\rm n} \nabla^2 {\bm v}_{\rm n}.
\end{equation}
By averaging the equations for the laminar normal flow between two parallel plates along the $xz$-plane, the authors obtained\footnote{This equation appearing in Ref. \cite{khomenko} should lack a factor $1/\rho_{\rm n}$ in the first term of the right hand side.}
\begin{equation}
  \frac{\partial V_{\rm n}(y)}{\partial t} = \frac{dP}{dx} + \frac{\mathcal F_{\rm ns}(y)}{\rho_{\rm n}} + \nu_{\rm n} \frac{\partial ^2 V_{\rm n}(y)}{\partial y^2},
\end{equation}
where $V_{\rm n}$ is the mean normal fluid velocity.
Here, ${\mathcal F}_{\rm ns}$ is a stream-wise projection of the mutual friction force.

\begin{figure}
  \centering
  \includegraphics[width=1.0\textwidth]{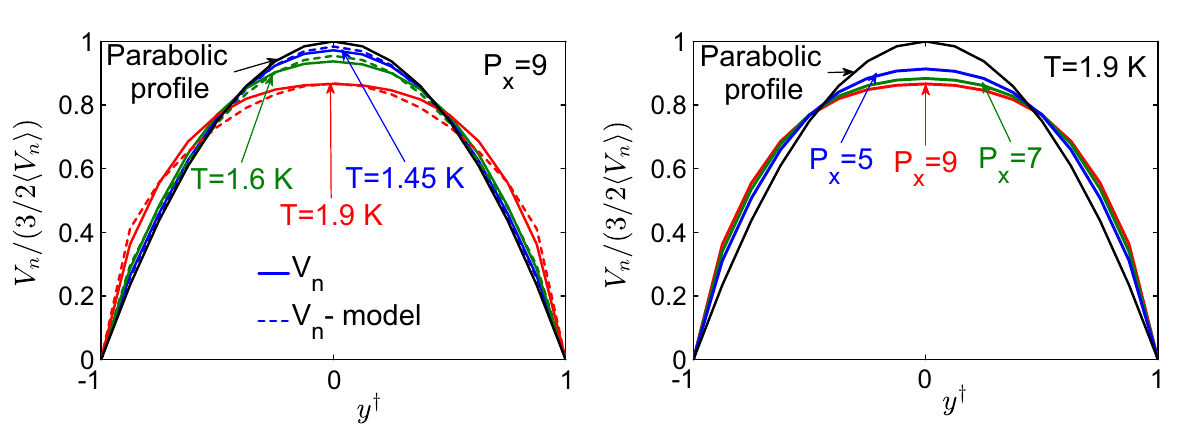}
  \caption
  {
  Results of the numerical study using the two coupled fluid simulation for counterflow quantum turbulence between parallel two palates by Khomenko, Mishra, and Pomyalov \cite{khomenko}.
  ({\it left})
  Mean velocity $V_{\rm n}$ of normal fluid component versus $y^{\dagger} = y/h$ at the effective pressure gradient $P_x = 9$ with different temperatures.
  The dashed lines are the model profiles obtained by their analysis (we skip those details).
  ({\it right})
  $P_x$-dependence of the velocity profile of the mean normal flow.
  [Reprinted figure with permission of Springer from D. Khomenko, P. Mishra, and A. Pomyalov, Coupled Dynamics for Superfluid $^4$He in a Channel, J. Low Temp. Phys., doi:10.1007/s10909-016-1718-2 (2016). Copyright by the Springer Science+Business Media New York 2016.]
  (Color figure online)
  }
  \label{khomenko1.pdf}
\end{figure}

Khomenko {\it et al.} numerically calculated the VF model and the averaged Navier--Stokes equation to investigate the normal fluid velocity profile in the channel.
The simulations were performed between two parallel plates $4h \times 2h \times 2h$ with $h=0.05 ~\rm{cm}$.
Periodic boundary conditions are applied in the $x$- and $z$-directions, and solid boundary conditions are applied in the $y$-direction.
The effective pressure gradient $P_x = dP/dx$ is a parameter of the system and induces normal flow and superflow through the counterflow condition $\rho_{\rm n} \langle {\bm V}_{\rm n} \rangle + \rho_{\rm s} \langle {\bm V}_{\rm s} \rangle = 0$, where $\langle \cdot \rangle$ indicates spatial averaging over the whole volume.
The temperature is $1.45 ~\rm{K}$, $1.6 ~\rm{K}$, and $1.9 ~\rm{K}$, and the effective pressure gradient is $P_x = 5$, $7$, and $9$.

The results obtained by their simulation are shown in Fig. \ref{khomenko1.pdf}, where $y^\dagger = y/h$.
The temperature-dependence of the normal fluid velocity profile at $P_x = 9$ is shown in Fig. \ref{khomenko1.pdf}({\it left}).
The $P_x$-dependence of the normal fluid velocity profile at $T=1.9 ~\rm{K}$ is shown in Fig. \ref{khomenko1.pdf}({\it right}).
The profile becomes flatter in the central region with $T$ and $P_x$ because the mutual friction forces become stronger with increasing $T$ and $P_x$.
These profiles exhibit a different type of flattening than the velocity profiles recently visualized by Marakov {\it et al.} \cite{marakov} for experimental conditions close to the laminar-turbulent transition in the normal component, and their laminar model does not reproduce these conditions.

\subsubsection{Comparison of $\gamma$ in Vinen's equation}
The studies \cite{baggaley15,yui15,khomenko} reviewed above confirmed the steady state condition Eq. (\ref{Lvns}) of Vinen's equation:
\begin{equation}
  L^{1/2} = \gamma (\overline{v}_{\rm ns} - v_0),
\end{equation}
where $\gamma$ is a temperature dependent parameter, $\overline{v}_{\rm ns}$ is a mean counterflow velocity, and $v_0$ is a constant\footnote{Here we introduce a critical velocity $v_0$ sustaining a vortex tangle.}.
It is interesting that the relationship is satisfied even in spatially inhomogeneous systems even though Vinen's equation originally assumed a homogeneous system.
The values of the parameters are shown in Tab. \ref{tab:gamma}.
Baggaley {\it et al.} argued that $\gamma_{\rm p}$ obtained by their simulation between parallel plates are comparable to $\gamma_{\rm uni}$ obtained in homogeneous counterflow by Adachi {\it et al.} \cite{adachi} (with triply periodic boundary conditions and a constant counterflow velocity), as well as the experimental study of Tough \cite{tough}, where it is believed that the normal fluid is laminar.
The values of $\gamma_{\rm hp}$ obtained by the simulation in a duct are smaller than $\gamma_{\rm uni}$.
The authors argued that the normal fluid velocity is prescribed and the mass conservation $\rho_{\rm n} \overline{\bm v}_{\rm n} + \rho_{\rm s} \overline{\bm v}_{\rm s} = {\bm 0}$ is not satisfied, but the mass conservation $\rho_{\rm n} \overline{\bm v}_{\rm n} + \rho_{\rm s} {\bm v}_{\rm s,a} = {\bm 0}$ for the applied flow is satisfied in their simulation.
Recall that the superfluid velocity ${\bm v}_{\rm s}$ consists of ${\bm v}_{\rm s,\omega}$, ${\bm v}_{\rm s,b}$, and ${\bm v}_{\rm s,a}$, so that $\overline{\bm v}_{\rm s} \ne {\bm v}_{\rm s,a}$ when the vortex tangle produces mean flow through the Biot--Savart law.
The values of $\gamma_{\rm c}$ are obtained by the two-fluid coupled simulation between the two parallel plates by Khomenko {\it et al.} \cite{khomenko}, which adjusts the applied superflow ${\bm v}_{\rm s,a}$ to satisfy the mass conservation.
The values of $\gamma_{\rm c}$ are comparable with $\gamma_{\rm uni}$, so that the simulation shows that counterflow quantum turbulence with the laminar normal flow is consistent with the T1 state.

\begin{table}
  \centering
  \caption
  {
    Parameter $\gamma$ of the steady state condition Eq. (\ref{Lvns}) of  Vinen's equation obtained by the 3D simulations of thermal counterflow.
    $\gamma_{\rm exp}$ for the experimental value for the T1 state \cite{tough};
    $\gamma_{\rm uni}$ for the uniform counterflow by Adachi {\it et al.} \cite{adachi};
    $\gamma_{\rm p}$ for the Poiseuille normal flow between the parallel two plates by Baggaley {\it et al.} \cite{baggaley15};
    $\gamma_{\rm hp}$ ($\gamma_{\rm tf}$) for the Hagen--Poiseuille (tail-flattened) normal flow in a duct by Yui and Tsubota \cite{yui15};
    $\gamma_{\rm c}$ for the two-fluid coupled simulation between the parallel two plates by Khomenko {\it et al.} \cite{khomenko}.
    Here, the experimental values \cite{tough} are comparable $\gamma_{\rm uni}$.
  }
  \label{tab:gamma}
  \begin{tabular}{ccccccc}
    \hline
    \hline
    $T$ & $\gamma_{\rm exp}$ &  $\gamma_{\rm uni}$ & $\gamma_{\rm p}$ & $\gamma_{\rm hp}$ & $\gamma_{\rm tf}$ & $\gamma_{\rm c}$  \\
    $(\rm{K})$ & $(\rm{s/cm^2})$ & $(\rm{s/cm^2})$ & $(\rm{s/cm^2})$ & $(\rm{s/cm^2})$ & $(\rm{s/cm^2})$ & $(\rm{s/cm^2})$  \\
    \hline
    $1.3$        & $59$ & $53.5$   &      $67.9$     & $31$    & ---   & ---       \\
    $1.45$      & --- & ---       &      ---             & ---        & ---   & $83$   \\
    $1.6$        & $93$ & $109.6$ &      $83.6$     & $47$    & ---   & $114$  \\
    $1.9$        & $133$ & $140.1$ &      $105.7$   & $103$  & 176 & $165$  \\
    \hline
    \hline
  \end{tabular}
\end{table}

\subsubsection{Two-dimensional simulation}
Galantucci {\it et al.} studied the two-fluid coupled dynamics in counterflow quantum turbulence of $^4$He in channels \cite{galantucci}.
They used a 2D simulation of quantized vortices with the Navier--Stokes equation for the normal fluid component, and they investigated the velocity profiles realized by the two-fluid coupled dynamics.

The Navier--Stokes equation for a normal fluid is written by \cite{donnelly}
\begin{equation}
  \frac{\partial {\bm v}_{\rm n}}{\partial t} + ({\bm v}_{\rm n} \cdot \nabla) {\bm v}_{\rm n} = \frac{\nabla P}{\rho_{\rm n}} + \nu_{\rm n} \nabla ^2 {\bm v}_{\rm n} - \frac{\rho_{\rm s}}{2\rho} \nabla ({\bm v}_{\rm n} - {\bm v}_{\rm s})^2 + \frac{1}{\rho_{\rm n}} \tilde{\bm F}_{\rm ns}
\end{equation}
with incompressible condition $\nabla \cdot {\bm v}_{\rm n} = 0$, where $\tilde{\bm F}_{\rm ns}$ is the coarse-grained mutual friction force.
The normal fluid velocity ${\bm v}_{\rm n}$ is decomposed in two solenoidal fields:
\begin{equation}
  {\bm v}_{\rm n} = {\bm v}_{\rm n}^{\rm p} + {\bm v}_{\rm n}',
\end{equation}
where ${\bm v}_{\rm n}^{\rm p} = (u_{\rm n}^{\rm p}, v_{\rm n}^{\rm p})$ is the Poiseuille flow, and ${\bm v}_{\rm n}'=(u'_{\rm n},v'_{\rm n})$ originates from the back reaction of the quanitized vortex tangle on the normal fluid component.
The vorticity-stream function $\Psi'$ for ${\bm v}_{\rm n}'$ is defined as
\begin{equation}
  {\bm v}_{\rm n}' = \left( \frac{\partial \Psi'}{\partial y}, - \frac{\partial \Psi'}{\partial x} \right).
\end{equation}
Hence, Galantucci {\it et al.}  obtained the two scalar equations
\begin{equation}
  \nabla^2 \Psi' = - \omega'_{\rm n}
\end{equation}
and
\begin{equation}
  \frac{\partial \omega'_{\rm n}}{\partial t} + \left( u_{\rm n}^{\rm p} + \frac{\partial \Psi'}{\partial y} \right) \frac{\partial \omega'_{\rm n}}{\partial x} - \frac{\partial \Psi'}{\partial x} \left( \frac{\partial \omega'_{\rm n}}{\partial y} - \frac{d^2 u_{\rm n}^{\rm p}}{dy^2} \right)
  = \nu_{\rm n} \nabla^2 \omega'_{\rm n} + \frac{1}{\rho_{\rm n}} \left( \frac{\partial \tilde{F}^y}{\partial x} - \frac{\partial \tilde{F}^x}{\partial y} \right),
\end{equation}
where $\tilde{\bm F}_{\rm ns} = (\tilde{F}^x, \tilde{F}^y)$.

\begin{figure}
  \centering
  \includegraphics[width=1.0\textwidth]{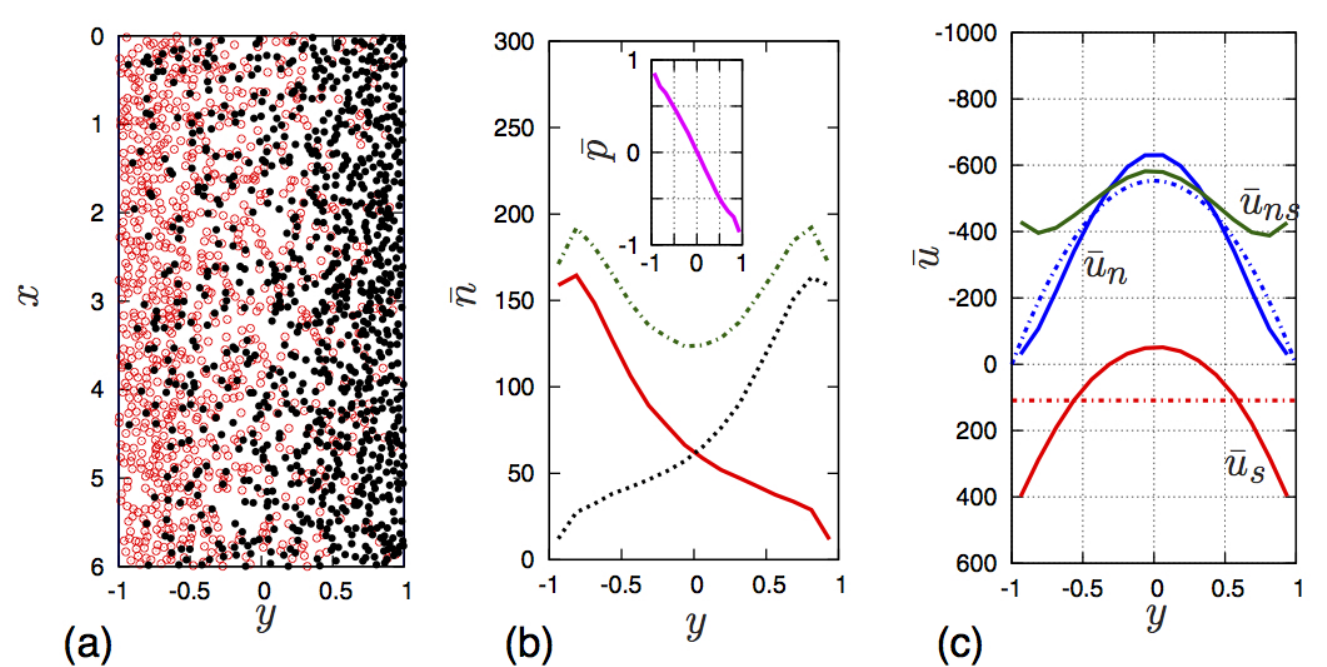}
  \caption
  {
  Results of a numerical study using 2D simulations for counterflow QT in a channel by Galantucci, Sciacca, and Barenghi \cite{galantucci}.
  (a) Vortex distribution in the statistically steady state, where red empty (black filled) circles indicate positive (negative) vortices.
  (b) Coarse-grained profiles of vortex density.
  Positive (negative) vortex density is indicated by a solid red line (dashed black line), and total vortex density $\overline{n}$ is indicated by a dot-dashed green line.
  The inset shows the profile of polarization magnitude $\overline{p}(y)$.
  (c) Profiles of superfluid velocity $\overline{u}_{\rm s}$, normal fluid velocity $\overline{u}_{\rm n}$, and counterflow velocity $\overline{u}_{\rm ns}$.
  Red and blue dot-dashed lines indicate the initial laminar profiles of the superfluid and normal fluid component, respectively.
  [Reprinted figures with permission from 
  \href{https://doi.org/10.1103/PhysRevB.92.174530}
  {L. Galantucci, M. Sciacca, and C. F. Barenghi, Phys. Rev. B {\bf 92}, 174530 (2015)}.
  Copyright (2015) by the American Physical Society.]
  (Color figure online)
  }
  \label{gala1.pdf}
\end{figure}

Galantucci {\it et al.} performed numerical simulations of 2D quantized vortices with the above equation for counterflow QT in a channel to investigate the spatial distributions of the superfluid and normal fluid velocities.
Their results are shown in Fig. \ref{gala1.pdf}.
Figure \ref{gala1.pdf}(a) shows a snapshot of the quantized vortices in the statistically steady state.
We can see polarization of the quantized vortices.
Figure \ref{gala1.pdf}(b) shows the spatial distribution of the quantized vortices.
The quantized vortices concentrate near the channel walls.
These results are similar with the {\it inverse distribution of quantized vortices}, which appears in the 3D simulations reviewed above.
The coarse-grained polarization vector $\overline{\bm p}(y)$ was introduced, defined by \cite{jou}
\begin{equation}
  \overline{p} (y) = \frac{\overline{\omega}_{\rm s}(y)}{\kappa \overline{n} (y)} = \frac{\overline{n}^{+} (y) - \overline{n}^{-} (y)}{\overline{n}^{+}(y) + \overline{n}^{-} (y)} \hat{\bm z},
\end{equation}
where $\overline{n}^{+}$ ($\overline{n}^{-}$) is the positive (negative) vortex density profile, and $\hat{\bm z}$ is the unit vector along the $z$-direction.
Note that $\overline{\bm p} = {\bm 0}$ when QT is spatially homogeneous.
The value of $\overline{p}$ in the statistically steady state is shown in the inset of Fig. \ref{gala1.pdf}(b).
This polarized pattern directly arises from the vortex-point equation of motion, where the friction term containing $\alpha$ depends on the polarity of the vortex.

Figure \ref{gala1.pdf}(c) shows the velocity profiles of the superfluid and normal fluid components, and the counterflow velocity profile in the statistically steady state.
The superfluid velocity profile obeys a parabolic profile
\begin{equation}
  \overline{u}_{\rm s} (y) \sim y^2,
\end{equation}
which is different from that in the study of a logarithmic velocity profile described later \cite{yui15b}.
The profile originates when the superfluid velocity profile mimics the normal fluid velocity.
This mimicking is supported by analytical results obtained via simple models \cite{barenghi02}, and numerical studies that observed the normal fluid and superfluid velocity matching and vorticity locking \cite{barenghi02,samuels93,barenghi97,kivotides06}.
The normal fluid velocity is suppressed near the channel walls and amplified in the central region, which is different from the behavior in the 3D simulation between the two parallel plates by Khomenko {\it et al.} \cite{khomenko}.
Furthermore, they studied the transient state from the initial state to the steady state and found that the normal fluid velocity distribution becomes quasi-parabolic profile in the transient state, which is approximately uniform in the central region.
Galantucci {\it et al.} concluded that their numerical model predicts the shape of the profile of normal fluid that has been experimentally observed in channels using laser-induced fluorescence of metastable helium molecules \cite{marakov}.
However, in the 2D simulation, we must inject quantized vortices into the system, and the results of simulation will change based on the choice of method of injection.

\subsection{Logarithmic velocity profile for quantum turbulence} \label{loglaw}
In this subsection, we review a study of mean velocity profile QT between two parallel plates \cite{yui15b}.
The study found that the mean flow of the superfluid component obeys a logarithmic profile of the distance from the solid wall, {\it i.e.}, the log-law.
The log-law is famous in field of a classical turbulence, which is confirmed by theoretical analysis \cite{landau,davidson}, experiments \cite{marusic,hutchins}, and numerical studies \cite{lee}.
We briefly describe the basics of the logarithmic velocity profile in classical turbulence, followed by our numerical studies in quantum turbulence.

\subsubsection{Logarithmic velocity profile in classical turbulence}
The logarithmic velocity profile can be obtained using dimensional analysis \cite{landau,davidson}.
Let us consider the statistically steady state of turbulent flow along a solid surface, as shown in Fig. \ref{surface.eps}.
The flow direction is the $x$-axis, and the solid surface is the $xz$-plane, so that $y$ is the distance from the solid surface.
Here, the $y$ and $z$ components of the mean velocity are zero:
\begin{equation}
  \overline{u}_{x} = u(y), ~\overline{u}_{y} = \overline{u}_{z} = 0,
\end{equation}
where the overline indicates time averaging.

\begin{figure}
  \centering
  \includegraphics[width=0.5\textwidth]{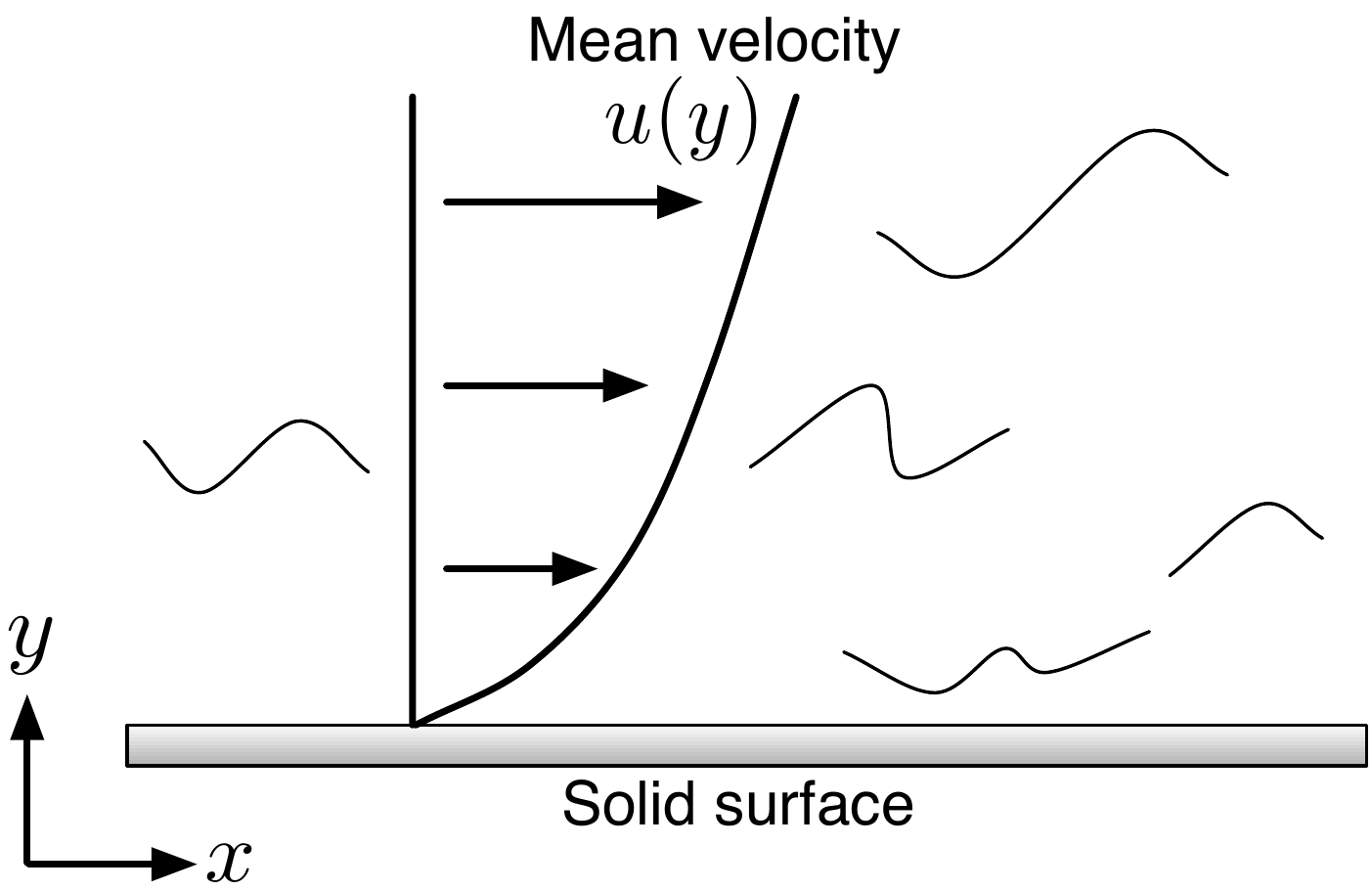}
  \caption
  {
  Turbulent flow along a solid surface.
  The flow direction is the $x$-axis, and the solid surface is the $xz$-plane, so that $y$ is the distance from the solid surface.
  }
  \label{surface.eps}
\end{figure}

The total shear stress $\sigma$ between the layers of fluid is sum of two shear stresses, namely, a viscous stress $\tau$ and a Reynolds stress $\tau^{\rm R}$.
The viscous stress comes from the viscous term in the Navier--Stokes equation.
The Reynolds stress comes from the advection term, which shows momentum transfer by turbulent fluctuations between the layers of fluid.
Here, vortices tend to drift toward the turbulence core region in the center of the channel.
For the statistically steady state, $\sigma$ is a constant independent of $y$, and equal to the frictional stress on the solid surface.

For large distances $y$, the viscosity $\nu$ is unimportant.
In this region, the value of the velocity gradient at each point must be determined only by constant parameters such as the density $\rho$, shear stress $\sigma$, and distance $y$.
The dimensions of these quantities are, respectively, $\rm [M L^{-3}]$, $\rm [M L^{-1} T^{-2}]$, and $\rm [L]$.
On the other hand, the dimensions of $du/dy$ are $\rm [T^{-1}]$.
Hence, $du/dy$ should take the form
\begin{equation}
  \frac{du}{dy} = \frac{\sqrt{\sigma / \rho}}{by},
\end{equation}
where $b$ is a numerical constant, which cannot be determined by dimensional analysis, and the experimental value is \cite{marusic}
\begin{equation}
  b \sim 0.4.
\end{equation}
We introduce the convenient notation
\begin{equation}
  u_{*} = \sqrt{\sigma / \rho},
\end{equation}
which is a characteristic velocity for the turbulent flow.
Thus, we obtain the logarithmic velocity profile
\begin{equation}
  u = \frac{u_{*}}{b} (\log y + c),
\end{equation}
where $c$ is a constant of integration.
This expression is valid for large distances $y$, so that we cannot use the boundary condition at the solid surface for this formula to determine the constant $c$.

For small distances $y$, the viscosity $\nu$ becomes important.
We denote the order of magnitude of these distances by $y_0$, and the velocity is of the order of $u_{*}$, so that the Reynolds number is $Re \sim u_{*} y_0 / \nu$.
The viscosity becomes important when $Re$ is on the order of unity.
Thus, we obtain
\begin{equation}
  y_0 \sim \frac{\nu}{u_{*}},
\end{equation}
which determines the width $y_0$ of the viscous region.
In the region $y < y_0$, the Reynolds stress is negligible. The shear stress $\sigma$ becomes the viscous stress $\tau = \rho \nu du/dy$, so that
\begin{equation}
  u = \frac{\sigma y}{\rho \nu} = \frac{u_{*} ^2 y }{\nu}.
\end{equation}
This region is called a viscous sublayer.
By connecting the viscous sublayer and the logarithmic velocity profile at $y = y_0$, we can determine the constant $c$ of integration, so that
\begin{equation}
  u = \frac{u_*}{b} \log \left( \frac{y}{y_0} \right).
\end{equation}
For large distances, this formula determines the mean velocity distribution in the turbulent flow along the solid surface.

\subsubsection{Numerical study in quantum turbulence}
We describe a numerical study for the logarithmic velocity profile in quantum turbulence \cite{yui15b}.
This study investigated a mean velocity profile in a superfluid boundary layer, motivated by numerical studies for spatially inhomogeneous thermal counterflow \cite{baggaley13,baggaley15,yui15,khomenko15} and visualization experiments \cite{marakov,bewley,guo09,guo10,mantia,paoletti,zhang}.
As a result, they found that the logarithmic velocity profile could appear in wall-bounded QT.

We consider the case of Fig. \ref{setup.eps}(a); the normal fluid component flows between two parallel plates, and the superfluid component has no external flow, {\it i.e.}, ${\bm v}_{\rm s,a}=0$.
The flow direction is along the $x$-axis.
The solid boundaries are at $y/D=0$ and $2$ with the half-width $D$ of the channel.
The normal flow is prescribed to be a Poiseuille profile:
\begin{equation}
  {\bm v}_{\rm n} = v_{\rm n,0} \left[ 1- \left( \frac{y-D}{D} \right)^2 \right] \hat{\bm x},
\end{equation}
where $v_{\rm n,0}$ is a normalization factor and $\hat{\bm x}$ is a unit vector in the $x$-direction.

\begin{figure}
\begin{center}
  \includegraphics[width=0.9\textwidth]{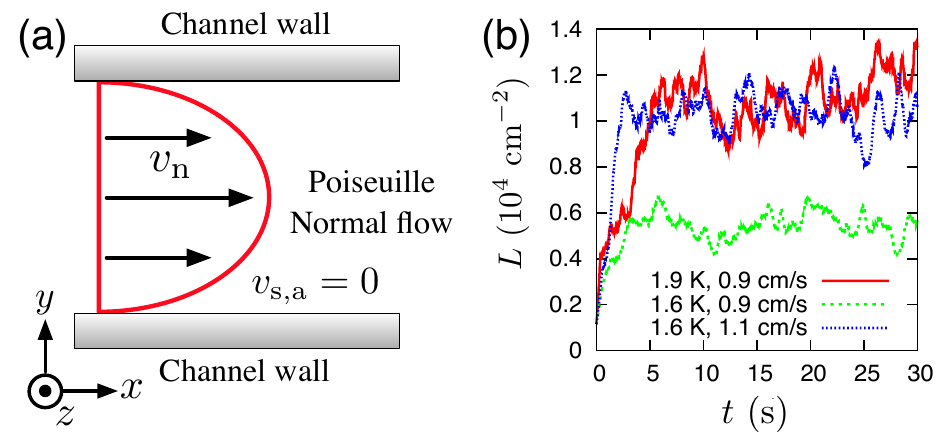}
  \caption
  {
     (a) Schematic of the numerical simulation of the VF model by Yui {\it et al.} \cite{yui15b}.
     The laminar flow of normal fluid moves between parallel two plates, whereas superfluid has no external flows.
     The flow direction is the $x$-axis.
     Periodic boundary conditions are applied along the $x$- and $z$-directions, and a solid boundary condition is applied along the $y$-direction.
     (b) Vortex line density in pure normal flow as a function of time.
     The QT reaches a statistically steady state.
     [Reprinted figure with permission from 
  \href{https://doi.org/10.1103/PhysRevB.92.224513}
  {S. Yui, K. Fujimoto, and M. Tsubota, Phys. Rev. B {\bf 92}, 224513 (2015)}.
  Copyright (2015) by the American Physical Society.]
     (Color figure online)
  }
  \label{setup.eps}
\end{center}
\end{figure}

\begin{figure}
\begin{center}
  \includegraphics[width=0.6\textwidth]{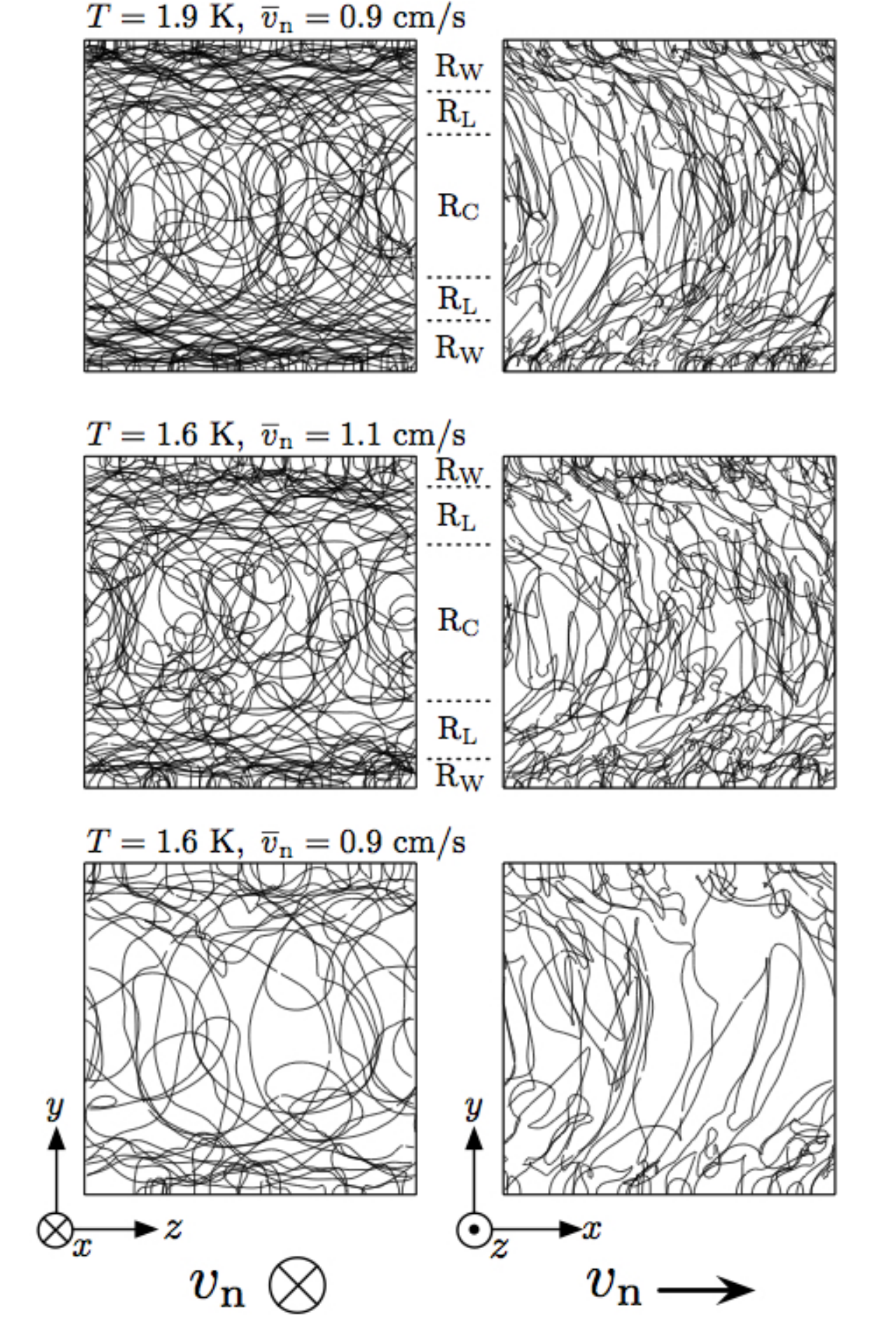}
  \caption
  {
  Snapshots of the vortex tangles in the statistically steady state obtained by numerical simulation as shown in Fig. \ref{setup.eps}.
  The left (right) figures are the stream-wise (side) views.
  Here, ${\rm R}_{\rm L}$ indicates the region of the logarithmic velocity profile; ${\rm R}_{\rm C}$ and ${\rm R}_{\rm W}$ are the rest regions.
  The region ${\rm R}_{\rm W}$ near the wall tends to have the high curvature vortices.
  [Reprinted figure with permission from 
  \href{https://doi.org/10.1103/PhysRevB.92.224513}
  {S. Yui, K. Fujimoto, and M. Tsubota, Phys. Rev. B {\bf 92}, 224513 (2015)}.
  Copyright (2015) by the American Physical Society.]
  }
  \label{tangle.eps}
\end{center}
\end{figure}

The QT in the pure normal flow reaches a statistically steady state.
Fig. \ref{setup.eps}(b) shows the vortex line density $L$ versus time.
The value of $L$ increases from the initial value and then fluctuates around some constant values.
The value of $L$ in the statistically steady state increases with $T$ or $\overline{v}_{\rm n}$.
The reason is that the mutual friction term $\alpha {\bm s}' \times ({\bm v}_{\rm n} - {\bm v}_{\rm s})$, which injects energy to QT by stretching quantized vortices, becomes stronger with $T$ or $\overline{v}_{\rm n}$.
In the statistically steady state, an spatially inhomogeneous QT appears as shown in Fig. \ref{tangle.eps}.
The vortex line density increases near the walls.
This structure is similar to the {\it inverse distribution of quantized vortices}, which appears in the numerical simulations of the inhomogeneous QT in thermal counterflow.
The inhomogeneous structure becomes clearer with $T$ or $\overline{v}_{\rm n}$ since the mutual friction becomes stronger, which is the origin of the structure.

\begin{figure}
\begin{center}
  \includegraphics[width=0.9\textwidth]{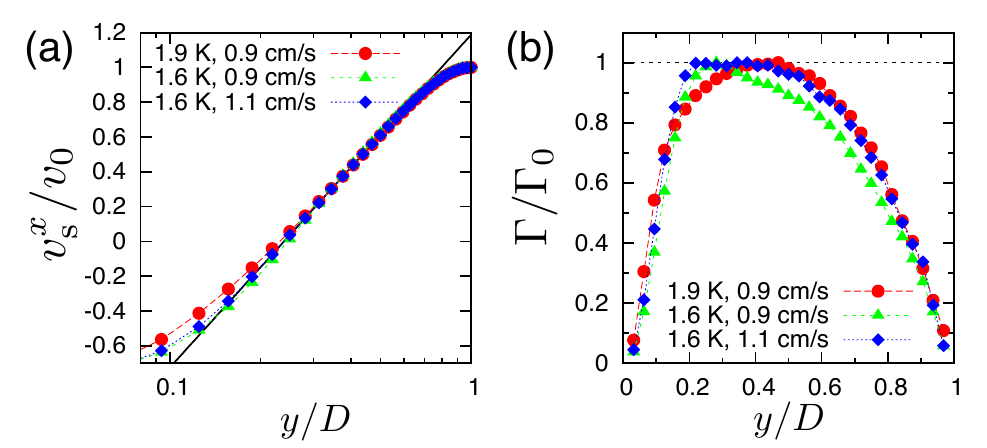}
  \caption
  {
    Results obtained by numerical simulation as shown in Fig. \ref{setup.eps}.
    (a) The flow direction component $v_{\rm s}^{x}$ of the superfluid velocity ${\bm v}_{\rm s}$ as a function of the distant $y/D$ from the wall.
    The wall is $y/D=0$, and the center of the channel.
    The solid line is obtained by fitting Eq. (\ref{eq:log}) to the data of run3($T=1.6 ~\rm{K}, \overline{v}_{\rm n} = 1.1 ~\rm{cm/s}$).
    (b) Verification of another form of $\Gamma \equiv y(dv_{\rm s}^{x} /dy) = \rm{const}.$ of the logarithmic velocity profile.
    The logarithmic profile are found at run1($T=1.9 ~\rm{K}$, $\overline{v}_{\rm n} = 0.9 ~\rm{cm/s}$) and run3($T=1.6 ~\rm{K}$, $\overline{v}_{\rm n}=1.1 ~\rm{cm/s}$), since they have higher vortex line density.
    [Reprinted figure with permission from 
  \href{https://doi.org/10.1103/PhysRevB.92.224513}
  {S. Yui, K. Fujimoto, and M. Tsubota, Phys. Rev. B {\bf 92}, 224513 (2015)}.
  Copyright (2015) by the American Physical Society.]
    (Color figure online)
  }
  \label{velocity.eps}
\end{center}
\end{figure}

In the statistically steady state, the log-law of quantum turbulence appears.
Fig. \ref{velocity.eps}(a) shows the mean value of the flow direction component ${\bm v}_{\rm s}^{x}$, which is the superfluid velocity ${\bm v}_{\rm s}$.
Here, $v_0$ is the value of $v_{\rm s}^{x}$ at the center $y/D = 1$ of the channel, and the values are averaged temporally over the statistically steady state and spatially over the $x$- and $z$-directions.
All of the data may show the log-law because we can see that they are parallel to the logarithmic profile
\begin{equation}
  \overline{v}_{\rm s}^{x} (y) = \frac{v_{\rm q}^{*}}{\kappa_{\rm q}} \left[ \log \left( \frac{y}{D} \right) + c \right],
  \label{eq:log}
\end{equation}
where $v_{\rm q}^{*}$ is a characteristic velocity, $\kappa_{\rm q}$ is the Karman constant for QT, and $c$ is a constant.
However, we should carefully determine whether the log-law appears.
We should check a differential form of the log-law
\begin{equation}
  \Gamma(y) \equiv y \frac{d \overline{v}_{\rm s}^{x} (y)}{dy} = \frac{v_{\rm q}^{*}}{\kappa_{\rm q}}.
\end{equation}
This form is obtained by differentiating Eq. (\ref{eq:log}).
In the log-law region, $\Gamma$ is constant value independent of $y$.
Fig. \ref{velocity.eps}(b) shows $\Gamma(y)/\Gamma_0$ as a function of $y/D$, where $\Gamma_0$ is the maximum value of $\Gamma$.
Here, we define the log-law region as $0.95 \leq \Gamma/\Gamma_0 \leq 1.00$.
The results are shown in Table \ref{tab:loglaw}, where the log-law region is written by $y_1 < y < y_2$.
For $\overline{v}_{\rm n} = 0.9 ~\rm{cm/s}$, the log-law cannot be found at $T=1.9 ~\rm{K}$, but $T=1.6 ~\rm{K}$.
The reason that the log-law does not appear at run2 ($T=1.6 ~\rm{K}$, $\overline{v}_{\rm n} = 0.9 ~\rm{cm/s}$) may be low vortex line density.
Indeed, for the same temperature $T=1.6 ~\rm{K}$, run3 ($T=1.6 ~\rm{K}$, $\overline{v}_{\rm n} = 1.1 ~\rm{cm/s}$) with higher vortex line density obtains a logarithmic velocity profile.

\begin{table}
  \centering
  \caption
  {
    Results of numerical simulation for the log-law in quantum turbulence \cite{yui15b}.
    The mean value $v_0$ of the superfluid velocity at the center of the channel.
    The log-law region $y_1 < y < y_2$ for the bottom half of the channel and the width of the region $y_2/y_1$.
    The characteristic velocity $v_{\rm q}^{*} /\kappa_{\rm q}$ and the constant $c$.
  }
  \label{tab:loglaw}
  \begin{tabular}{ccccccccc}
    \hline
    \hline
            &        $T$        &        $\overline{v}_{\rm n}$    & $v_0$ & $ y_1/D$   &   $ y_2/D$    &   $y_2/y_1$   &  $v_{\rm q}^{*}/\kappa_{\rm q}$   &   $c$    \\
            &    $(\rm{K})$    &        $(\rm{cm/s})$               & $(\rm{cm/s})$ &         ---       &          ---        &         ---          &  $(\rm{cm/s})$     &   ---   \\
    \hline
    run1        &        $1.9$        &        $0.9$     & $0.189$ &    $0.31$        &    $0.57$        &   $1.84$        &     $0.153$             & $1.45$  \\
    run2        &        $1.6$        &        $0.9$     & $0.080$ &    ---               &     ---               &   ---               &    ---                     & ---  \\
    run3        &        $1.6$        &        $1.1$     & $0.159$ &    $0.18$        &    $0.54$        &   $3.00$        &     $0.132$           & $1.43$  \\
    \hline
    \hline
  \end{tabular}
\end{table}

The width $y_2/y_1$ of the logarithmic velocity profile is shown in Table \ref{tab:loglaw}.
A typical numerical study \cite{lee} of classical turbulence obtains the width $y_2/y_1 = 2.37$ as the largest width.
The widths of the superfluid log-law in Table \ref{tab:loglaw} are comparable with numerical simulations of classical turbulence.

By fitting Eq. (\ref{eq:log}) to the data of Fig. \ref{velocity.eps}(a), the values of $v^{*}_{\rm q}/{\kappa}_{\rm q}$ and $c$ are obtained.
Here, the fitting range is $y_1 < y < y_2$.
The results are shown in Table. \ref{tab:loglaw}.
The Karman constant $\kappa_{\rm q}$ for quantum turbulence cannot be determined using these quantities since the value $v^{*}_{\rm q}$ is unknown.
If we build up a theory for the log-law of QT, the value of $v^{*}_{\rm q}$ will become clear.

In Fig. \ref{tangle.eps}, the log-law region, which is determined by $y_1 < y < y_2$ for the bottom half of the channel in Table \ref{tab:loglaw}, is indicated by ${\rm R}_{\rm L}$.
The region ${\rm R}_{\rm W}$ near the wall, which is defined as $0 < y < y_1$, tends to have high curvature vortices.
The central region ${\rm R}_{\rm C}$, which is defined as $y_2<y<D$, tends to show lower vortex line density.
According to their analysis, the vortex tangle drifts toward the walls, which is the opposite direction as that of classical turbulence, and the anisotropy becomes larger in the log-law region.
Hence, Yui {\it et al.} argue that the log-law requires superfluid momentum flux toward the walls.
Recall that, in classical turbulence, the log-law is realized by Reynolds stress $\tau^{\rm R}$, which is produced when a flow directional momentum is steadily transferred toward the wall by turbulence.
From the analogy between the classical and superfluid log-law turbulence, we expect that they could be caused by a similar mechanism even though the origins of the dissipation are different.

\section{QT in ultracold atomic BECs}
In this section, we introduce theoretical and experimental researches for QT in ultracold atomic BECs, which are a different type of quantum fluid than the superfluid helium addressed in the last section. 

\subsection{Background}
As explained in Section 2, QT has been primarily studied in superfluid helium because this system was unique for the research of the quantum hydrodynamics. However, the realization of the ultracold atomic BECs in 1995 changed this situation. 

In this system, the atomic species are captured by a magnetic or optical trapping potential and the application of cooling techniques such as laser and evaporative cooling leads to highly degenerate atomic gases. As a result, Bose--Einstein condensation occurs, enabling the study of quantum hydrodynamics in ultracold atomic gases \cite{Stringari,Pethick}. In fact, nucleation of a quantized vortex \cite{Inouye01,Neely10}, quantized vortex lattice formation \cite{lattice1,lattice2}, superfluidity and phase slip \cite{superfluidity1,superfluidity2,superfluidity3}, and the Kibble--Zureck mechanism \cite{KZ0,KZ1,KZ2} have been experimentally and theoretically studied. 

From the perspective of QT research, ultracold atomic BECs have two distinct features: (i) highly controllable systems, and (ii) realization of multi-component BECs, which lead to novel phenomena for quantum hydrodynamics not addressed in the superfluid helium.

We can control the spatial dimension of the ultracold atomic gas owing to (i), and it is possible to study the dimensional dependence of QT. In classical turbulence (CT), CT in 3D systems is well known to be different from that in 2D systems  because of the existence of the inverse cascade \cite{Frisch,davidson}, which poses a question whether the 2D nature of QT appears. In superfluid helium, this kind of study has never been investigated, but feature (i) of ultracold gases allows us to approach this important theme.

Feature (ii) further enriches the variety of QT. By making use of a mixture of different kinds of atomic species or the internal degrees of freedom of atomic hyperfine spin, we can realize multi-component BECs \cite{Stringari,Pethick,KU,Sta}. This kind of system provides a unique stage for studying turbulence in the mixture of quantum fluids, where not only the superfluid velocity field but also other fields such as the spin density vector field plays an important role, and various topological defects can be nucleated. Thus, this system has the possibility of opening new avenues in QT research.

In the following sections, we describe experimental and theoretical studies for QT in ultracold atomic BECs. First of all, in Sec. 3.2, the experimental results for QT are explained, where we introduce the recent results for not only 3D and 2D QT but also QT in a spinor BEC. After the experimental results, we describe theoretical studies addressing vortex turbulence, weak wave turbulence, dimensionality of turbulence, and spin turbulence in ultracold atomic BECs. Section 3.3 describes the studies for the 3D QT in single-component BECs, in which we explain vortex turbulence and weak wave turbulence. The subsection for the vortex turbulence treats the Kolmogorov $-5/3$ power law from the perspective of decomposition of kinetic energy spectrum into incompressible and compressible ones, while, as for the weak wave turbulence, we explain two possibilities of weak wave turbulence in single-component BECs and show existence of inverse cascade by using the Fj\o rft argument. In Sec. 3.4, 2D QT is explained, and we briefly review the inverse cascade in the 2D CT. Then, we introduce the recent studies finding the inverse cascade and dynamical formation of the Onsager vortex. Finally, in Sec. 3.5, turbulence in multi-component BECs, especially spinor BECs is described. 

\subsection{Experimental studies of QT in ultracold atomic BECs}
The experiments for QT in the ultracold atomic BECs are in the early stages, so there are not many works. Most of the experiments focus on QT in single-component BECs, but, recently, turbulence in spinor BECs have been investigated. In this section, we review their experimental studies. 

\begin{figure}[t]
\begin{center}
\includegraphics[keepaspectratio, width=11cm,clip]{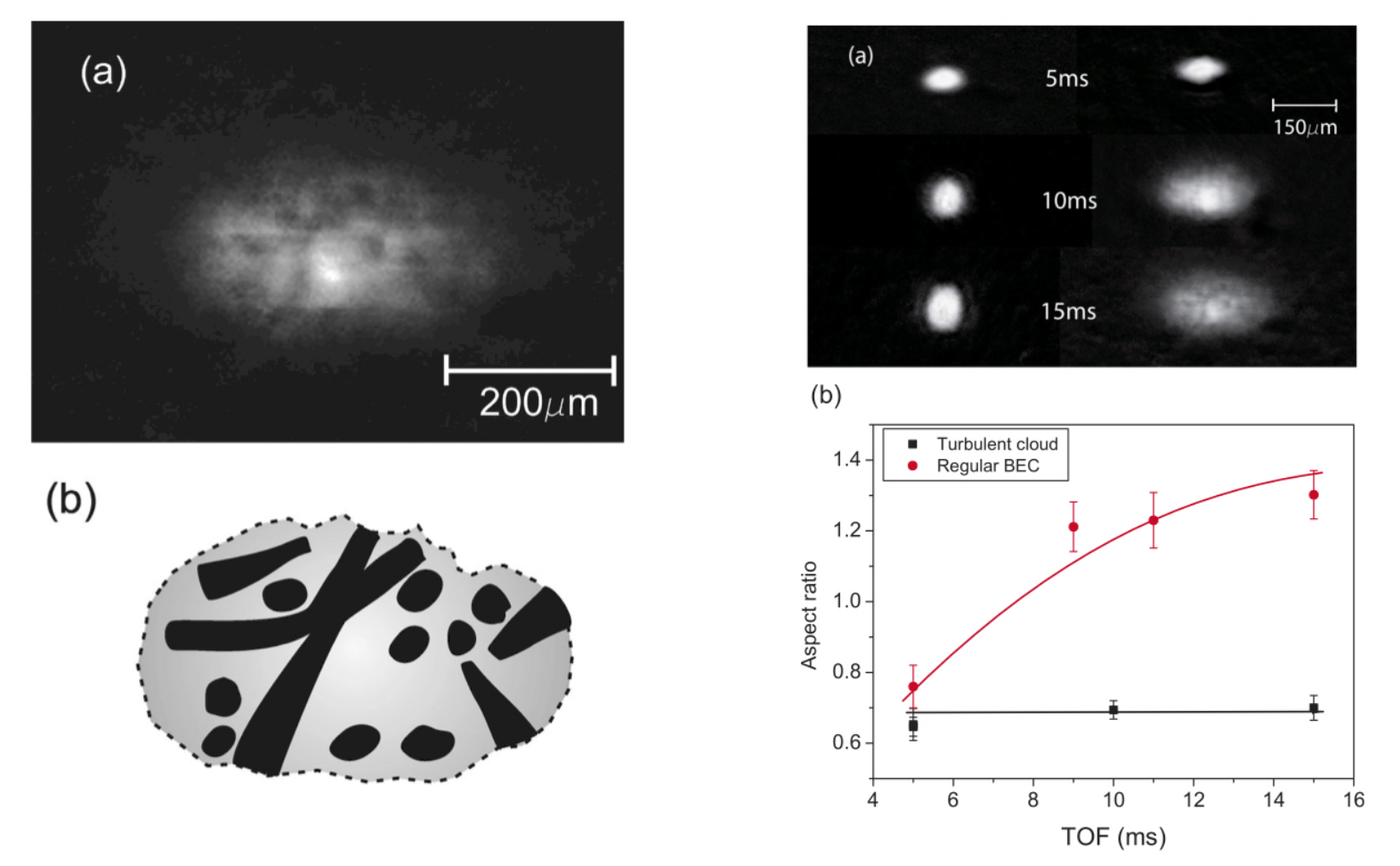}
\caption{Experimental results in \cite{Henn09}. The left figures are (a) the density profile in QT observed by the time-of-flight expansion and (b) the distribution of quantized vortices inferred from (a), where the vortices are depicted by the black region. The right figures are (a) the time-of-flight expansion dynamics for the stationary (left) and turbulent (right) states, and (b) the time-development for the aspect ratio of the size of the atomic gas. When the system is not turbulent, the ratio is reversed due to the uncertainty principle of the position and the momentum. On the other hand, in QT, the ratio tends to remain at its initial value. 
(Color figure online)
[Reprinted figure with permission from \href{https://doi.org/10.1103/PhysRevLett.103.045301}{E. A. L. Henn $et~al.$, Phys. Rev. Lett. {\bf 103}, 045301 (2009)}. Copyright (2009) by the American Physical Society.]
\label{Brazil}} 
\end{center}
\end{figure}

\subsubsection{3D QT in single-component BECs}

The first pioneering experiments for QT were performed by Henn $et~al.$ \cite{Henn09}, where 3D QT was investigated in a single-component BEC.They prepared a single atomic BEC with $^{87}{\rm Rb}$ and obtained the turbulent state by shaking and rotating the trapping potential. The left profile (a) in Fig. {\ref{Brazil}} is the density distribution in the turbulence observed by the usual time-of-fight method. The sketch (b) is the position of quantized vortices estimated by this density distribution, which shows that some vortices are nucleated and the vortex distribution is disturbed. 

Furthermore, they observe peculiar dynamics characteristic of turbulence in the time-of-fight expansion. The quantity of focus is an aspect ratio for the size of the atomic cloud. In the expansion of a stationary state, this aspect ratio inverts over time because of the uncertainty principle of position and momentum. However, in QT, this aspect ratio tends to retain its initial value. The right figure (a) in Fig. {\ref{Brazil}} is the time-of-fight dynamics of the density distribution for both stationary (left) and turbulent states (right), from which they obtained the time-development of the aspect ratio and find distinct difference between the stationary and turbulent states, as shown in the right graph (b) of Fig. {\ref{Brazil}}. 

\begin{figure}[t]
\begin{center}
\includegraphics[keepaspectratio, width=11cm,clip]{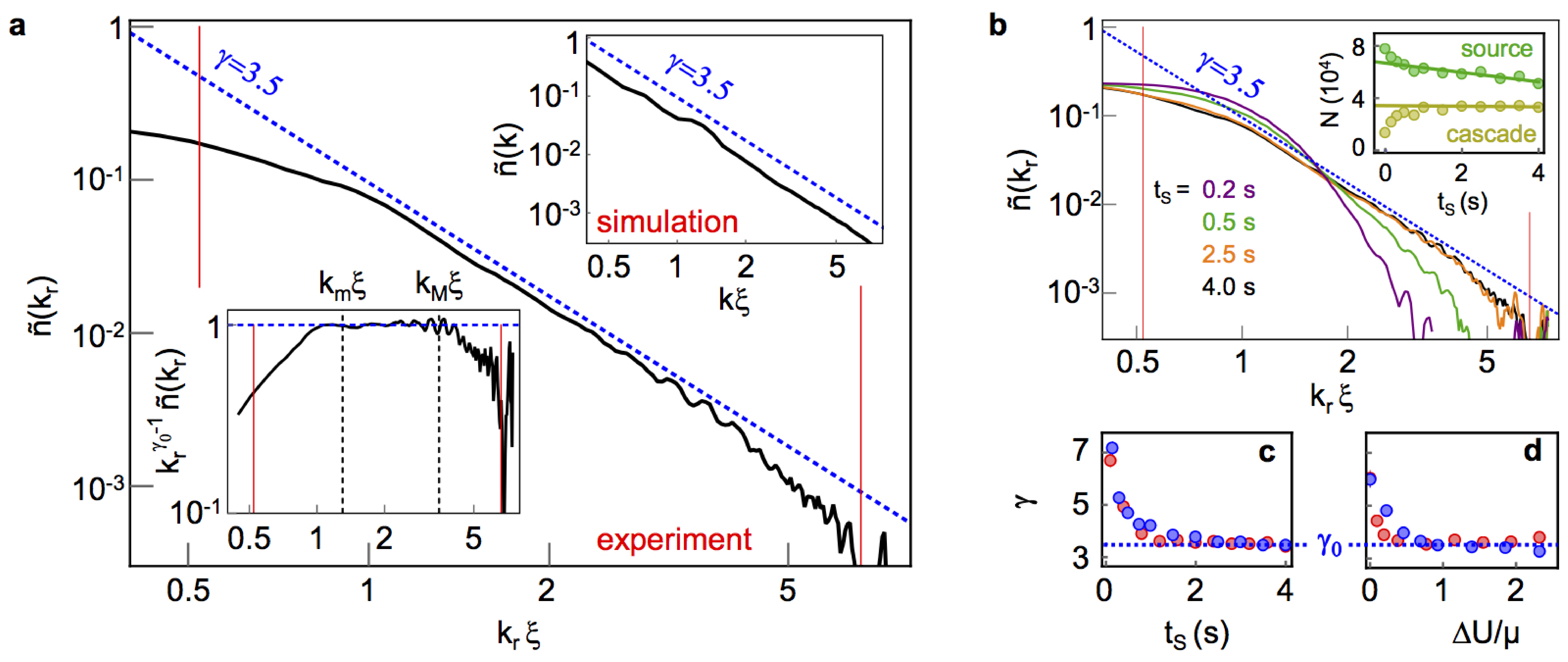}
\caption{Experimental results in \cite{Nir16}. Figure (a) is the momentum distribution obtained by the conventional time-of-flight method, where $-3.5$ power law behavior is observed. This result is consistent with the simulation shown in the upper right inset. The lower left inset is the same momentum distribution, but the vertical axis is $k^{\gamma_0}n(k)$. The  value of $\gamma_0$ is obtained in Figs. (c) and (d). Figure (b) is the time-development of the momentum distribution. The inset shows the time development of the particle number in the source region and cascade region. [Reprinted figure with permission from N. Navon, A. L. Gaunt, R. P. Smith, and Z. Hadzibabic: 
Nature 539, 72 (2016). Copyright (2016) by the Nature Publishing Group] (Color figure online) \label{Cambridge}} 
\end{center}
\end{figure}

At present, this anomalous behavior of the aspect ratio is considered to be attributed to the velocity field induced by the quantized vortices \cite{Caracanhas}. In fact, a numerical calculation shows that a few vortices tend to suppress the inversion of the ratio \cite{Tsuchitani}. However, to our knowledge, no numerical calculations have been computed for the time-of-fight dynamics in the turbulence, so a quantitative comparison between the experimental and numerical studies has not been performed. After this work,  they investigated the phase diagram for the turbulent state and the excitation strength \cite{Seman}, and the momentum distribution \cite{momentum1}. 

Navon $et~al.$ also studied QT in 3D systems captured by a uniform box potential \cite{Nir16}. The method of generating turbulence is oscillation of the box potential, where they temporally change the slope of the line connecting the edges. Their observation focuses on the momentum distribution obtained by the time-of-fight, which is similar method in the experiment of Ref. \cite{momentum1}, confirming direct cascade behavior from low- to high-wave number region and observing formation of the $-3.5$ power law in the momentum distribution as shown in Fig. \ref{Cambridge}. Furthermore, they perform numerical calculations of the GP equation, obtaining results that exhibited good agreement with the experiment. Note that their observation of the direct cascade is not based on a kinetic energy spectrum for the velocity field, which is an important quantity in CT, because in ultracold atomic gases it is much difficult to detect the superfluid velocity. 

To our knowledge, this $-3.5$ power law is not sufficiently understood. The $-3.5$ power is similar to an exponent $-3$ predicted in the weak wave turbulence with four-wave interaction \cite{wt2}. However, according to their numerical calculation, the vortices are present, so that it is unclear whether the weak wave turbulence theory is applicable to this system. At present, we do not know which this system is, the vortex turbulence or the weak wave turbulence. Thus, the theoretical derivation of this power law and detailed numerical studies will be important future works.

\begin{figure}[t]
\begin{center}
\includegraphics[keepaspectratio, width=11.5cm,clip]{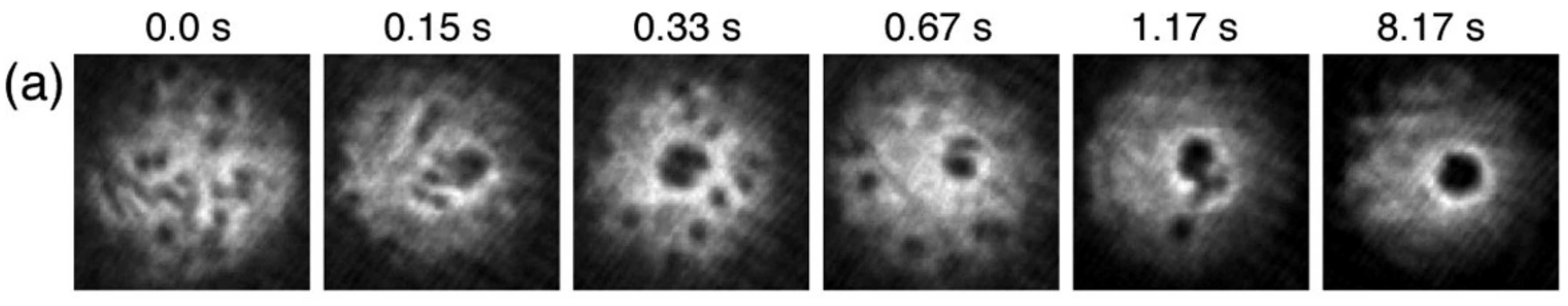}
\caption{Experimental results in \cite{Neely13}. These figures are the time-development of the density distribution in 2D QT. Each distribution is pictured by the time-of-fight with different samples. After the generation of the initial vortex distribution by making use of the obstacle potential, they investigate the decay of QT. In the early stage of the dynamics, the vortices seem to be randomly distributed, but as time passes, a large density hole is formed. In this state, persistent current is realized.  
[Reprinted figure with permission from \href{https://doi.org/10.1103/PhysRevLett.111.235301}{T. W. Neely $et~al.$, Phys. Rev. Lett. {\bf 111}, 235301 (2013)}. Copyright (2013) by the American Physical Society.]
\label{America}} 
\end{center}
\end{figure}

\subsubsection{2D QT in single-component BECs}
The above experiments addressed 3D QT, but Neely $et~al.$ investigated 2D QT in an atomic BEC, reporting 2D nature of QT \cite{Neely13}. 

In classical fluids, the features of the turbulence depend on the spatial dimension. In 3D systems, the kinetic energy is transferred from low- to high-wave number region (energy direct cascade). However, in 2D system, the enstrophy is transferred from low- to high-wave number region (enstrophy direct cascade), and the kinetic energy is transferred from high- to low-wave number region (energy inverse cascade). This dependence comes from the conservation of the enstrophy in the non-viscous limit of the two-dimensional Navier--Stokes equation \cite{Frisch,davidson}. 
The theoretical details of this inverse cascade are discussed in Sec. 3.4. 

To reveal the features of 2D QT in analogy with CT, Neely $et~al.$ generates many quantized vortices in ultracold atomic gases with $^{87}{\rm Rb}$ \cite{Neely13}. In this study, the vortices are nucleated by rotating the trapping potential with fixed Gaussian obstacles located in the center. After the stirring, in order to observe the decay of 2D QT, the condensate freely developed in an optional hold time and they turned off the obstacle at the end of the experiment. Their striking observation is the formation of a large density hole accompanying the persistent current as shown in Fig. \ref{America}, which is consistent with the inverse cascade of 2D CT.  This behavior seems to be result of energy transport from high- to low-wave number region. However, as discussed in their paper, to clearly confirm the inverse cascade, it is necessary to observe the energy flux. Thus, further works may be needed to confirm whether the inverse cascade occurs in this system. Note that, in the experiment, the conservation of angular momentum is approximately satisfied and they excite the directed angular momentum. This fact and the inverse cascade are consistent with the formation of the persistent current.

\begin{figure}[t]
\begin{center}
\includegraphics[keepaspectratio, width=12cm,clip]{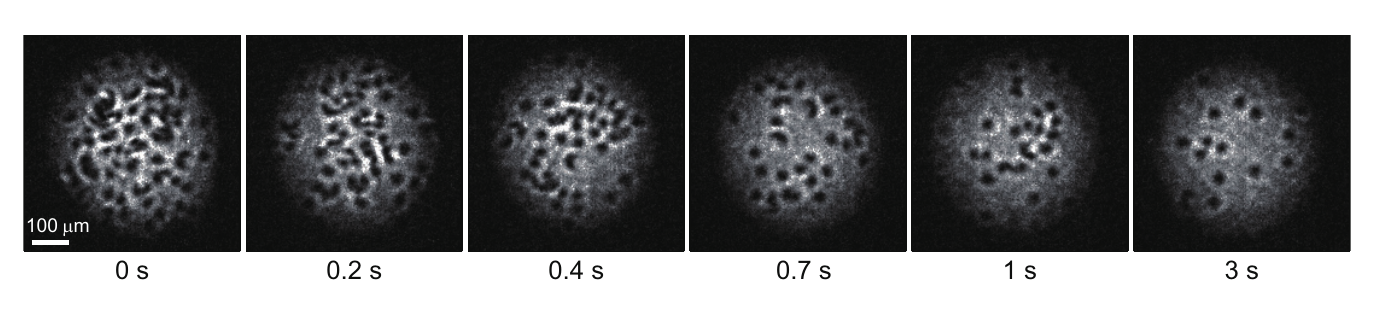}
\caption{Experimental results in \cite{Kwon14}. Figure shows the time-development of the density distribution. In this QT, more than 50 quantized vortices are initially nucleated by utilizing a repulsive Gaussian potential. These figures correspond to the states after the disappearance of the Gaussian potential. Over time, the vortex number decreases through vortex annihilation and vortex escape from the trap.  [Reprinted figure with permission from \href{https://doi.org/10.1103/PhysRevA.90.063627}{W. J. Kwon $et~al.$, Phys. Rev. A {\bf 90}, 063627 (2014)}. Copyright (2014) by the American Physical Society.]
\label{Korea}} 
\end{center}
\end{figure}

Kwon $et~al.$ also studied 2D QT \cite{Kwon14}. This experiment used $^{23}{\rm Na}$, generating more than 50 quantized vortices. Figure \ref{Korea} is the time-development of the density profiles of 2D QT, where the initial state with many vortices is prepared by utilizing a repulsive Gaussian obstacle. In this preparation, the obstacle is fixed at the center of the atomic gas, and they move the trapping potential in one direction, turning off the obstacle potential. From Fig. \ref{Korea}, it follows that the number of vortices temporally decreases and formation of large density hole observed in \cite{Neely13} does not appear. 

To characterize these decay dynamics, they count the vortex number and investigate the mechanism of disappearance of the quantized vortices. Two mechanisms are identified: one is the annihilation of quantized vortices, and the other is vortex escape from the atomic gas. Thereafter, Stagg $et~al.$ performed numerical calculations of the dissipative GP equation, obtaining results consistent with the experiment \cite{Stagg15}.  

\subsubsection{QT in spinor BECs}
Seo $et~al.$ and Seji $et~al.$ began studying turbulence in the spinor BEC \cite{Seo16,Seji17}. The atomic species is $^{23}{\rm Na}$, so that the system is a spin-1 spinor BEC with antiferromagnetic interaction. Quenching of the external field is used to generate the QT in this system. In this procedure, they initially applied the external field such that the polar state is stable, then decreasing the field and change the value of the quadratic Zeeman coefficient. Thus, the system exhibits instability, and many half-quantized vortices are nucleated. This is the first experimental realization of turbulence in multi-component BECs. This type of turbulence could be a new target of theoretical and numerical study for QT. 

\subsection{Theoretical studies of QT for single-component BECs in 3D systems}
Before these experimental works, QT in ultracold atomic BECs was theoretically studied by using the GP equation. 
At early stages, the incompressible kinetic energy spectrum was investigated, being confirmed that the Kolmogorov $-5/3$ power law appears in 3D QT where many quantized vortices are nucleated and the vortex turbulence is realized. The pioneering work for this direction is done by Nore $et~al.$, who performed numerical calculations for decaying QT \cite{Nore}. Originally, this study focused on superfluid helium, but this result is important for understanding QT in ultracold atomic BECs. After this study, some theoretical works studied QT in ultracold atomic BECs, finding the same Kolmogorov $-5/3$ law \cite{Parker,KT05,KT05_2,KT07,Bor11,Bor12}. 

Another type of turbulence exists apart from vortex turbulence, that is, weak wave turbulence \cite{wt1,wt2}. In weak wave turbulence, linear waves weakly interact with each other, leading to a turbulent cascade associated with power law behaviors. This kind of turbulence was studied in ultracold atomic BECs \cite{Dyachenko92,Zakharov05,Nazarenko06,Proment09,FT15}. Here, we comment on Ref. \cite{Dyachenko92}, which originally did not treat the BECs but consider the weak wave turbulence described by the GP model. 

In this subsection, we review both vortex and weak wave turbulence for single-component BECs in 3D systems.

\subsubsection{Single-component GP equation}
The theoretical model for QT in single-component BECs is the GP equation, which well describes the dynamics of ultracold atomic BECs when the temperature is much lower than the Bose-Einstein condensation transition temperature \cite{Stringari,Pethick}: 
\begin{eqnarray}
i \hbar \frac{\partial}{\partial t} \psi(\bm{r},t) = -\frac{\hbar^2}{2M} {\bm \nabla}^2 \psi(\bm{r},t) + V_{\rm trap}(\bm{r})  \psi(\bm{r},t) + g \rho(\bm{r},t)  \psi(\bm{r},t), 
\label{1GP}
\end{eqnarray}
where $\rho=|\psi|^2$ is the local density distribution, $g$ is the interaction parameter expressed by $4\pi \hbar^2 a/M$ with the s-wave scattering length $a$, and $V_{\rm trap}$ is the trapping potential. 

This equation can be rewritten in the canonical form given by
\begin{eqnarray}
i \hbar \frac{\partial}{\partial t} \psi (\bm{r},t) = \frac{\delta E[\psi (\bm{r},t),\psi ^*(\bm{r},t)] }{\delta \psi^*(\bm{r},t)} \label{1GP2}, 
\end{eqnarray}
where the energy functional $E[\psi (\bm{r},t),\psi(\bm{r},t)^*]$ is expressed by
\begin{eqnarray}
E[\psi (\bm{r},t),\psi(\bm{r},t)^*] = \int \Bigl[ \frac{\hbar^2}{2M} | \bm{\nabla} \psi(\bm{r},t) |^2 + V_{\rm trap}(\bm{r}) \rho(\bm{r},t) + \frac{g}{2}\rho(\bm{r},t)^2 \Bigl]d\bm{r}.
\end{eqnarray}
This energy $E$ is a conserved quantity in the dynamics because of $[E,E]=0$ with the bracket $[A,B] = \int \{(\delta A/\delta \psi)(\delta B/\delta \psi^*)-(\delta B/\delta \psi)(\delta A/\delta \psi^*) \} d\bm{r}$, where $\delta (\cdot) /\delta \psi$ is the functional derivative. The total particle number $N$ is another conservation quantity, defined by
\begin{eqnarray}
N[\psi (\bm{r},t),\psi(\bm{r},t)^*] = \int \rho(\bm{r},t) d\bm{r},  
\end{eqnarray}
from which it follows that $[E,N]=0$ is satisfied, thus leading to the conservation of the total particle number. 

In CT with the Kolmogorov $-5/3$ power law, some energy dissipation mechanism is necessary for the establishment of statistically steady turbulence. For this reason, some papers add phenomenological dissipation terms to the GP equation, the detail of which is explained in each subsection. 

Finally, we describe the quantization of circulation. Let us express the wavefunction as 
\begin{eqnarray}
\psi (\bm{r},t) = \rho(\bm{r},t) {\rm exp}[i \phi (\bm{r},t)]. 
\label{madelung}
\end{eqnarray}
Then, substituting Eq. (\ref{madelung}) to the GP equation (\ref{1GP}), we obtain the continuity equation of the local density expressed by 
\begin{eqnarray}
\frac{\partial}{\partial t}\rho (\bm{r},t) + {\bm \nabla} \cdot  [\rho(\bm{r},t) \bm{v}(\bm{r},t)] = 0, 
\end{eqnarray}
where the superfluid velocity field is defined as $\bm{v}(\bm{r},t)=\hbar \bm{\nabla} \phi(\bm{r},t) /M$.
Therefore, we calculate the circulation $\Gamma$ with a closed loop $C$, finding that the single-valueness leads to 
\begin{eqnarray}
\Gamma &=& \int _{C} \bm{v}(\bm{r},t) \cdot d\bm{r} \nonumber \\
 &=& \kappa n \hspace{5mm} (n=0,\pm 1, \pm 2, \cdots), 
\end{eqnarray}
from which it follows that the circulation is quantized with the quantum circulation $\kappa = h/M$. 

\subsubsection{Vortex turbulence of single-component BECs in 3D systems}
We review QT in 3D systems with many quantized vortices.
Most studies for QT focus on the kinetic energy spectrum corresponding to the spatial correlation 
function of the velocity field because, in CT, this quantity exhibits the celebrated Kolmogorov $-5/3$ 
power law.  However, there is a distinct difference between CT with the Kolmogorov law and QT in ultracold atomic BECs:
incompressibility. Originally, the Kolmogorov law is discussed in incompressible fluids \cite{Frisch,davidson}, so we cannot easily expect this manifestation of this law in ultracold atomic BECs since this system is a compressible quantum fluid. To resolve this problem, Nore $et~al.$ performed decomposition of the superfluid velocity field into compressible and incompressible parts by utilizing the Helmholtz decomposition theorem \cite{Nore}. 

We first describe this decomposition. The system is assumed to be a $d$-dimensional box with periodic 
boundary conditions. We denote the system size and volume as $L$ and $V=L^d$, respectively. For this setup, the wave number is 
expressed by the discrete number $k_{j} = 2\pi m_{j}/L$ with the integer $m_j$ and the spatial direction $j$.

In the GP model, the kinetic energy per unit volume is expressed by 
\begin{eqnarray}
E_{\rm k}(t) &=& \frac{\hbar^2}{2MV} \int |\nabla \psi(\bm{r},t) |^2 d\bm{r} = E_{\rm v}(t) + E_{\rm q}(t), 
\end{eqnarray}
\begin{eqnarray}
E_{\rm v}(t) = \frac{M}{2V} \int \rho(\bm{r},t) \bm{v}(\bm{r},t)^2 d\bm{r}, 
\end{eqnarray} 
\begin{eqnarray}
E_{\rm q}(t) = \frac{\hbar^2}{2MV} \int \Bigl[ \bm{\nabla} \sqrt{\rho(\bm{r},t)} \Bigl]^2 d\bm{r}, 
\end{eqnarray} 
where $E_{\rm v}(t)$ and $E_{\rm q}(t)$ are the energy for the superfluid velocity field and for quantum pressure, respectively. 
In QT, $E_{\rm v}$ is often called the kinetic energy, and we follow this convention in what follows. 
To derive the expression of the kinetic energy spectrum, an effective velocity field $\bm{A}(\bm{r},t) = \sqrt{\rho(\bm{r},t)} \bm{v}(\bm{r},t)$ is introduced. Then, we can obtain
\begin{eqnarray}
E_{\rm v}(t) &=& \frac{M}{2V} \int \bm{A}(\bm{r},t)^2 d\bm{r} =  \frac{M}{2}\sum _{\bm{k}} |\tilde{\bm{A}}(\bm{k},t)|^2,  
\end{eqnarray} 
where $\tilde{\bm{A}}(\bm{k},t)$ is the Fourier component of $\bm{A}(\bm{r},t)$ which is defined by 
$\mathcal{F}[A(\bm{r},t)]= \int \bm{A}(\bm{r},t) {\rm exp}(-i\bm{k}\cdot \bm{r})d\bm{r}/V $. 
The relation $\tilde{\bm{A}}(\bm{k},t) = \tilde{\bm{A}}(-\bm{k},t)^*$, which is obtained by noting that $\bm{A}(\bm{r},t)$ is real, is used to derive the above expression. Then, we can define the kinetic energy spectrum: 
\begin{eqnarray}
\mathcal{E}_{\rm v}(k,t) =\frac{M}{2\triangle k}\sum _{\Omega(\bm{k_1},k)} |\tilde{\bm{A}}(\bm{k}_1,t)|^2,  
\end{eqnarray} 
where the notation $\Omega(\bm{k_1},k)$ means the summation for $k-\triangle k/2 \leq |\bm{k}_1| < k+\triangle k/2 $ with $\triangle k = 2\pi /L$. This quantity corresponds to the following correlation function $C_{\rm v}(\bm{r})$ defined by 
\begin{eqnarray}
C_{\rm v}(\bm{r},t) =\frac{1}{V} \int \bm{A}(\bm{x}+\bm{r},t) \cdot \bm{A}(\bm{x},t) d\bm{x}, 
\end{eqnarray} 
because the correlation function is expressed by
\begin{eqnarray}
C_{\rm v}(\bm{r},t) = \sum_{\bm{k}} |\tilde{\bm{A}}(\bm{k},t)|^2 {\rm exp}(i\bm{k}\cdot \bm{r}), 
\end{eqnarray}  
from which the kinetic energy spectrum is equivalent to the Fourier component of the spatial correlation function of the effective velocity field when the turbulence is isotropic. 

Finally, we apply the Helmholtz decomposition theorem to the effective velocity field $\bm{A}(\bm{x},t)$, obtaining
\begin{eqnarray}
\bm{A}(\bm{r},t) =  {\bm A}_0(t) + \bm{A}_{\rm c}(\bm{r},t) + \bm{A}_{\rm i}(\bm{r},t), \label{Ekin_v}
\end{eqnarray} 
\begin{eqnarray}
\bm{A}_{\rm c}(\bm{r},t) &=& \sum _{\bm{k} \neq 0 }  \tilde{\bm{A}}_{\rm c}(\bm{k},t)  {\rm exp}(i\bm{k}\cdot \bm{r}) \nonumber \\
&=& \sum _{\bm{k} \neq 0} \frac{ \bm{k}\cdot \tilde{\bm{A}}(\bm{k},t) }{k^2}\bm{k} {\rm exp}(i\bm{k}\cdot \bm{r}) \label{Ekin_c}, 
\end{eqnarray} 
\begin{eqnarray}
\bm{A}_{\rm i}(\bm{r},t) &=& \sum _{\bm{k} \neq 0 }  \tilde{\bm{A}}_{\rm i}(\bm{k},t)  {\rm exp}(i\bm{k}\cdot \bm{r}) \nonumber \\
&=& \sum_{\bm{k} \neq 0 } \biggl\{ \tilde{\bm{A}}(\bm{k},t) -  \frac{\bm{k}\cdot \tilde{\bm{A}}(\bm{k},t) }{k^2}\bm{k}  \biggl\} 
{\rm exp}(i\bm{k}\cdot \bm{r})
 \label{Ekin_i}. 
\end{eqnarray} 
These fields $\bm{A}_{\rm c}(\bm{r},t)$ and $\bm{A}_{\rm i}(\bm{r},t)$ satisfy the relations ${\rm rot} \bm{A}_{\rm c}(\bm{r},t) = 0$ and ${\rm div} \bm{A}_{\rm i}(\bm{r},t) = 0$, being the compressible and incompressible parts, respectively. The constant ${\bm A}_0 (t)$ is the Fourier component of $\bm{A}(\bm{r},t) $ for the zero wave number. The derivation of Eqs. (\ref{Ekin_c}) and (\ref{Ekin_i}) is described in Appendix C. 

Using the orthogonality property $\tilde{\bm{A}}_{\rm c}(\bm{k},t) \cdot \tilde{\bm{A}}_{\rm i}(\bm{k},t) = 0$, we can decompose the kinetic energy spectrum into the compressible and incompressible spectra as follows: 
\begin{eqnarray}
\mathcal{E}_{\rm v}(k,t) = \mathcal{E}_0(t) \delta_{\bm{k},\bm{0}} + \bigl( \mathcal{E}_{\rm cv}(k,t) + \mathcal{E}_{\rm iv}(k,t) \bigl)(1-\delta_{\bm{k},\bm{0}}), 
\end{eqnarray} 
\begin{eqnarray}
\mathcal{E}_{0}(t) =\frac{M}{2 \triangle k}\sum _{\Omega(\bm{k_1},0)} |{\bm{A}}_{0}(t)|^2,  
\end{eqnarray} 
\begin{eqnarray}
\mathcal{E}_{\rm cv}(k,t) =\frac{M}{2 \triangle k}\sum _{\Omega(\bm{k_1},k)} |\tilde{\bm{A}}_{\rm c}(\bm{k}_1,t)|^2,  
\end{eqnarray} 
\begin{eqnarray}
\mathcal{E}_{\rm iv}(k,t) =\frac{M}{2\triangle k}\sum _{\Omega(\bm{k_1},k)} |\tilde{\bm{A}}_{\rm i}(\bm{k}_1,t)|^2.
\end{eqnarray} 
From the definitions of Eq. (\ref{Ekin_i}), $\mathcal{E}_{\rm iv}(k,t)$ is found to be the correlation function of the effective velocity field $\bm{A}_{\rm i}(\bm{r},t)$, possessing information of rotational flow induced by quantized vortices. 

To numerically calculate these spectra, the superfluid velocity field is calculated by using the expression 
\begin{eqnarray}
\rho(\bm{r},t) \bm{v}(\bm{r},t) = \frac{\hbar}{2Mi} [ \psi(\bm{r},t)^*\bm{\nabla} \psi(\bm{r},t) - \psi(\bm{r},t)\bm{\nabla} \psi(\bm{r},t)^* ] . \nonumber 
\end{eqnarray}
Then, we can calculate the effective velocity field and apply the Fourier transformation. Substituting the Fourier component into Eqs. (\ref{Ekin_v}) -- (\ref{Ekin_i}), we obtain each spectra.  

\begin{figure}[t]
\begin{center}
\includegraphics[keepaspectratio, width=11cm,clip]{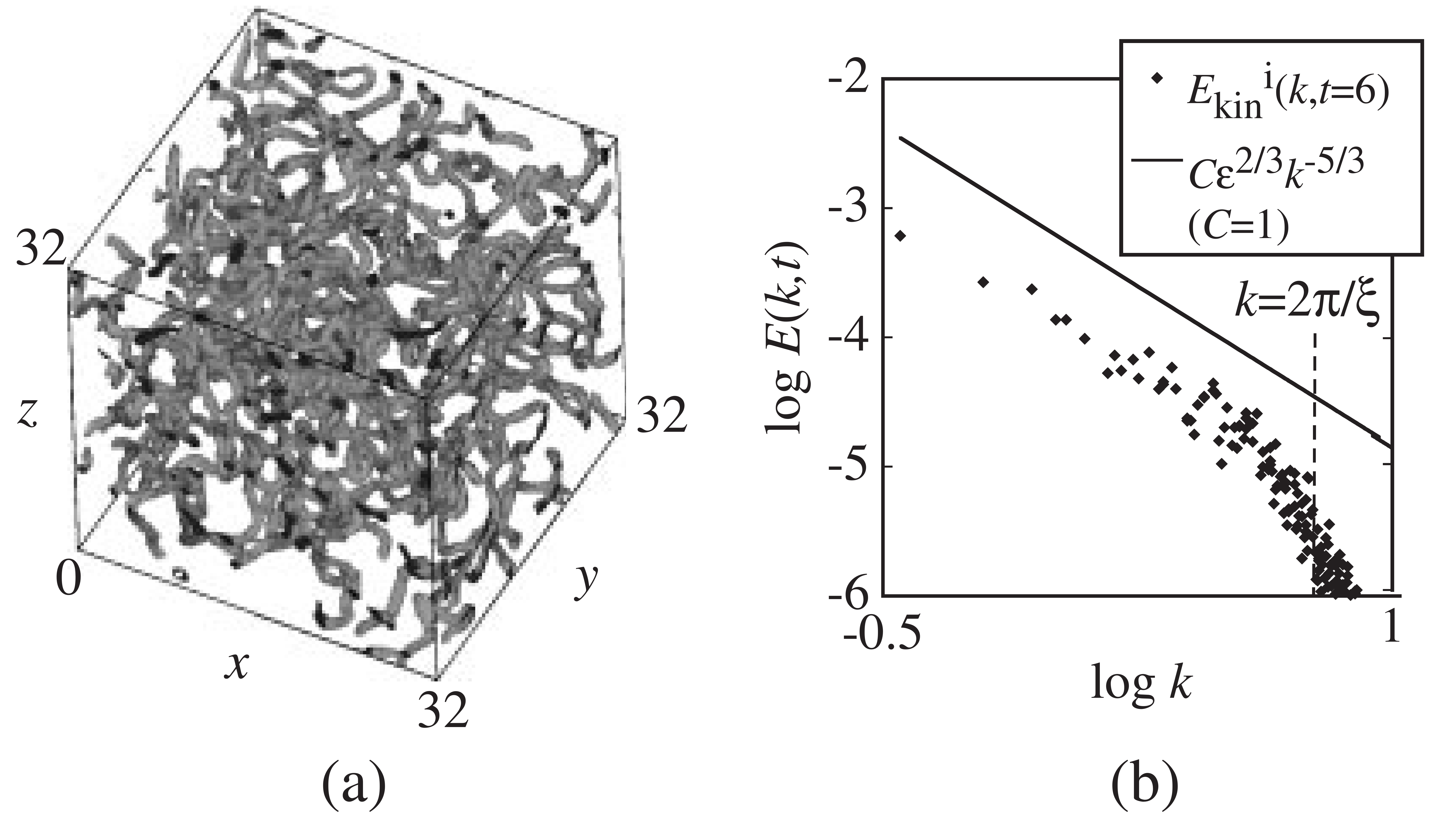}
\caption{Numerical results for the decaying turbulence. Figure (a) is the vortex configuration at a fixed time obtained by the GP equation with phenomenological dissipation. The initial state is the flat density with a random phase. The incompressible kinetic energy spectrum corresponding to (a) is shown in Fig. (b), which is averaged by 20 initial states with different noises. This graph clearly shows the Kolmogorov $-5/3$ power law.  [Reprinted figure with permission from \href{http://journals.jps.jp/doi/pdf/10.1143/JPSJ.74.3248}{M. Kobayashi and M. Tsubota,
J. Phys. Soc. Jpn. {\bf 74}, 3248 (2005)}. Copyright (2005) by the Physical Society of Japan.]
\label{decay_turbulence}} 
\end{center}
\end{figure}

Such a decomposition was done by Nore $et~al.$ for the first time in QT generated by the GP equation \cite{Nore}. Their numerical calculation is performed in an initial state with the Taylar--Green vortex to investigate the kinetic energy spectra. Their method for the generation of QT is based on the instability of the initial state, and the decaying QT is studied. Then, the incompressible kinetic energy spectrum was found to exhibit the Kolmogorov $-5/3$ power law in a limited time period. After this study, Parker $et~al.$ also confirmed the Kolmogorov law in QT in a rotating system \cite{Parker}, and Kobayashi $et~al.$ introduced a phenomenological dissipation term into the GP equation and obtained the Kolmogorov law \cite{KT05}. Figures \ref{decay_turbulence} (a) and (b) are the vortex configuration and the incompressible kinetic energy spectrum in Ref. \cite{KT05}, where the decaying QT is generated by the initial state with random phase and the Kolmogorov law is confirmed. Furthermore, Kobayashi $et~al.$ generated the stationary steady state QT using the dissipative GP equation and obtained the Kolmogorov law \cite{KT05_2,KT07}.  Recently, QT was investigated in terms of a non-thermal fixed point, and the Kolmogorov law was also observed \cite{Bor11,Bor12}. The numerical method for solving the GP equation by pseudo-spectral calculations was well described in Ref. \cite{KT05_2}. 

Finally, we comment on the effective velocity field $\bm{A}(\bm{r},t)$. Because this field is weighted by the root of the density profile $\rho(\bm{r},t)$, the incompressible kinetic energy does not investigate the correlation function of the very superfluid velocity. To eliminate the effect of the density profile, we need to calculate the kinetic energy spectrum with $\bm{v}(\bm{r},t)$ instead of $\bm{A}(\bm{r},t)$. However, the divergence of the velocity field at the core of the quantized vortex must be considered. For this divergence, the spectrum of $\bm{v}(\bm{r},t)$ exhibits numerically unstable behavior. To overcome this issue, one candidate is introduction of a spatial cutoff of the velocity field when we calculate the spectrum \cite{KU16}. Such a calculation has been recently performed in single-component and spin-2 spinor BECs, and in the case of the single-component BEC, the Kolmogorov $-5/3$ power law is confirmed.  

\subsubsection{Weak wave turbulence of single-component BECs in 3D systems}
Until the preceding subsection, we review the vortex turbulence where the quantized vortex tangle is formed. 
Here, another type of turbulence, i.e., weak wave turbulence, in single-component BECs is reviewed \cite{Dyachenko92,Zakharov05,Nazarenko06,Proment09,FT15}. In this class of turbulence, linear waves with various wave numbers are excited \cite{wt1,wt2}, but because the wave amplitudes are small, the interaction between the waves is weak. To address the weak wave turbulence in ultracold atomic BECs, the meaning of the linear wave in the GP equation is important. To begin with, we explain this linear wave. In the GP model, two kinds of linear wave are possible: one is a wave around no-condensate and 
the other is a wave around a strong condensate. 

In the former case of the no-condensate, the wave function is expressed by 
\begin{eqnarray}
\psi(\bm{r},t) = \delta \psi(\bm{r},t), \label{no_condensate}
\end{eqnarray} 
where $\delta \psi(\bm{x},t)$ is the fluctuation around the no-condensate \cite{Dyachenko92,Nazarenko06,Proment09}. Substituting Eq. (\ref{no_condensate}) into Eq. (\ref{1GP}), we obtain
\begin{eqnarray}
i \hbar \frac{\partial}{\partial t} \delta \psi(\bm{r},t) = -\frac{\hbar^2}{2M} {\bm \nabla}^2 \delta \psi(\bm{r},t) + g |\delta \psi(\bm{r},t)|^2  \delta \psi(\bm{r},t), 
\label{1GP_fluc}
\end{eqnarray}
where we consider a uniform system without trapping potentials. 

When we neglect the nonlinear term, the fluctuation behaves as a free particle: 
\begin{eqnarray}
\delta \psi(\bm{r},t) = \sum_{\bm{k}} C(\bm{k}) {\rm exp}(i\bm{k}\cdot \bm{r} - i \epsilon_{\rm f}(k)t/\hbar). 
\end{eqnarray} 
with the dispersion relation $\epsilon_{\rm f}(k) = \hbar^2k^2/2M$ and the weight function of an initial condition $C(\bm{k})$.

However, because we consider the nonlinear term in Eq. (\ref{1GP_fluc}), the weight function generally has time dependence. 
Then, we can expand the wave function as 
\begin{eqnarray}
\delta \psi(\bm{r},t) = \sum_{\bm{k}} C(\bm{k},t) {\rm exp}(i\bm{k}\cdot \bm{r}),  \label{free_expand}
\end{eqnarray} 
and substitution of Eq. (\ref{free_expand}) into Eq. (\ref{1GP_fluc}) leads to the equation of $C(\bm{k},t)$:
\begin{eqnarray}
i\hbar \frac{\partial}{\partial t}C(\bm{k},t) &=&  \epsilon_{\rm f}(k) C(\bm{k},t) \nonumber \\
&+& g\sum_{\bm{k}_1,\bm{k}_2,\bm{k}_3} C(\bm{k}_1,t)C(\bm{k}_2,t)C(\bm{k}_3,t)^* \delta(\bm{k}_1 + \bm{k}_2 - \bm{k}_3 - \bm{k}),  \label{four_eq}
\end{eqnarray} 
Therefore, in the case of the wave around the no-condensate, the linear wave is free-particle-like, and the interaction between the waves is the four-wave interaction. 

On the other hand, in the latter case with the strong condensate, the situation changes drastically \cite{Dyachenko92,Zakharov05,Nazarenko06,Proment09,FT15}. In this case, the wave function is expanded as 
\begin{eqnarray}
\psi(\bm{r},t) = \psi_{0}(t) (1+ \delta \phi(\bm{r},t)), \label{condensate1}
\end{eqnarray}
\begin{eqnarray}
\psi_{0}(t) = \frac{1}{V} \int \psi(\bm{r},t) d\bm{r}, \label{condensate2}
\end{eqnarray}
where the strong condensate $\psi_{0}(t)$ is the Fourier component of $\psi(\bm{r},t)$ with zero wave number, and 
$\delta \phi(\bm{r},t)$ is the fluctuation around the strong condensate. 

We substitute Eq. (\ref{condensate1}) into the GP equation (\ref{1GP}) in the uniform system, deriving equations for $\psi_0$ and 
$\bar{\phi} (\bm{k},t) = \mathcal{F}[\delta  \phi (\bm{r},t)]$:
\begin{eqnarray}
i\hbar \frac{\partial}{\partial t} \psi _0 =  g \rho_0 \psi_0 \Big[  1 +  \sum_{\bm{k}_1} \big(  2|\bar{\phi} (\bm{k}_1)|^2 + \bar{\phi} (\bm{k}_1)\bar{\phi} (-\bm{k}_1) \big)  \Big] ,  \label{GP_condensate}
\end{eqnarray}
\begin{eqnarray}
i\hbar \psi_0 \frac{\partial}{\partial t} \bar{\phi}(\bm{k}) = &-& i\hbar \bar{\phi}(\bm{k}) \frac{\partial}{\partial t} \psi_0 + \frac{\hbar^2 k^2} {2m} \psi_0 \bar{\phi}(\bm{k}) + g \rho_0 \psi_0 \Big[  2\bar{\phi}(\bm{k})+ \bar{\phi}^{*}(-\bm{k})  \nonumber \\ 
&+&  2\sum_{\bm{k}_1 \bm{k}_2} \bar{\phi}^{*}(\bm{k}_1)\bar{\phi}(\bm{k}_2)\delta(\bm{k}+\bm{k}_1-\bm{k}_2) \nonumber \\
&+&  \sum_{\bm{k}_2 \bm{k}_3} \bar{\phi}(\bm{k}_2)\bar{\phi}(\bm{k}_3)\delta(\bm{k}-\bm{k}_2-\bm{k}_3)    \nonumber \\
&+& \sum_{\bm{k}_1,\bm{k}_2,\bm{k}_3} \bar{\phi}(\bm{k}_1,t)\bar{\phi}(\bm{k}_2,t)\bar{\phi}(\bm{k}_3,t)^* \delta(\bm{k}_1 + \bm{k}_2 - \bm{k}_3 - \bm{k}) \Big],   \label{GP_fluctuation1}
\end{eqnarray}
with $\rho_0(t) = |\psi_{0}(t)|^2$.

When we neglect the nonlinear term, e.g., the second- and third-order of $\delta \phi(\bm{r},t)$, we can solve these equations. The solution of $\psi_{0}(t)$ is given by 
\begin{eqnarray}
\psi_{0}(t) = \sqrt{n_0} {\rm exp}(-i\mu t/\hbar), \label{condensate3}
\end{eqnarray}
with $\mu = g n_0$, where $n_0$ is the initial value of $|\psi_{0}(t=0)|^2$ corresponding to the initial particle number of the strong condensate. 
The fluctuation has a solution: 
\begin{eqnarray}
\bar{\phi}(\bm{k},t) = u(k)B(\bm{k},t) + v(k)B^{*}(-\bm{k},t),  \label{Bogo_trans1}
\end{eqnarray}
\begin{eqnarray}
B(\bm{k},t) = B(\bm{k},0) {\rm exp}(-i\epsilon _{\rm b}(k)t/\hbar ), \label{Bogo_trans2}
\end{eqnarray}
\begin{eqnarray}
u(k) = \sqrt{ \frac{1}{2} \Big( \frac{\epsilon _{0}(k) + g n_{0}}{\epsilon_{\rm b}(k)} +1 \Big)  },   \label{Bogo_trans3}
\end{eqnarray}
\begin{eqnarray} 
v(k) = - \sqrt{ \frac{1}{2} \Big( \frac{\epsilon _{0}(k) + g n_{0}}{\epsilon_{\rm b}(k)} -1 \Big) },   \label{Bogo_trans4}
\end{eqnarray}
\begin{eqnarray}
\epsilon _{\rm b}(k) = \sqrt{\epsilon _{0}(k)(\epsilon _{0}(k) + 2g n_{0})}.  \label{Bog_dis}
\end{eqnarray}
The dispersion relation shows the linear $k$ dependence in the low-wave number region and quadratic $k$ dependence in the high-wave number region. This kind of wave is called the Bogoliubov excitation or Bogoliubov wave.

When we include the nonlinear terms, the time-development is different from Eq. (\ref{Bogo_trans2}) \cite{FT15}. In this case, we need to derive the time-development 
equation of $B(\bm{k},t)$. When we neglect the third order of $\delta \phi(\bm{r},t)$, the equation is given by 
\begin{eqnarray}
i\hbar \frac{\partial}{\partial t} B(\bm{k})  =  \frac{\partial H}{\partial B^{*}(\bm{k})},  \label{Bogo_equ}
\end{eqnarray}
\begin{eqnarray}
H_2 = \sum _{\bm{k}_{1}} \epsilon _{\rm b}(k_1) |B(\bm{k}_{1})|^{2},  \label{Hamiltonian4}
\end{eqnarray}
\begin{eqnarray}
H_3= & &\sum _{\bm{k}_1, \bm{k}_2, \bm{k}_3} \delta(\bm{k}_1 - \bm{k}_2 - \bm{k}_3)  V(\bm{k}_1, \bm{k}_2, \bm{k}_3)  \nonumber \\
&\times& \Big( B^{*}(\bm{k}_1)B(\bm{k}_2)B(\bm{k}_3) + B(\bm{k}_1) B^{*}(\bm{k}_2)B^{*}(\bm{k}_3)  \Big) \nonumber \\
&+& \sum _{\bm{k}_1, \bm{k}_2, \bm{k}_3} \delta(\bm{k}_1 + \bm{k}_2 + \bm{k}_3)  W(\bm{k}_1, \bm{k}_2, \bm{k}_3) \nonumber \\ 
&\times& \Big( B^{*}(\bm{k}_1)B^{*}(\bm{k}_2)B^{*}(\bm{k}_3) + B(\bm{k}_1)B(\bm{k}_2)B(\bm{k}_3) \Big),    \label{Hamiltonian5}
\end{eqnarray}
where $V$ and $W$ are the interaction functions for Bogoliubov waves.  These functions are given by 
\begin{eqnarray}
V(\bm{k}_1, \bm{k}_2, \bm{k}_3) = g \rho_{0} & \Big( & u(\bm{k}_1) u(\bm{k}_2) u(\bm{k}_3) + v(\bm{k}_1) v(\bm{k}_2) u(\bm{k}_3) \nonumber \\ 
&+& v(\bm{k}_1) u(\bm{k}_2)v(\bm{k}_3) + v(\bm{k}_1) v(\bm{k}_2) v(\bm{k}_3) \nonumber \\ 
&+& u(\bm{k}_1) v(\bm{k}_2) u(\bm{k}_3) + u(\bm{k}_1) u(\bm{k}_2) v(\bm{k}_3) \Big),  \label{Interaction1}
\end{eqnarray}
\begin{eqnarray}
W(\bm{k}_1, \bm{k}_2, \bm{k}_3) = g \rho_{0} \Big( u(\bm{k}_1) v(\bm{k}_2) v(\bm{k}_3) + v(\bm{k}_1) u(\bm{k}_2) u(\bm{k}_3) \Big).  \label{Interaction2}
\end{eqnarray}
This equation is essentially different from Eq. (\ref{four_eq}) corresponding to the case of the wave around no-condensate. First, the dispersion relation $\epsilon _{\rm b}(k)$ is not free-particle-like. Second, the interaction between the Bogoliubov wave is the three-wave interaction, which is very important in application of the weak wave turbulence theory. 

Originally, these weak four-wave and three-wave turbulence in the GP model were studied in optical turbulence \cite{Dyachenko92}. 
However, in the advent of the ultracold atomic BECs, studies of such weak wave turbulences are performed in the field of the BECs. As for the four-wave case, Nazarenko $et~al.$ and Proment $et~al.$ investigated the correlation function of the wave function in 2D and 3D systems \cite{Nazarenko06,Proment09}. As for the three-wave case, Proment $et~al.$ and Fujimoto $et~al.$ studied the feature of weak three-wave turbulence in 3D systems \cite{Proment09,FT15}.  Proment $et~al$ investigated the transition between the weak three-wave and four-wave turbulence, which is also studied in 2D system \cite{Nazarenko06}. Fujimoto $et~al.$ investigated two kinds of correlation in addition to the correlation function of the wave function. In the following, we show the detailed results for the four-wave and three-wave cases. \\

\noindent{\it --four-wave case--}\\
\begin{figure}[t]
\begin{center}
\includegraphics[keepaspectratio, width=12cm,clip]{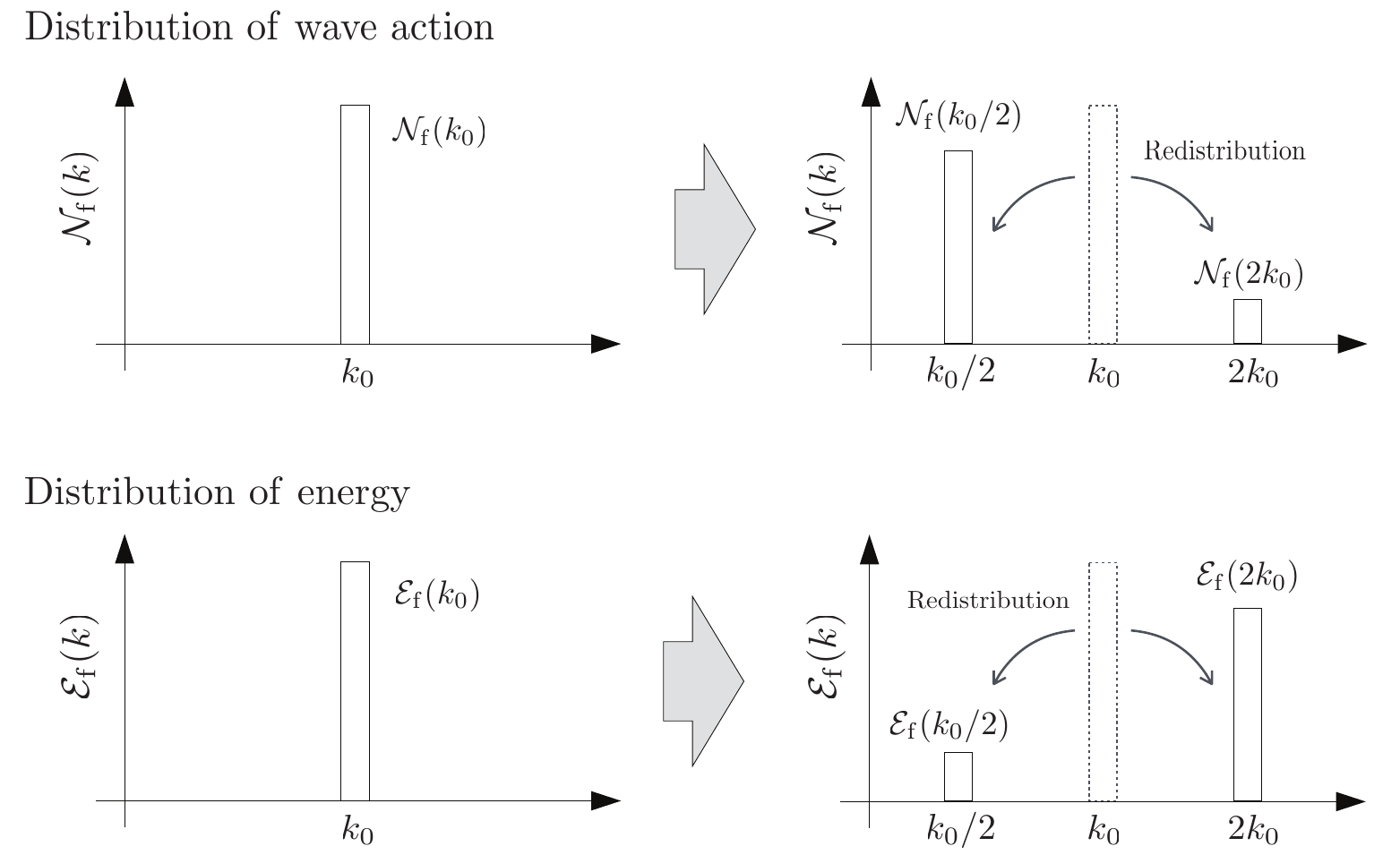}
\caption{Fj\o rft argument \cite{Fjo}. The initial state has a wave with wave number $k_0$. After time-development, the initial wave of $k_0$ is redistributed into two waves with wave numbers $k_0 /2$ and $2k_0$. Then, the conservation laws restrict the distribution of $\mathcal{E}_{\rm f}(k)$ and $\mathcal{N}_{\rm f}(k)$, leading to direct cascade for the energy and inverse cascade for the action (see text). }
\label{inverse_cascade}
\end{center}
\end{figure}

The striking feature of the four-wave interaction is the existence of two conserved quantities. 
From Eq. (\ref{four_eq}) and the weak wave turbulence theory, the energy $E_{\rm f}$ and the action $N_{\rm f}$ of linear waves are conserved quantities. The definition of these are given by 
\begin{eqnarray}
E_{\rm f} = \sum_{\bm{k}} \epsilon_{\rm f}(k) |C(\bm{k},t)|^2 = \triangle k \sum_{\bm{k}} \mathcal{E}_{\rm f}(k), \label{inverse1}
\end{eqnarray}
\begin{eqnarray}
N_{\rm f} = \sum_{\bm{k}}  |C(\bm{k},t)|^2 =\triangle k  \sum_{\bm{k}} \mathcal{N}_{\rm f}(k), \label{inverse2}
\end{eqnarray}
where a relation $\mathcal{E}_{\rm f}(k) = \epsilon_{\rm f}(k) \mathcal{N}_{\rm f}(k)$ holds. 
The conservation laws lead to the direct energy cascade and the inverse action cascade, 
which can be understood by the usual Fj\o rtoft argument as follows \cite{wt1,wt2,Fjo}. 

\begin{figure}[t]
\begin{center}
\includegraphics[keepaspectratio, width=11cm,clip]{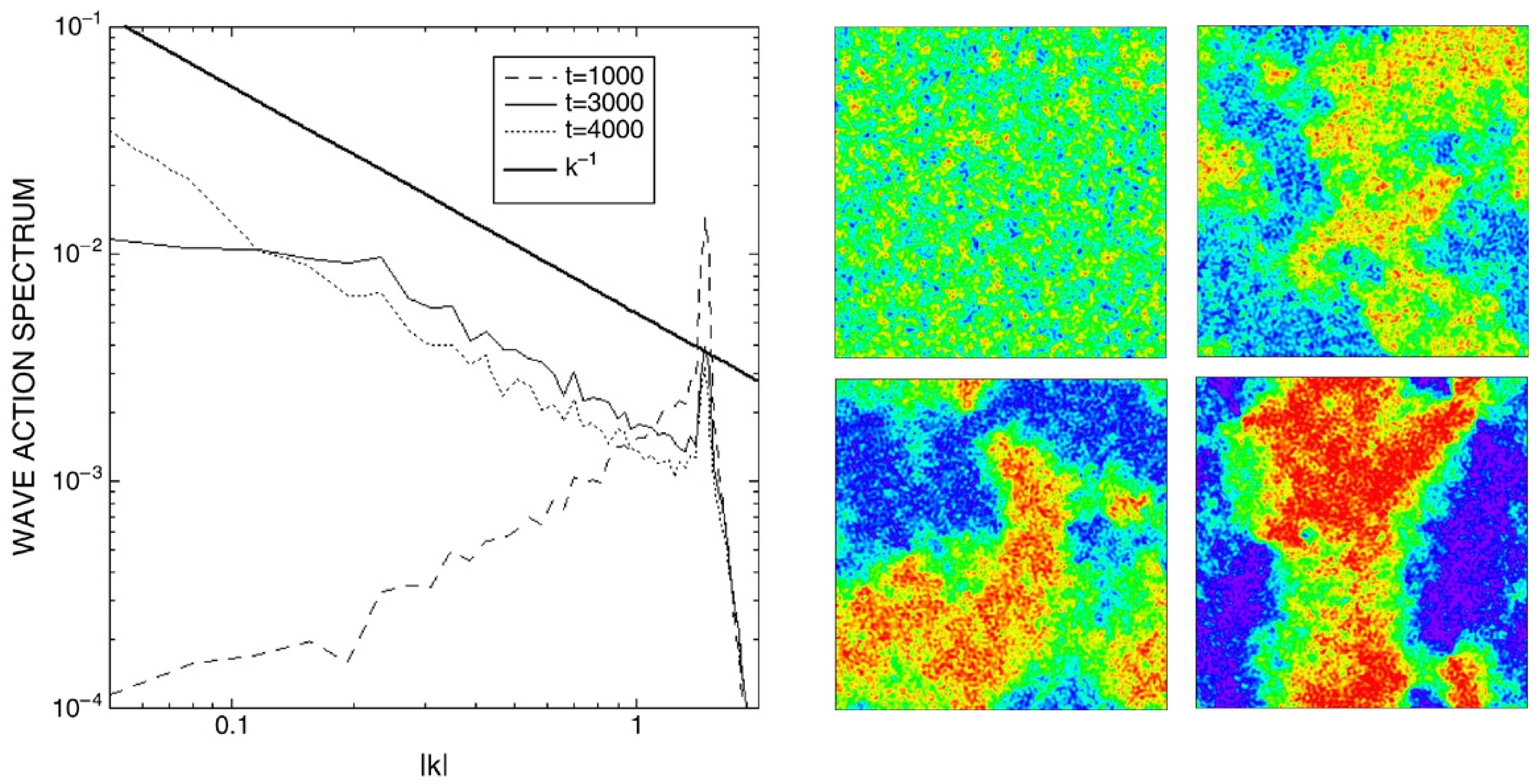}
\caption{Correlation function (left) and time-development of ${\rm Re}[\psi(x,y,t)]$ (right) in 2D weak wave turbulence with the four-wave interaction. In the right figure, the distributions are shown at the normalized times of $2500$, $5000$, $7500$, and $10000$. These results show that the high-wave number mode of the wave function is transferred into the low-wave number mode. [With permission from N. Nazarenko and M. Onorato, Physica D 219, 1-12, Elsevier, Copyright (2006)] (Color figure online)}
\label{fourwave_spe}
\end{center}
\end{figure}

We now consider the initial state shown in Fig. \ref{inverse_cascade}, where the linear-wave is excited initially only at a wave number $k_0$. From this state, the system is supposed to evolve, and the linear wave is assumed to be transferred into two waves with two wave numbers $k_0 /2$ and $2k_0$. Then, the conservation laws of Eqs. (\ref{inverse1}) and (\ref{inverse2}) lead to 
\begin{eqnarray}
\mathcal{E}_{\rm f}(k_0) = \mathcal{E}_{\rm f}(k_0/2) + \mathcal{E}_{\rm f}(2k_0), 
\end{eqnarray}
\begin{eqnarray}
\mathcal{N}_{\rm f}(k_0) = \mathcal{N}_{\rm f}(k_0/2) + \mathcal{N}_{\rm f}(2k_0). 
\end{eqnarray}
Solving these coupled equations using the relation $\mathcal{E}_{\rm f}(k) = \epsilon_{\rm f}(k) \mathcal{N}_{\rm f}(k)$, we can derive
$\mathcal{N}_{\rm f}(k_0/2) = 4\mathcal{N}_{\rm f}(k_0)/5$, 
$\mathcal{N}_{\rm f}(2k_0) = \mathcal{N}_{\rm f}(k_0)/5$, 
$\mathcal{E}_{\rm f}(k_0/2) = \mathcal{E}_{\rm f}(k_0)/5$, and
$\mathcal{E}_{\rm f}(2k_0) = 4 \mathcal{E}_{\rm f}(k_0)/5$. 
Figure \ref{inverse_cascade} shows the image of the distribution corresponding to this solution, which 
shows that the linear wave energy is transferred from the low- to the high-wave-number region and vice versa for the wave action. Therefore, the existence of the two conserved quantities induces the direct and inverse cascade. 
This discussion is called the Fj\o rft argument \cite{Fjo}. 

The feature of four-wave system is numerically confirmed in the 2D system of the GP model by Nazarenko $et~al.$ \cite{Nazarenko06}. In this subsection, although we focus on 3D QT, the numerical results are shown for the 2D system.  Figure \ref{fourwave_spe} shows the spectrum $\sum _{\Omega(\bm{k}_1,k)} |\psi(\bm{k}_1,t)|$ and the time-development of the spatial distribution for the real part of the wave function ${\rm Re}[\psi(x,y,t)]$. These graphs show low wave number condensation similar to the inverse cascade. However, in the 2D case, there is a problem concerning the derivation of the power exponent with the turbulent cascade in the weak wave turbulence theory, where the Kolmogorov-Zakharov power exponent for the inverse cascade is irrelevant and that for the direct cascade is same as the exponent of thermodynamic equilibrium. The detail of this issue is described by Ref.\cite{wt2,Dyachenko92,Nazarenko06}.\\

\noindent{\it --three-wave case--}\\
Apart from the weak wave turbulence with the four-wave interaction, the case with the three-wave interaction in 3D systems is studied analytically and numerically \cite{Dyachenko92,Zakharov05,Proment09,FT15}. In this turbulence, the three-wave interaction breaks the conservation of the wave action, so the inverse cascade is not expected. Here, according to Ref. \cite{FT15}, we review this kind of turbulence.

This paper addresses the 3D system, and the following three spectra are investigated:
\begin{eqnarray}
C_{\rm b} (k,t) =\frac{1}{\triangle k}  \sum _{\Omega(\bm{k}_1,k)}  \langle | B (\bm{k}_1,t) |^{2} \rangle ,  \label{bogo_correlation1}
\end{eqnarray}
\begin{eqnarray}
C_{\rm w} (k,t) = \frac{1}{\triangle k} \sum _{\Omega(\bm{k}_1,k)}  \langle |\tilde{\psi} (\bm{k}_1,t) |^{2}  \rangle ,  \label{wave_correlation1}
\end{eqnarray}
\begin{eqnarray}
C_{\rm d} (k,t) = \frac{1}{\triangle k} \sum _{\Omega(\bm{k}_1,k)}  \langle |\tilde{\rho} (\bm{k}_1,t) |^{2} \rangle  ,  \label{density_correlation1}
\end{eqnarray}
with the Fourier components of the density distribution $\tilde{\rho}(\bm{k},t) = \mathcal{F}[\rho(\bm{r},t)]$.
The bracket indicates an ensemble average. The physical meaning of these spectra is that $C_{\rm b}(k,t)$ is the Bogoliubov wave distribution in the wave number space, and $C_{\rm w}(k,t)$  and $C_{\rm d}(k,t)$ are two-point spatial correlation functions for the wave function and density distribution. 

\begin{figure}[t]
\begin{center}
\includegraphics[keepaspectratio, width=12cm,clip]{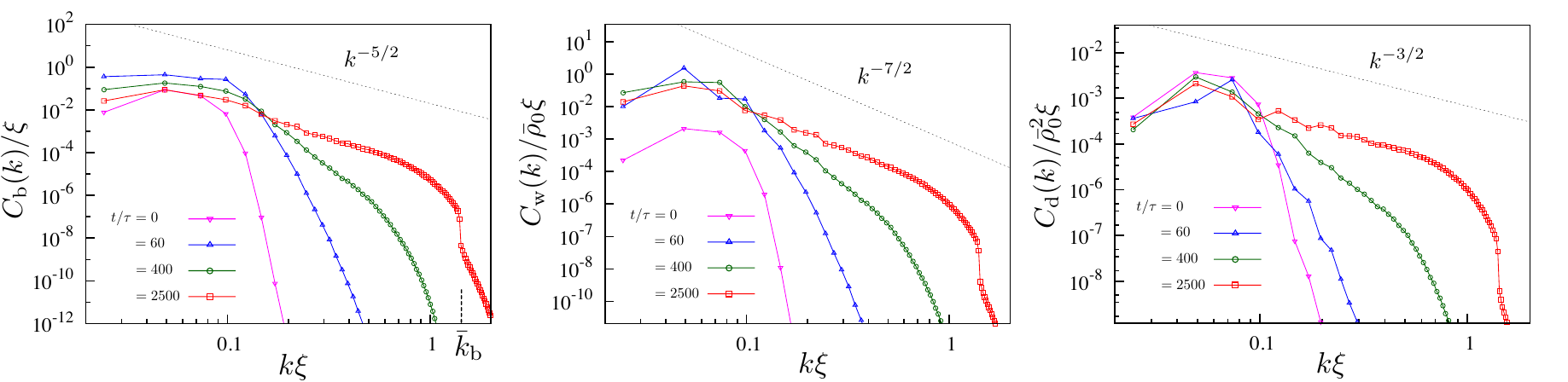}
\caption{Time-development of three spectra in the weak wave turbulence with the three-wave case.
The Bogoliubov wave distribution $C_{\rm b} (k,t)$ (left), the correlation function of the wave function $C_{\rm w} (k,t)$ (middle), and the density correlation function $C_{\rm d} (k,t)$ (right) are shown. At the late stage, these spectra exhibit good agreement with analytical result of Eqs. (\ref{bogo_correlation2})--(\ref{density_correlation2}).
[Reprinted figure with permission from \href{https://doi.org/10.1103/PhysRevA.93.039901}{K. Fujimoto and M. Tsubota, Phys. Rev. A {\bf 93}, 039901 (2016)}. Copyright (2016) by the American Physical Society.] (Color figure online)
\label{Bogo_spe}} 
\end{center}
\end{figure}

The application of the weak wave turbulence theory leads to the power laws given by 
\begin{eqnarray}
C_{\rm b} (k) \propto k^{-5/2}, 
\label{bogo_correlation2}
\end{eqnarray}
\begin{eqnarray}
C_{\rm w} (k) \propto k^{-7/2}, 
\label{wave_correlation2}
\end{eqnarray}
\begin{eqnarray}
C_{\rm d} (k) \propto k^{-3/2}. 
\label{density_correlation2}
\end{eqnarray}

To confirm these power laws, the numerical calculation of the GP equation in the 3D system is performed. In this calculation, the weak wave turbulence is generated by an unstable initial state with random phase. Then, each spectra are computed in the decaying turbulence. The results are shown in Fig. (\ref{Bogo_spe}), exhibiting good agreement with Eqs. (\ref{bogo_correlation2})--(\ref{density_correlation2}).

Finally, we comment on the experimental observation for the density correlation function. 
As shown in Fig. \ref{projection}, the experimental distribution is integrated along the direction of laser injection. This can change the power exponent of density correlation function. To consider this problem, we derive the power exponent for the correlation function of $\rho_{\rm exp}(x,y,t)$, which is defined by
\begin{eqnarray}
\rho_{\rm exp}(x,y,t) = \int \rho(x,y,z,t) dz,  \label{density_exp1}
\end{eqnarray}

\begin{figure}[t]
\begin{center}
\includegraphics[keepaspectratio, width=10cm,clip]{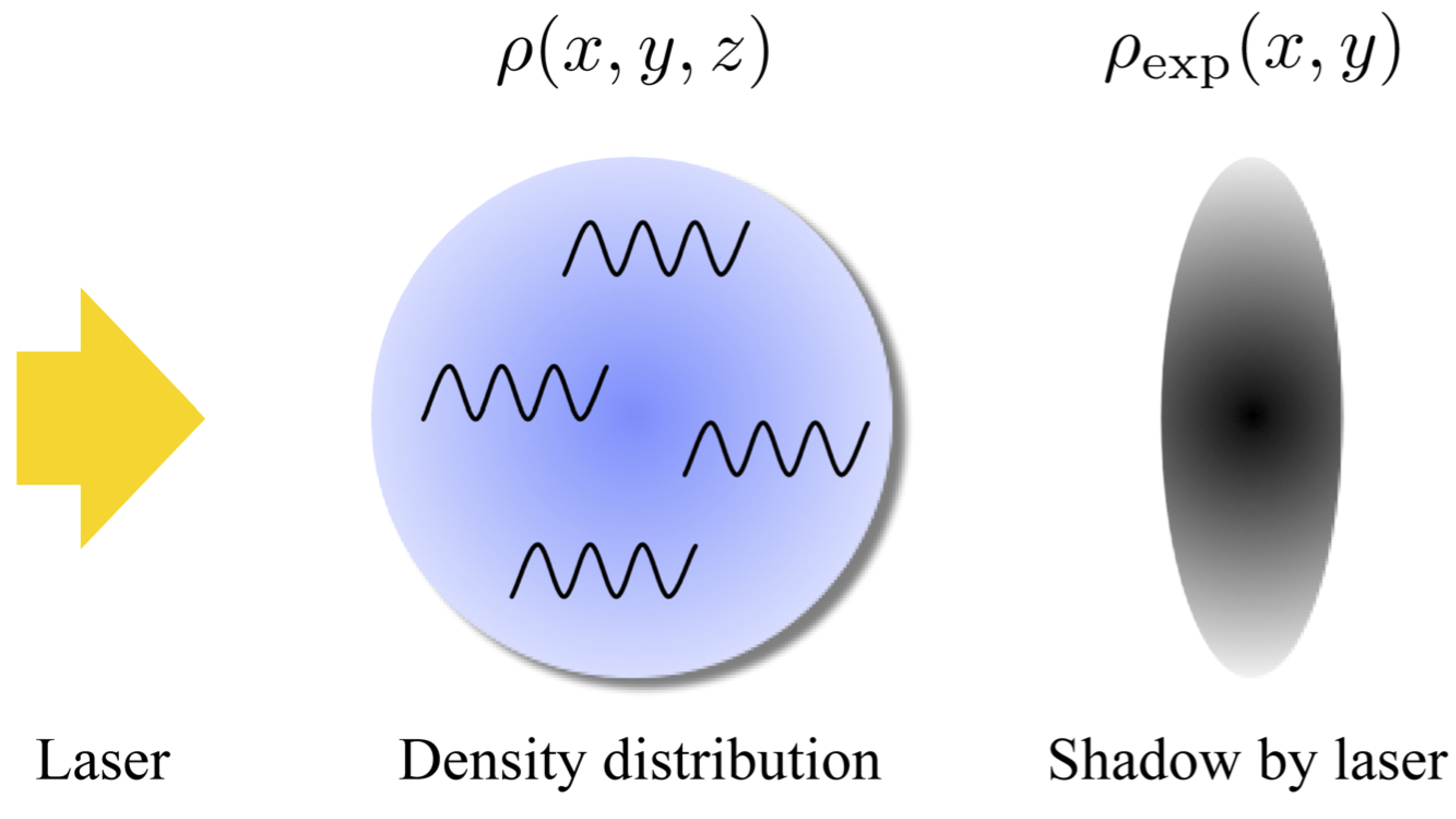}
\caption{Projected density distribution. In typical experiments with ultracold atomic gases, the density distribution is observed by the time-flight-expansion, where the atomic gas is released from the trapping potential and laser light is injected. Then, as shown in the figure, we can obtain the shadow of the atomic gas, which is the experimental observable $\rho_{\rm exp}(x,y)$. This is the integrated density distribution along the laser injection direction. (Color figure online)} 
\label{projection}
\end{center}
\end{figure}

\begin{eqnarray}
C_{\rm d,exp}(k,t) = \frac{1}{\Delta k}\sum _{\bm{k}_1 \in \Omega_k} \langle | \bar{\rho}_{\rm exp}(k_{1,x},k_{1,y},t) |^2 \rangle, 
\label{2d_correlation}
\end{eqnarray}
with $\bar{\rho}_{\rm exp} (k_x,k_y,t) = \mathcal{F}[\rho_{\rm exp}(x,y,t)]$. 

Then, considering the isotropy in the wave number space, the power exponent of $C_{\rm d,exp}(k,t)$ is derived. First, we focus on Eq. (\ref{density_exp1}), which means that this integral calculation corresponds to the Fourier transformation for the $z$-direction with $k_z=0$. Thus, the relation $\bar{\rho}_{\rm exp}(k_x,k_y,t) =  \bar{\rho}(k_x,k_y,k_z=0,t)$ is satisfied. 
As a result, the isotropy in the wave number space and Eq. (\ref{density_correlation2}) lead to 
\begin{eqnarray}
\langle |\bar{\rho}_{\rm exp}(k_x,k_y)|^2 \rangle &\propto & (k_{x}^{2}+k_{y}^{2})^{-7/2} \nonumber \\
 &\propto & k^{-7/2},  \label{density_exp3}
\end{eqnarray}
where we use $\langle |\bar{\rho}(k_x,k_y,k_z)|^2 \rangle \propto (k_{x}^{2}+k_{y}^{2}+k_{z}^{2})^{-7/2}$ derived by Eq. (\ref{density_correlation2}) and the Jacobian in the 3D system. 

Hence, performing the summation of Eq. (\ref{2d_correlation}), we can derive
\begin{eqnarray}
C_{\rm d,exp}(k) &\propto & k^{-7/2} \times k \nonumber \\
&\propto & k^{-5/2}. 
\end{eqnarray}
When the isotropy assumption in the wave number space is valid, this $-5/2$ power law is equivalent to Eq. (\ref{density_correlation2}), which is the important law to confirm whether the weak wave turbulence with the three-wave interaction is realized in the experiments.

\subsection{Theoretical studies of QT for single-component BECs in 2D systems}
The strong point of experiment with ultracold atomic gases is the high controllability of various kinds of system parameters. 
For this reason, we can set the 2D system, which enables us to study 2D QT. 
In this subsection, we describe the 2D CT and show recent works for QT in the 2D systems. 

\subsubsection{Brief review of 2D classical turbulence}
Let us consider the classical incompressible fluid system obeying the two-dimensional Navier—Stokes equation. As described in the following, this system is known to show the direct and inverse cascades. The key point is the existence of two conserved quantities, namely kinetic energy and enstrophy in the inviscid limit \cite{Frisch,davidson}.

The system is supposed to be a periodic 2D square system whose linear size and area are denoted as $L$ and $S=L^2$, respectively. The Navier--Stokes equation is given by 
 \begin{eqnarray}
\frac{\partial}{\partial t} \bm{u} (\bm{r},t) + (\bm{u}(\bm{r},t) \cdot \bm{\nabla}) \bm{u}(\bm{r},t)  = -\frac{1}{\rho} \bm{\nabla}P(\bm{r},t) +\nu \triangle \bm{u}(\bm{r},t),  
\end{eqnarray}
with the velocity field $\bm{u} (\bm{r},t)$, pressure $P(\bm{r},t)$, mass density $\rho$, and viscosity coefficient $\nu$. The pressure is determined by the incompressible condition ${\rm div}  \bm{u} (\bm{r},t)=0$.
We define the kinetic energy $E_{\rm class}$ of $\bm{u} (\bm{r},t)$ per unit area as
 \begin{eqnarray}
 E_{\rm class}(t) = \frac{\rho}{2S} \int \bm{u} (\bm{r},t)^2 d\bm{r}. 
\end{eqnarray}
Then, the kinetic energy spectrum is expressed by 
\begin{eqnarray}
\mathcal{E}_{\rm class}(k,t) = \frac{\rho}{2\triangle k} \sum_{\Omega(\bm{k}_1,k)} |\tilde{\bm{u}} (\bm{k}_1,t)|^2,  
\end{eqnarray}
with $\tilde{\bm{u}} (\bm{k},t) = \mathcal{F}[\bm{u}(\bm{r},t)]$. This satisfies the relation $E_{\rm class}(t) = \triangle k \sum_{k}\mathcal{E}_{\rm class}(k,t)$. In the inviscid limit, this kinetic energy $E_{\rm class}(t)$ is conserved, which is true in both 
2D and 3D systems. 

When the spatial dimension of the system is two, another conserved quantity appears. To explain this, we derive the equation of motion for the vorticity $\bm{\omega}(\bm{r},t) = {\rm rot}\bm{v}(\bm{r},t)$. 
In the 3D case, the equation becomes
\begin{eqnarray}
\frac{\partial}{\partial t} \bm{\omega} (\bm{r},t) + (\bm{u}(\bm{r},t) \cdot \bm{\nabla}) \bm{\omega}(\bm{r},t)  =
 (\bm{\omega}(\bm{r},t) \cdot \bm{\nabla}) \bm{u}(\bm{r},t) + \nu \triangle \bm{\omega}(\bm{r},t). 
\label{omega3d}
\end{eqnarray}
On the other hand, in the 2D system, the first term on the right hand side of Eq. (\ref{omega3d}) vanishes because 
the product of $\bm{\omega} \cdot \bm{\nabla}= \omega_z(x,y) \partial/ \partial z$ and $\bm{u} = (u_x(x,y),u_y(x,y),0)$ is zero due to the $z$-derivative. Thus, in the 2D system, the equation of the vorticity is given by
\begin{eqnarray}
\frac{\partial}{\partial t} \bm{\omega} (\bm{r},t) + (\bm{u}(\bm{r},t) \cdot \bm{\nabla}) \bm{\omega}(\bm{r},t)  =
 \nu \triangle \bm{\omega}(\bm{r},t).
\label{omega2d}
\end{eqnarray}
The physical meaning of the disappearance of $ (\bm{\omega}(\bm{r},t) \cdot \bm{\nabla}) \bm{u}(\bm{r},t)$ is the impossibility of the stretching of vortex in the 2D system. For this reason, only in the 2D system, the enstrophy $W_{\rm class}(t)$ is conserved in the inviscid limit. The definition of this quantity is given by 
 \begin{eqnarray}
 W_{\rm class}(t) = \frac{\rho}{2S} \int \bm{\omega} (\bm{r},t)^2 d\bm{r}. 
\end{eqnarray}
We can easily verify that this time derivative of $ W_{\rm class}(t)$ is zero when the system obeys Eq. (\ref{omega2d}) with $\nu = 0$.
For the following explanation, we define the spectrum of the enstrophy as follows:
\begin{eqnarray}
\mathcal{W}_{\rm class}(k,t) = \frac{\rho}{2\triangle k} \sum_{\Omega(\bm{k}_1,k)} |\tilde{\bm{\omega}} (\bm{k}_1,t)|^2,  
\end{eqnarray}
with the Fourier component of the vorticity $\tilde{\bm{\omega}} (\bm{k},t) = \mathcal{F}[\bm{\omega}(\bm{r},t)]$. 
Noting the relation $\tilde{\bm{\omega}} (\bm{k},t) = i\bm{k} \times \tilde{\bm{u}} (\bm{k},t)$ and the incompressible condition $\bm{k}\cdot \tilde{\bm{u}} (\bm{k},t)=0$, we can obtain
 \begin{eqnarray}
 \mathcal{W}_{\rm class}(k,t) = k^2 \mathcal{E}_{\rm class} (k,t). \label{enstrophy_spe}
\end{eqnarray}

This situation resembles the weak wave turbulence with the four-wave interaction in the GP model as described in Fig. \ref{inverse_cascade}. In the present case, Eq. (\ref{enstrophy_spe}) is satisfied, so the kinetic energy corresponds to the action of the wave and the enstrophy corresponds to the wave energy. Therefore, according to the previous section (Fj\o rft argument), the kinetic energy and enstrophy can exhibit the inverse and direct cascade, respectively. This inverse cascade causes energy transfer from high- to low-wave number region, which are considered to lead to the formation of large-scale vortices (vortex clusters). 

In this dual cascade, the kinetic energy spectrum shows two kinds of power laws:
\begin{eqnarray}
\mathcal{E}_{\rm class}(k) \propto \left\{ \begin{array}{ll}
k^{-5/3} & (k<k_{\rm f} );  \\
k^{-3} & (k_{\rm f}<k), \\
\end{array} \right.
\end{eqnarray}
where the $-5/3$ and $-3$ power laws correspond to the energy inverse cascade and the direct enstrophy cascade, respectively. 
The parameter $k_{\rm f}$ is the wave number with the energy injection. The derivation of the $-5/3$ power law is same as that in 3D CT. On the other hand, the $-3$ power law can be obtained by slightly modifying the derivation of the $-5/3$ power law. As discussed above, the $-3$ power law corresponds to the enstrophy cascade, which means that this law has constant enstrophy flux. This fact and the usual dimensional analysis in the Kolmogorov 41 theory lead to the $-3$ power exponent. Indeed, the dimension of velocity is $[L]/[T]$ with length dimension $[L]$ and time dimension $[T]$, so that the kinetic energy spectrum $\mathcal{E}_{\rm class}$ and enstrophy flux $\eta$ per unit area have the following dimensions:
 \begin{eqnarray}
[\mathcal{E}_{\rm class}] = \frac{[M][L^3]}{[T^2]}, \label{inverse}
\end{eqnarray}
 \begin{eqnarray}
[\eta] = \frac{[M]}{[T^3]}, 
\end{eqnarray}
with the mass dimension $[M]$. By eliminating $[T]$ from Eq. (\ref{inverse}), we can derive the $-3$ power law.  

In summary, in the 2D system obeying the Navier—Stokes equation, the kinetic energy and the enstrophy are conserved in the inviscid limit, which induces the dual cascade for kinetic energy and enstrophy. 

\subsubsection{Inverse cascade of 2D QT}
There are many studies for 2D QT  \cite{Bor11,Bor12,Horng,Num10,Sch12,Reeves12,Bradley12,White12,Reeves13,Billam14,Simula14,Billam15,Simula16,Yu16}, some of which discuss the possibility of the inverse cascade in QT. To our understanding, the existence of the inverse cascade in QT is controversial. In the GP model, quantized vortices can disappear through pair annihilation, 
so enstrophy is not a conserved quantity. Thus, it seems that the inverse cascade in QT may not occur. 
In fact, Numasato $et~al.$ have argued for the non-existence of the inverse cascade \cite{Num10}. However, some studies obtain evidence of the inverse cascade in QT \cite{Horng,Reeves13}. 
The result of Ref. \cite{Num10} has already been reviewed in Ref. \cite{TKT}; in this subsection, we describe the other study showing the evidence of the inverse cascade in QT \cite{Reeves13}. 

\begin{figure}[t]
\begin{center}
\includegraphics[keepaspectratio, width=11.5cm,clip]{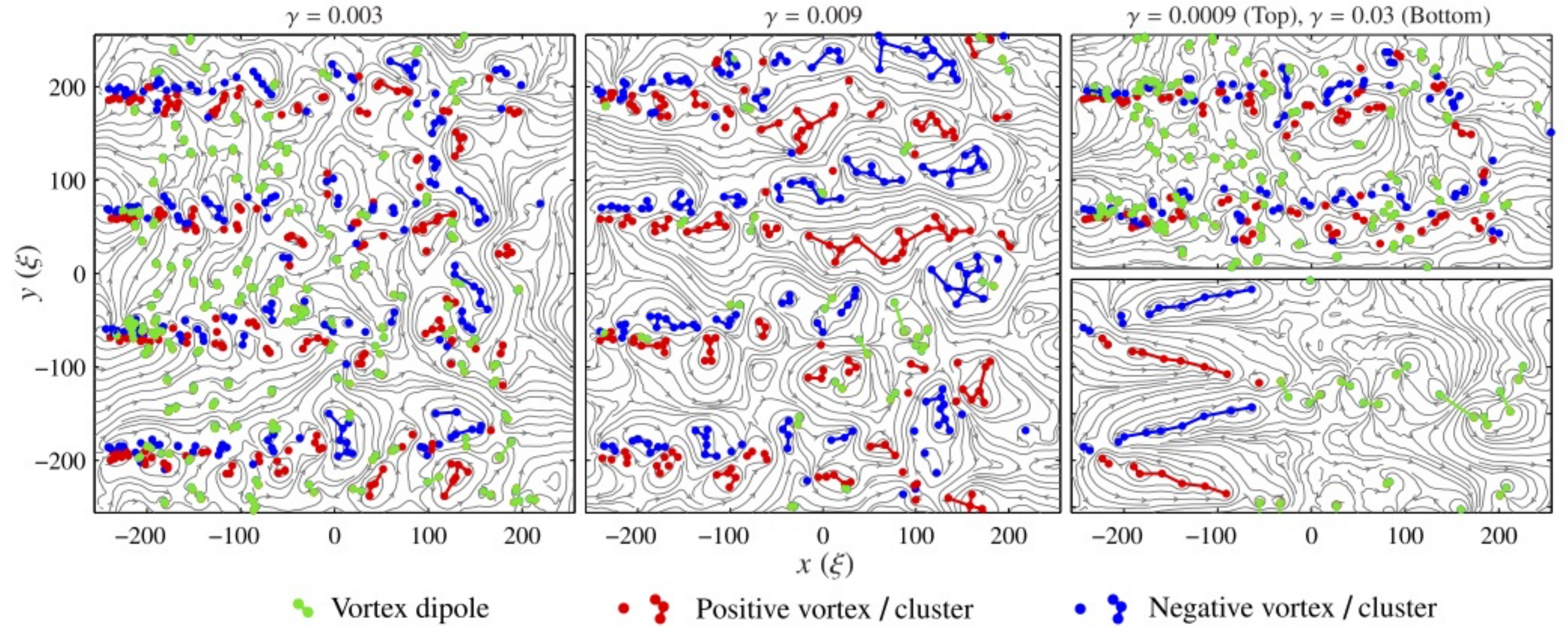}
\caption{Vortex configuration and streamline in 2D QT. This result is obtained by the damped GP equation with $\gamma = 0.003$ (left), $\gamma = 0.009$ (middle), $\gamma = 0.0009$ (right top), and $\gamma = 0.03$ (right bottom). Using the cluster-finding algorithm, numerical calculations classify the vortex structure; green, red, and blue points correspond to the vortex dipole, positive vortex/cluster, and negative vortex/cluster, respectively.  
[Reprinted figure with permission from \href{https://doi.org/10.1103/PhysRevLett.110.104501}{M. T. Reeves $et~al.$, Phys. Rev. Lett. {\bf 110}, 104501 (2013)}. Copyright (2013) by the American Physical Society.]
(Color figure online)} 
\label{2DQT1}
\end{center}
\end{figure}

Numerical evidence of the inverse cascade in QT is reported by Reeves $et~al.$ \cite{Reeves13}, who investigated vortex clustering, the energy spectra, and spectral condensation with the damped GP equation defined by 
\begin{eqnarray}
i\hbar \frac{\partial}{\partial t} \psi(\bm{r},t) = (1-i\gamma)(\mathcal{L}-\mu) \psi(\bm{r},t), 
\end{eqnarray}
\begin{eqnarray}
\mathcal{L} = -\frac{\hbar^2}{2M}\biggl(  \frac{\partial^2}{\partial x^2} + \frac{\partial^2}{\partial y^2} \biggl) + V(x,y) + g_2 |\psi(\bm{r},t)|^2 \psi(\bm{r},t). 
\end{eqnarray}
Here, $g_2$ and $\gamma$ are the interaction coefficient in the 2D system and the phenomenological dissipation parameter, respectively. 

The method for generating QT is the application of a velocity field in the $+x$-direction while setting obstacle potentials at four points. Figure \ref{2DQT1} is the vortex configuration and the streamlines obtained by this excitation. To determine whether the clustering of vortices occurs, they develop a cluster-finding algorithm, which leads to the classification of three kinds of vortex structures: (i) vortex dipole (green), (ii) positive vortex/cluster (red), and (iii) negative vortex/cluster (blue), as shown in Fig. \ref{2DQT1}. Based on this algorithm, the tendency of vortex clustering occurs in the case with $\gamma = 0.009$. However, in other cases, strong evidence of the vortex clustering is not clearly confirmed. 

To investigate this behavior from the perspective of the inverse energy cascade, the kinetic energy spectrum is calculated. In the case with $\gamma=0.009$, the spectrum is found to be consistent with the $-5/3$ power law predicted by the inverse cascade. Furthermore, the spectral condensation induced by the flux of inverse cascade is confirmed through the rapid growth of the condensation in the low-wave number region. From these numerical results, the inverse cascade was demonstrated in 2D QT. 

After this study, a method used for calculating the energy flux is suggested by same group, where they utilized the point-vortex model \cite{Billam15}. Using this method, this group confirmed negative energy flux in a different setup. 

\subsubsection{Formation of the Onsager vortex of 2D QT}
\begin{figure}[t]
\begin{center}
\includegraphics[keepaspectratio, width=11cm,clip]{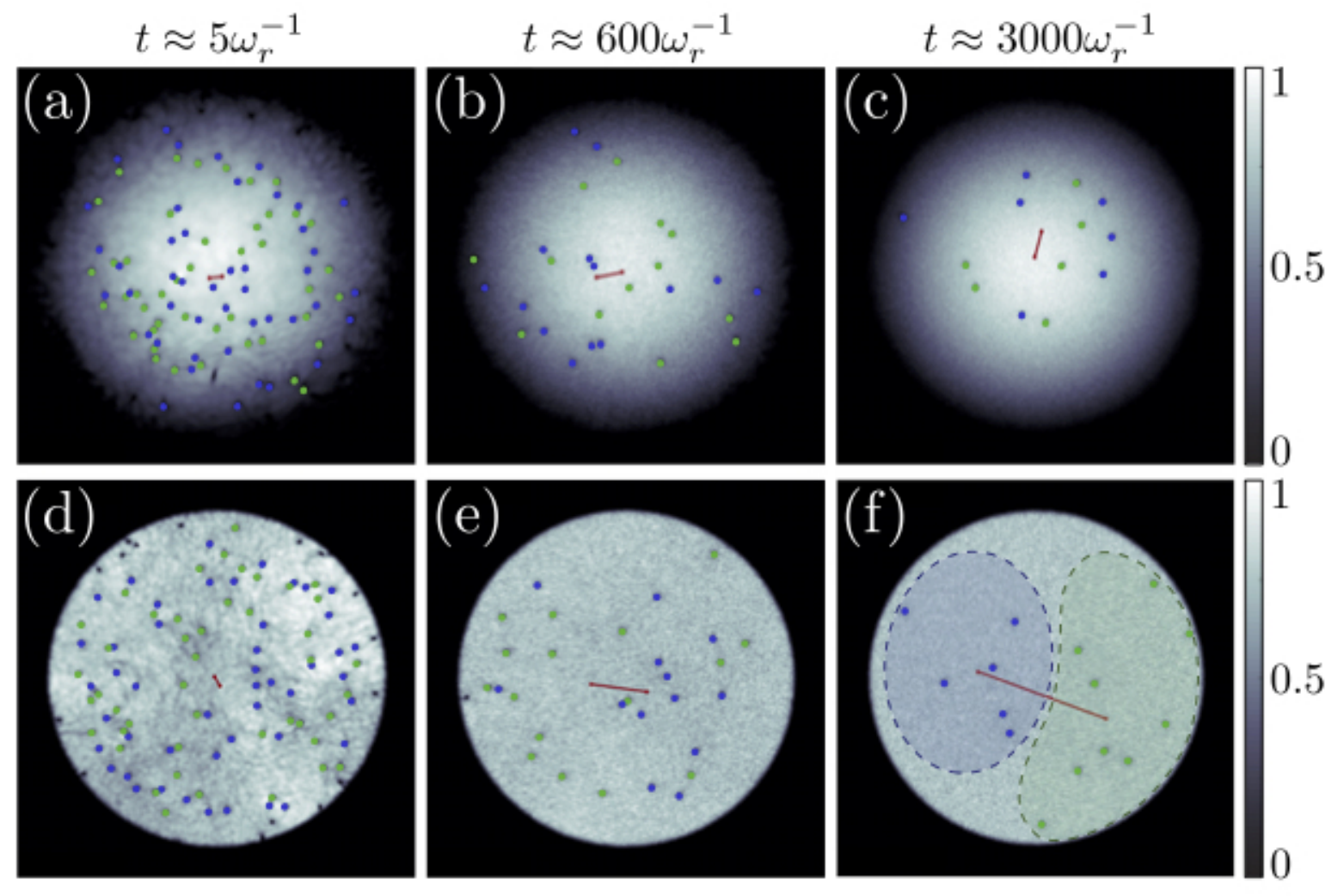}
\caption{Density profiles and vortex configurations in the harmonic trap (top) and the disk-shaped flat trap (bottom). Green and blue points denote the vortices and antivortices, respectively, and the red line represents the dipole moment for the vortex distribution. This result is obtained by numerical calculation of the GP equation. These graphs show that the Onsager vortex (negative temperature state) is dynamically formed in the disk-shaped flat trap. [Reprinted figure with permission from \href{https://doi.org/10.1103/PhysRevA.93.043614}{A. J. Groszek $et~al.$, Phys. Rev. A {\bf 93}, 043614 (2016)}. Copyright (2016) by the American Physical Society.] (Color figure online)} 
\label{Onsager}
\end{center}
\end{figure}

As another topic related to the inverse cascade, there is the Onsager vortex which is a state with negative temperature. This kind of vortex clustering is theoretically investigated by the GP equation and Monte Carlo calculation \cite{Billam14,Simula14,Simula16,Yu16}. Recently, Simula $et~al.$ found that such a negative temperature state is dynamically formed through the evaporative heating of vortex when the disk-shaped flat potential is used \cite{Simula14}  \footnote[1]{Note that they do not introduce any dissipation term into the GP equation, focusing on the dynamics in the isolated system.}. In this mechanism, the annihilation of vortices increases the energy per vortex, leading to the phase transition to the negative temperature phase. On the other hand, when the usual harmonic trapping potential is used, this formation cannot be confirmed. Figure \ref{Onsager} is the density profile and vortex configuration in both harmonic (top) and disk-shaped flat (bottom) potentials, exhibiting distinct differences in the vortex clustering. To understand this behavior, as discussed in Ref. \cite{Simula16}, the energy per vortex and the dependence of vortex dipole energy on the separation are important. 

\subsection{Theoretical studies of turbulence in spinor BECs}
From the perspective of quantum hydrodynamics, one of the most characteristic features of ultracold atomic BECs is the realization of multi-component BECs \cite{Stringari,Pethick}. Such systems are prepared by utilizing different atomic species or the internal degrees of freedom of atoms. For example, in the former case, $\rm{Yb}$ and $\rm{Rb}$ are used to realize two-component BECs. In the latter case, $^{87}{\rm Rb}$ with hyperfine spin $F=1$ is used, and a spin-1 spinor BECs is realized, which is a mixture of BEC with different magnetic quantum numbers $m_F = 1, 0, -1$ \cite{KU,Sta}. 

This system provides a novel stage for investigating turbulence in multi-component BECs. 
In contrast to conventional QT, this system has many degrees of freedom, e.g., internal spin states, which leads to various topological defects such as spin domain wall, half quantized vortices, $Z_2$ vortex, and monopoles \cite{KU}. Then, the multi-component BEC enriches the physics of QT because new types of problem such as the role of various topological defects and behavior of spin correlation in turbulence are posed.  

Note that the superfluid $^3{\rm He}$, which has A and B phases, is a similar system \cite{Volhardt}. In particular, the order parameter of the A phase has a similar structure to the spinor BEC, but due to the difficulty of the experimental and theoretical studies, QT in this system has not been investigated. Thus, QT in multi-component BEC has the possibility to exhibit exotic properties not shown in QT of 
superfluid helium. 

In this section, as an example of QT in multi-component BEC, we review the turbulence in ferromagnetic spinor BECs. In this turbulence, not only the velocity field but also the spin density vector field is disturbed, so we call it spin turbulence. Though QT in the binary \cite{Berl06,Take10,Bezett10,Ishino11,Karl13,Koba14,Vill14} and spin-2 spinor BECs \cite{KU16,spin2QT} have been studied, we do not address these in this review.

\subsubsection{Spin-1 spinor GP equation}
We consider a system being comprised of $N_{\rm t}$ bosons with hyperfine spin $F=1$. 
When the system temperature is much lower than the Bose--Einstein condensation temperature, we can neglect the effect of the thermal component. Then, three kinds of macroscopic wave functions $\psi_{m}(\bm{r},t)~(m=1,0,-1)$ well describe the static and the dynamical properties of the system. The equation of motion for these wave functions \cite{Ohmi98,Ho98} is given by 
\begin{eqnarray}
i\hbar \frac{\partial}{\partial t} \psi _{m}(\bm{r},t) =  \Bigl(-\frac{\hbar ^2 }{2M} \nabla ^2 + V_{\rm trap}(\bm{r}) + pm+qm^2 \Bigl) \psi _{m}(\bm{r},t) \nonumber \\
+ c_{0} \rho_{\rm t}(\bm{r},t) \psi _{m}(\bm{r},t) 
+ c_{1} \sum_{n=-1}^{1}  \bm{F}(\bm{r},t) \cdot \bm{F} _{mn} \psi _{n}(\bm{r},t) ,  \label{spin-1_spinor1}
\end{eqnarray}
where the total density $\rho_{\rm t}(\bm{r},t)$ and the spin density vectors $\bm{F}(\bm{r},t)$ are defined by
\begin{eqnarray}
\rho_{\rm t}(\bm{r},t) = \sum_{m=-1}^{1}  \psi_{m}^*(\bm{r},t) \psi_{m}(\bm{r},t), \label{spin-1_spinor2}
\end{eqnarray}
\begin{equation}
F_{\mu}(\bm{r},t) =  \sum_{m,n=-1}^{1}  \psi _{m}^{*}(\bm{r},t) (F_{\mu})_{mn} \psi _{n}(\bm{r},t), \label{spin-1_spinor3}
\end{equation}
\begin{alignat}{3}
F_x=\frac{1}{\sqrt{2}}
\begin{pmatrix} 
0 & 1 & 0\\
1 & 0 & 1\\
0 & 1 & 0
\end{pmatrix}
\qquad
F_y=\frac{i}{\sqrt{2}}
\begin{pmatrix} 
0 & -1 & 0\\
1 &  0 & -1\\
0 &  1 & 0
\end{pmatrix}
\qquad
F_z=
\begin{pmatrix} 
1 & 0 & 0\\
0 & 0 & 0\\
0 & 0 & -1
\end{pmatrix}
.
\end{alignat}
The parameters $p$, $q$, $c_0$, and $c_1$ are the coefficients of the first Zeeman, the second Zeeman, the density-dependent interaction, and the spin-dependent interaction terms, respectively. This equation (\ref{spin-1_spinor1}) is called the spin-1 spinor GP equation. 

In the same way for Eq. (\ref{1GP2}),  Eq. (\ref{spin-1_spinor1}) is transformed into the canonical form:  
\begin{eqnarray}
i \hbar \frac{\partial}{\partial t} \psi_m (\bm{r},t) = \frac{\delta E_{\rm t}[\psi_m (\bm{r},t),\psi_m ^*(\bm{r},t)]}{\delta \psi_m^*(\bm{r},t)}, \label{energy_spin} 
\end{eqnarray}
where the energy functional $E_{\rm t}=E_{\rm f} + E_{\rm d} + E_{\rm s} $ is given by 
\begin{eqnarray}
E_{\rm f}[\psi_m (\bm{r},t),\psi_m(\bm{r},t)^*] = \int \sum_{m=-1}^{1} \Biggl[ && \frac{\hbar^2}{2M} | \bm{\nabla} \psi_m(\bm{r},t) |^2 + V_{\rm trap}
(\bm{r}) |\psi_m(\bm{r},t)|^2  \nonumber \\
&+& \bigl( pm +qm^2 \bigl) |\psi_m(\bm{r},t)|^2  \Biggl]d\bm{r},
\end{eqnarray}
\begin{eqnarray}
E_{\rm d}[\psi _m(\bm{r},t),\psi_m(\bm{r},t)^*] = \frac{c_0}{2} \int \rho_{\rm t}(\bm{r},t)^2 d\bm{r},
\end{eqnarray}
\begin{eqnarray}
E_{\rm s}[\psi _m(\bm{r},t),\psi_m(\bm{r},t)^*] = \frac{c_1}{2} \int \bm{F}(\bm{r},t)^2 d\bm{r}.
\end{eqnarray}
The first term of the energy functional is the free particle term $E_{\rm f}$, and second term is the density-dependent interaction term $E_{\rm d}$, which are similar terms in the single-component GP equation (\ref{1GP}). The term characteristic of spinor system is third term, which is the spin-dependent interaction term. 
This allows the system to exchange the particle number for each magnetic component, so that each particle 
number is not conserved. Furthermore, the sign of this interaction coefficient $c_{1}$ is decisively important for the spin dynamics. When external fields are not applied, the system with positive $c_1$ has the polar phase as the ground state, which is non-magnetized state. On the other hand, the system with negative $c_1$ is called a ferromagnetic system, where the ground state is the ferromagnetic state. 

Finally, we comment on the conservation law in Eq. (\ref{spin-1_spinor1}). The energy functional of Eq. (\ref{energy_spin}) satisfies $[E_{\rm t},E_{\rm t}]_{\rm s}=0$, $[E_{\rm t},N_{\rm t}]_{\rm s}=0$, $[E_{\rm t},F_{{\rm t},z}]_{\rm s}=0$ with the bracket $[A,B]_{\rm s} = \int \sum_{m=-1}^{1}\{ (\delta A/\delta \psi_m)(\delta B/\delta \psi^*_m)-(\delta B/\delta \psi_m)(\delta A/\delta \psi^*_m) \}d\bm{r}$. Here, we use the following notations for the total density $N_{\rm t} = \int \rho_{\rm t} d\bm{r}$ and the total spin density vector $F_{{\rm t},\mu} = \int F_{\mu} d\bm{r}~(\mu=x,y,z)$. Thus, the total energy, particle number, and the $z$-component of the spin density vector are the conserved quantities. When the external field is not applied ($p,q=0$), the Zeeman terms disappear and the relations $[E_{\rm t},F_{{\rm t},x}]_{\rm s}=0$ and $[E_{\rm t},F_{{\rm t},y}]_{\rm s}=0$ are satisfied. Thus, $F_{{\rm t},x}$ and $F_{{\rm t},y}$ become the conserved quantities.  

\subsubsection{Numerical study for spin turbulence in a spin-1 ferromagnetic spinor BEC}
We describe spin turbulence in a spin-1 ferromagnetic spinor BEC based on Eq. (\ref{spin-1_spinor1}). 
This type of study was performed by Fujimoto $et~al.$ for the first time  \cite{FT12a,FT12b}. 

The system is assumed to be a 2D system without external fields ($p=0, q=0$). The system size is $L\times L$, and the boundary condition is periodic. To generate spin turbulence, a counterflow state was used as the initial state, where the $m=1$ and $m=-1$ components flow in the opposite direction. The mathematical expression is given by
\begin{equation}
\begin{pmatrix} 
\psi _{1} \\
\psi _{0} \\
\psi _{-1}
\end{pmatrix}
= \sqrt{\frac{\rho_{{\rm t}0}}{2}}
\begin{pmatrix} 
{\rm{exp}}\biggl[i \frac{M}{2 \hbar}\bm{V}_{\rm R}\cdot \bm{r}  \biggl] \\
0 \\
{\rm{exp}}\biggl[-i \frac{M}{2 \hbar}\bm{V}_{\rm R}\cdot \bm{r}  \biggl]
\end{pmatrix}, 
\label{2initial_state_spinor}
\end{equation}
where $\bm{V}_{\rm R}=V_{\rm R} \hat{{\bm e}}_x$ and $\rho_{{\rm t}0}$ are the relative velocity of the counterflow and the bulk density.
This initial state is unstable, leading to spin turbulence. In the numerical calculation, the parameters are set to be $L=128\xi_{\rm d}$, $|c_0/c_1|=20$, and $V_{\rm R}/C_{\rm s} \sim 0.785$ with the bulk density $\rho_{{\rm t}0} = N_{\rm t}/L^2$,  characteristic time scale $\tau_{\rm d}=\hbar/c_0\rho_{{\rm t}0}$, density coherence length $\xi_{\rm d}=\hbar/\sqrt{2Mc_0\rho_{{\rm t}0}}$, and sound velocity $C_{\rm s}=\xi_{\rm d}/\tau_{\rm d} = \sqrt{c_0 \rho_{{\rm t}0}/2M}$.

\begin{figure}[t]
\begin{center}
\includegraphics[keepaspectratio, width=12cm,clip]{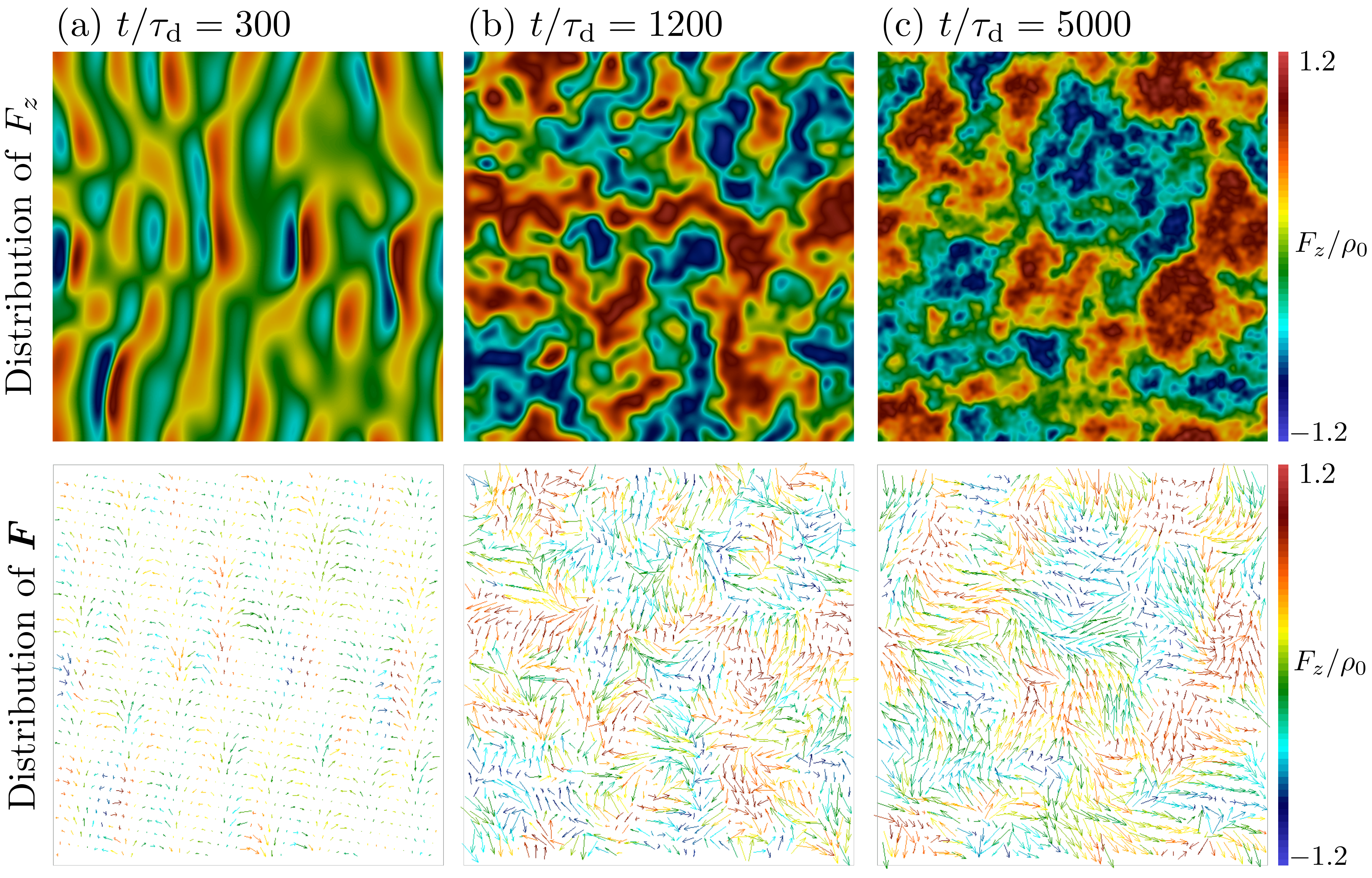}
\caption{Spatial-temporal distribution of $F_z$ ($upper$) and $\bm{F}$ ($lower$) at $t/\tau_{\rm d}={\rm (a)}300$, ${\rm (b)}1200$, and ${\rm (c)}5000$. The color distribution of lower figures shows $F_z$. (Color figure online) \label{SD}} 
\end{center}
\end{figure}

The spatial-temporal distribution of the spin density vector is shown in Fig. \ref{SD}. In the early stage, counterflow instability occurs, generating the stripe structure of $F_z$ shown in Fig. \ref{SD} (a). The property of the instability can be understood by solving the Bogoliubov-de Gennes equation. The solution gives imaginary parts of the eigenvalues, which describe the wavenumber of the stripe and the growth rate of $F_z$. 
As time passes, the stripe structure collapses, and finer spin structure is generated as shown in Figs. \ref{SD} (b) and (c).

To investigate the properties of the spin turbulence, the spectrum of the spin-dependent interaction 
energy per unit mass is considered, which is defined as
\begin{eqnarray}
\mathcal{E}_{\rm s}(k,t) = \frac{c_1}{2 N_{\rm t} M\triangle k} \sum_{\Omega(\bm{k}_1,k)} |\tilde{\bm{F}}(\bm{k}_1,t)|^2, \label{spin-dependent_interaction_energy}
\end{eqnarray}
with the Fourier component $\tilde{\bm{F}}(\bm{k},t) = \mathcal{F}[\bm{F}(\bm{r},t)]$ of the spin density vector.
This quantity is equivalent to a two-point spatial correlation function for the spin density vector 
in the wave number space, which contains information of the fluctuation of the spin density vector in the spin turbulence. 
Experimentally, the spin density vector can be observed by phase contrast imaging \cite{Berkeley08}. 

\begin{figure}[t]
\begin{center}
\includegraphics[keepaspectratio, width=12cm,clip]{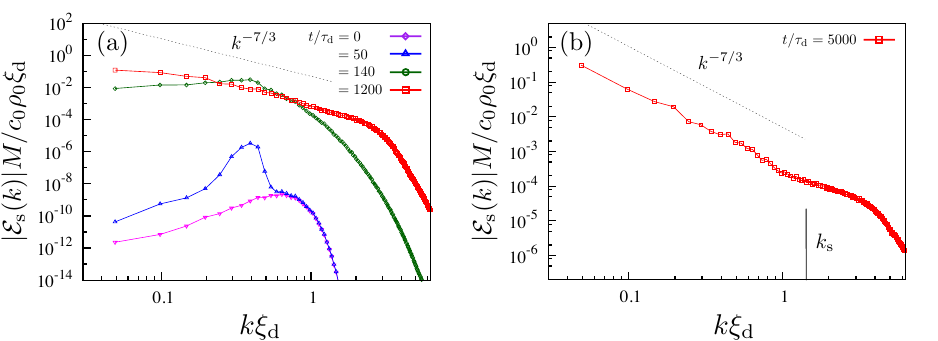}
\caption{Time-development for the spectrum of spin-dependent interaction energy．
This spectrum is averaged over 10 calculations with different initial noise values. 
The wave number $k_{\rm s}$ is defined by $2\pi/\xi_{\rm s}$ with the spin coherence length $\xi_{\rm s} = \hbar/ \sqrt{2M|c_1|\rho_{{\rm t}0}}$. (Color figure online) \label{ST}} 
\end{center}
\end{figure}

Figure \ref{ST} (a) is the numerical result for the time-development of this spectrum. In this calculation, the initial state is the polar state, which does not have a large amplitude of the spin density vector. However, over time, the counterflow instability leads to the growth of the spin density vector, and the spectrum begins to exhibit a peak near a wave number $k\xi_{\rm d} \sim 0.4$. This value corresponds to the most unstable wave number for the counterflow instability with the relative velocity $V_{\rm R}/C_{\rm s} \sim 0.785$, that is analytically obtained by solving the Bogoliubov--de Gennes equation. As time continues to increase, nonlinear coupling between different wave number modes is effective, and the spectrum at time $t/\tau_{\rm d}=1200$ in Fig. \ref{ST} (a) is realized. From this spectrum, it follows that the power law behavior begins to be formed. At a time $t/\tau_{\rm d}=5000$, the power law behavior becomes clear, as shown in Fig. \ref{ST} (b).  

Then, a new question is posed: what is the power exponent in the spectrum, and is this power law characteristic of the spin turbulence? The answer for this issue is described in the following subsection. 

\subsubsection{Derivation of the $-$7/3 power law}
The purpose of this subsection is the derivation for the power exponent of the spectrum in the spin turbulence by using the Kolmogorov-type dimensional scaling analysis. 
 
In this derivation, three assumptions are imposed: 
(1) the macroscopic wave function is expressed by the ferromagnetic state, 
(2) the total density is uniform, and
(3) the superfluid velocity is much smaller than the sound velocity.

From the assumption (1), we can express the wave function as
\begin{equation}
\begin{pmatrix} 
\psi _{1} \\
\psi _{0} \\
\psi _{-1}
\end{pmatrix}
= \sqrt{\rho_{\rm t}}{\rm e }^{i\phi} \hat{U}(\alpha,\beta,\gamma)
\begin{pmatrix} 
1 \\
0 \\
0
\end{pmatrix}
,
\label{2initial_state_spinor}
\end{equation}
where $\phi$ is the phase, and $\hat{U}(\alpha,\beta,\gamma)$ is the rotation operator in the spin space with Euler angles $\alpha$, $\beta$, and $\gamma$. The system under consideration has ferromagnetic interaction, and the excitation is not strong enough to break this state, so that assumption (1) can be justified. To confirm numerically whether this assumption is valid, it is suitable to calculate $\int |\bm{F}/\rho_{\rm t}| d\bm{r}/V$, which should be unity in the ferromagnetic state. The numerical calculation corresponding to Fig. \ref{ST} shows that this quantity is about $0.9$. As a result, we can derive the equation of spin density vector given by \cite{Lamacraft,Barnett,KK,YU12} 
\begin{eqnarray}
\frac{\partial}{\partial t} \rho_{\rm t} f_{\mu} +  \bm{\nabla} \cdot \biggl( \rho_{\rm t} f_{\mu} \bm{v} - \frac{\hbar}{2M}\epsilon_{\mu\nu\lambda} \rho_{\rm t}  f_{\nu} \bm{\nabla} f_{\lambda}  \biggl) = 0,   \label{scaling1}
\end{eqnarray}
with $f_{\mu} = F_{\mu}/\rho_{\rm t}$.

Next, we make use of assumption (2). This is justified by noting $|c_{0}/c_{1}| \gg 1$ in typical experiments, which implies that it is necessary to strongly excite the system to strongly modulate the density profile. The numerical calculation shows that the spatial average of the density fluctuation $|\delta \rho_{\rm t} (\bm{r},t)| = |\rho_{\rm t}(\bm{r},t) - \rho_{{\rm t}0}|$ is about $0.05\rho_{{\rm t}0}$, so this assumption also is valid. Thus, we use the approximation $\rho_{\rm t} \simeq \rho_{{\rm t}0}$, obtaining
\begin{eqnarray}
\frac{\partial}{\partial t} f_{\mu} +  \bm{\nabla} \cdot \biggl( f_{\mu} \bm{v} - \frac{\hbar}{2M}\epsilon_{\mu\nu\lambda} f_{\nu} \bm{\nabla} f_{\lambda}  \biggl) = 0.  \label{scaling2}
\end{eqnarray}

Finally, to use assumption (3), we transform Eq. (\ref{scaling2}) into the non-dimensional form where the space and time are normalized $t^{'} = t/\tau_{\rm d}$，$\bm{\nabla}' =  \xi_{\rm d} \bm{\nabla}$; 
\begin{equation}
\frac{\partial}{\partial t'} \bm{f} + \Bigl(\frac{\bm{v}}{C_{\rm s}} \cdot \bm{\nabla}' \Bigl) \bm{f} =  \bm{f} \times {\bm{\nabla}' }^{2} \bm{f}. \label{scaling3}
\end{equation}
Here, we use the incompressible condition $\bm{\nabla}\cdot \bm{v}=0$, which is derived by assumption (2) and the continuity equation for the particle number.

We consider the magnitude of superfluid velocity field. Generally, in the turbulence, vortices, solitons, and so on are nucleated. Such structures have velocity fields comparable to the sound velocity $C_{\rm s}$ near the core of these structures. However, it is expected that, except for the core region, the velocity field is much smaller than the sound velocity. In fact, the numerical calculation has $\int |\bm{v}| d\bm{r}/V \sim 0.15C_{\rm s}$. From this result and $|\bm{f}|\simeq 1$, the term with the velocity field in Eq. (\ref{scaling3}) is smaller than the nonlinear spin term. \footnote[1]{Due to the difference of the number of spatial derivatives, the amplitudes of two terms should depend on the wave number. The numerical calculation shows that the spatial average of the velocity field has about $0.15 C_{\rm s}$, so we can roughly estimate the wave number region $0.15 \lesssim k \xi_{\rm d}$ where the term with the velocity field in Eq. (\ref{scaling3}) is negligible.}. As a result, we can simplify Eq. (\ref{scaling3}) to 
\begin{equation}
\frac{\partial}{\partial t'} \bm{f} =  \bm{f} \times {\bm{\nabla}'}^{2} \bm{f}. \label{scaling4}
\end{equation}

Here, we note that the spin-dependent interaction energy is conserved when the ferromagnetic state assumption (1) is sustained.
This situation is similar to energy transfer in the wave number space, so that in the spin turbulence it has a possibility to have 
constant energy flux for the spin-dependent interaction energy. From this consideration, we apply the Kolmogorov-type scaling analysis \cite{Ottaviani91,Watanabe97} into Eq. (\ref{scaling4}).

First, we perform the scale transformation ($\bm{r} \rightarrow \alpha \bm{r}$, $t \rightarrow \beta t$) and require that 
Eq. (\ref{scaling4}) is invariant under this transformation.
Then, we find that when $\bm{f}$ is transformed into $\alpha ^{2} \beta ^{-1} \bm{f}$, this requirement is satisfied.
Hence, the scaling dimension of $\bm{f}$ is given by
\begin{equation}
\bm{f} \sim \Lambda _{\rm f} k^{-2} t^{-1}.\label{scale_transformation_1}
\end{equation}
Here, $\Lambda _{\rm  f}$ is a dimensional parameter．
This scaling dimension and the dimensional analysis leads to the estimation of the energy flux $\epsilon_{\rm s}$ per unit mass; 
\begin{align}
\epsilon_{\rm s} &\sim \frac{|c_1| (\rho_{{\rm t}0} \bm{f})^{2}}{M\rho_{{\rm t}0}t}  \nonumber \\
&\sim \frac{|c_1| \rho_{{\rm t}0} \Lambda_{\rm f}^2}{M}  k^{-4}t^{-3}.  \label{scale_transformation_2}
\end{align}

We assume the wave-number-independence of the flux in the same way as the Kolmogorov 41 theory. 
As a result, Eqs. (\ref{spin-dependent_interaction_energy})，(\ref{scale_transformation_1})，(\ref{scale_transformation_2}) lead to the $-7/3$ power law; 
\begin{align}
|\mathcal{E}_{s}(k)| &\sim \frac{|c_1| (\rho_{{\rm t}0} \bm{f})^{2}}{M \rho_{{\rm t}0} k} \nonumber \\ 
 &\sim  \Lambda _{\rm s}^{2/3} \epsilon _{\rm s}^{2/3} k^{-7/3}, 
\end{align}
with $\Lambda _{\rm  s} = \Lambda _{\rm  f} \sqrt{ |c_{1}| \rho_{{\rm t}0} / M}$.
Figure \ref{ST} shows the agreement with this $-7/3$ power law. Originally, this power law is theoretically derived and numerically confirmed by Fujimoto $et~al.$ \cite{FT12a}.

As described in the derivation of Eq. (\ref{scaling4}), the term with velocity field can be neglected in the region $0.15 \lesssim k \xi_{\rm d}$ by rough order estimation, which seems to be inconsistent with Fig. \ref{ST} (b) because the spectrum in the region $k \xi_{\rm d} \lesssim 0.15$ exhibits the $-7/3$ power law. This inconsistency may caused by the rough estimation, but the reason for this is not clear. 

Finally, we comment on the reason why the power exponent is different from the $-5/3$ exponent in the usual Kolmogorov power law. This difference comes from the difference in the number of the spatial derivative of the nonlinear terms. In the Navier—Stokes equation, the nonlinear term (inertial term) is the secondary term with the first spatial derivative, while Eq. (\ref{scaling4}) has the secondary term with the second spatial derivative as a nonlinear term. The $-7/3$ power law is the result that strongly reflects the property of this nonlinear term.


%
%

\section{Open problems of numerical studies of QT}
In this chapter, we discuss current open problems of numerical studies of QT. 

\subsection{Problems in QT of superfluid helium}
Most previous numerical studies have addressed the VF model at 0 K or under the prescribed normal fluid profile. However, it has recently become possible to simulate the coupled dynamics of the VFM model and the Navier--Stokes equation for the normal fluid, as described in 2.2.5. This trend opens lots of interesting directions.

\subsubsection*{Energy spectrum and cascade}
Investigating the universal law of the energy spectrum is a main task of the numerical studies of QT.  We consider {\it quasiclassical turbulence} at 0 K. In the region of wavenumbers $k$ smaller than $2\pi/\ell$ with the intervortex spacing $\ell$, the energy flows from small $k$ to large $k$ and the $-5/3$ spectrum is sustained, which is confirmed by several numerical studies. However, the inertial range of most numerical works is not wide, typically only one decade. Numerical simulations with wider inertial range would be preferable. In the region of wavenumbers $k$ larger than $2\pi/\ell$, the quantum nature of each vortex becomes important and the energy should be transferred by the Kelvin-wave cascade. The Kelvin-wave cascade and its statistical laws are described, for example, in Ref. \cite{Baggaley14}. It is not trivial how the energy spectra and cascade join in the intermediate region at $k \simeq 2\pi/\ell$ \cite{BarenghiPNAS2}. In order to numerically reveal the whole picture of the spectrum and the cascade, we need much bigger simulations covering a wider range of wavenumbers than the present level. 

The issue of the energy spectrum and cascade at finite temperatures is interesting and challenging. It is not clear how turbulence of a superfluid and normal fluid interact with each other. Although there is a theoretical proposal \cite{Babuin16}, numerical studies of the coupled dynamics of two fluids should contribute to this problem as well.

The Kolmogorov law comes from the self-similarity in wavenumber space, and is believed to be related to the Richardson cascade in real space \cite{Frisch}. Since vortices are not well defined in a viscous classical fluid, however, this picture is not clear. On the hand, a quantized vortex is a stable and definite topological defect, so investigating QT may connect the cascade process in the real and wavenumber spaces. Actually, some numerical works of the VF \cite{Alamri08,Baggaley1212} and GP \cite{Sasa11} models found the bundle-like structure of quantized vortices, which is thought to be indispensable for the cascade process in CT. 

\subsubsection*{Boundary conditions of superfluid and quantized vortices}
The boundary condition of a superfluid on a solid surface is $ {\bm v}_{\rm s} \cdot {\bm {\hat n}} = 0 $ with a unit vector $\bm {\hat n}$ perpendicular to the surface. The problem is what to do when quantized vortices are attached to the solid surface. As described in 2.2.2, this is trivial if the surface is assumed to be flat and smooth. When the surface is rough, however, this procedure is not easy. From the comparison between experiments and numerical simulations, it is necessary to consider the boundary condition properly. However, it is difficult to consider the rough surface condition in the simulation. The most difficulty is that we have no information of the surface roughness. It would be possible to consider the case in which there is a hemisphere pinning site on a flat surface\cite{schwarz85}. However, we do not know how many and how large pinning sites we need. In order to overcome this difficulty, we have to introduce some model with arbitrary parameters. Currently we have no definite criteria for making the model. 
The recent numerical simulation of the GP model finds superfluid boundary layer over the rough surface \cite{Stagg17}. This approach could be helpful in understanding this problem.

\subsubsection*{Thermal counterflow in a realistic channel}
Once the simulation of the coupled dynamics of two fluids becomes possible, the important target would be revealing the T1 and T2 states described in 2.1.1. Changing the aspect ratio of the channel cross-section, we should study how the turbulent state changes to the T3 state. Simulation under the rough surface condition is indispensable for this case as well.

\subsubsection*{QT generated by vibrating structures}
The simulation for QT created by vibrating structures, as described in 2.1.3 looks to capture the essence of the dynamics, but cannot explain the essence such as the critical velocity, hysteresis, and lifetime of QT properly. We should introduce the rough surface condition. The simulation of the coupled dynamics of two fluids could make sense in this case as well. 

\subsubsection*{Logarithmic velocity profile}
The numerical study \cite{yui15b} demonstrated a logarithmic velocity profile of QT, but did not propose a theory for a mean velocity profile.
To construct the theory, the momentum flux for a superfluid velocity may be investigated, which causes a logarithmic velocity profile of classical turbulence.
If the wider range of a logarithmic profile could be obtained, the logarithmic velocity profile of QT would be supported more strongly.

The simulation \cite{yui15b} was performed in a pure normal flow between two parallel plates.
Other types of flow will be studied to investigate the universality of the logarithmic velocity profile, {\it e.g.} a thermal counterflow or a pure superflow.
Additionally, the channel geometry may affect the mean velocity profile; a Karman constant could be changed in a duct from two parallel plates.
As a more advanced step, we should take into account the surface roughness to understand more realistic physics.

For numerical studies using the two-fluid coupled dynamics, the statistical law for velocity profile is an interesting topic, because the flow of normal fluid component can become laminar or turbulent, and the two fluids affect mutual velocity profiles: a laminar Poiseuille normal flow corresponds to the preceding study \cite{yui15b}.
When a normal flow becomes a tail-flattened laminar flow or turbulence, the mean velocity profile can change from the logarithmic velocity profile.
Additionally, normal flow can have different profiles from the logarithmic velocity profile, even though that is the turbulence of a viscous fluid.

Numerical studies should express their results in quantities that could more easily be observed by experiments.
Because superfluid velocity in QT cannot be easily observed, we may use quantities related to quantized vortices, ${\it e.g.}$ drift velocity of vortices, and express the logarithmic velocity profile using these quantities.

\subsection{Problems in QT of ultracold atomic gases}

\subsubsection*{Dimensionality of QT}
Recently, 2D turbulence in the GP model has been actively studied, and the 2D nature has been found \cite{Horng,Num10,Sch12,Reeves12,Bradley12,White12,Reeves13,Billam14,Simula14,Billam15,Simula16,Yu16}.  
For example, the dynamical clustering of quantized vortices, namely, Onsager vortex formation, has been numerically confirmed \cite{Simula14,Simula16}, which is consistent with behavior of the energy inverse cascade in 2D classical turbulence. However, this formation is understood by evaporative heating of vortex gas, so that the relationship between the inverse cascade and this heating process is unclear. The resolution of this issue is a challenging problem in 2D QT.

Additionally, in ultracold atomic gas, the spatial dimension is continuously changed by controlling the trapping frequency, so that 
we can investigate the connection between 2D and 3D QT. Then, the crossover or transition between 2D and 3D QT may become an important problem, where the question of whether the Onsager vortex formation occurs may be investigated while changing the trapping frequency. As another problem, 1D turbulence may be interesting, which has not been actively studied in the GP model. 

\subsubsection*{Finite temperature effect in QT}
In experiments of QT, there should be finite temperature effect, which is caused by interaction between condensate and non-condensate fractions. When the temperature is much lower than the transition temperature of the Bose--Einstein condensation, the non-condensate fraction is much smaller and the description of the GP model could be valid. However, when the temperature is not low, we cannot neglect the non-condensate fraction, where the kinetic energy spectrum in QT may be affected by this fraction because, in superfluid helium, the interaction between the superfluid and normal components induces the mutual friction and the vortex dynamics 
at finite temperature are drastically different from those at zero temperature. In fact, using the stochastic projected GP model \cite{SPGP} or Zaremba--Nikuni--Griffin method \cite{ZNG}, some reports have revealed the effect of finite temperature on vortex dynamics \cite{FV1,FV2}. However, at present, the relationship between QT and the finite temperature effect has not been investigated in ultracold atomic BECs. 

\subsubsection*{Application of closure method to the GP model}
In classical turbulence, analytical methods to approach a closure problem in the Navier--Stokes equation are highly developed. 
The application of this method to the GP model may lead to promising results. 

A challenging problem concerning the closure problem is the derivation of the $-7/2$ power law in the momentum distribution, which was observed 
in a recent experiment \cite{Nir16}. At present, as far as we know, this power exponent has never been derived analytically.
When this derivation is achieved, the physical meaning of this power law may be clear. 

Here, we comment on a paper addressing the closure problem of the GP model \cite{yoshida}. 
This paper uses the spectral closure approximation, deriving certain power exponents for the momentum distribution, but the obtained exponents are inconsistent with $-7/2$. 

\subsubsection*{QT in a binary BEC}
QT in a binary BEC gives us a platform for studying QT beyond superfluid helium, because this system has two kinds of BEC and binary QT is comprised of two different quantized vortices. 
Recently, studies of QT in this system have been carried out \cite{KU16,Berl06,Take10,Bezett10,Ishino11,Karl13,Koba14,Vill14,spin2QT}, in which correlation functions for the velocity field and spin density vector were investigated and some power laws were reported. 
However, these address binary QT in limited parameter regions, and QT in a mass imbalance system or QT with spin--orbit coupling have never been investigated.

\subsection{Common problems in superfluid helium and ultracold atomic gases}

\subsubsection*{Decay of QT}
Almost all topics in this article were about the statistically steady state. However, it is also important to study the decay of QT after turning off the excitation. For example, we suppose a statistically steady state of QT with the Kolmogorov spectrum. It would be interesting to study what happens after turning off the excitation, namely how the spectrum decays with the degeneration of vortices. Such a simulation would give important information regarding experiments, for example, on the dissipative mechanism in the zero-temperature limit \cite{Walmsley07} or classification of {\it semiclassical turbulence} and {\it ultraquantum turbulence} \cite{Walmsley08}.

\subsubsection*{Transition to turbulence as critical phenomena}
Most topics in this article were about the developed turbulence, but recently in the field of CT, some efforts have been devoted to understanding the transition between laminar flow and turbulence from the viewpoint of critical phenomena \cite{Kazumasa07,Sano16}. It would be interesting to numerically study the transition to QT.

\begin{acknowledgements}
We would like to acknowledge Brian P. Anderson and Ashton S. Bradley for critical reading of Sec. 3.2.2, and Wei Guo for preparing the experimental pictures of Fig. 3. M. T. was supported by JSPS KAKENHI Grant Numbers JP16H00807, JP26400366. K. F. was supported by Grant-in-Aid for JSPS Research Fellow Grant Number JP16J01683. S. Y. was supported by Grant-in-Aid for JSPS Research Fellow Grant Number JP16J10973.
\end{acknowledgements}

\section*{Appendix A: Derivation of Eq. (\ref{mutual friction})}
From Eq. (\ref{eq:drag}) and Eq. (\ref{eq:magnus}), the massless equation ${\bm f}_{\rm M} + {\bm f}_{\rm D} = {\bm 0}$ leads to
\begin{equation}
  {\bm s}' \times \left [ \rho_{\rm s} \kappa (\dot{\bm s} - \dot{\bm s}_0) + \gamma_0 {\bm s}' \times ( \dot{\bm s} - {\bm v}_{\rm n} ) - \gamma'_0 (\dot{\bm s} - {\bm v}_{\rm n}) \right ] = {\bm 0}.
  \label{eq:massless}
\end{equation}
By substituting Eq. (\ref{eq:vor_fric}) into Eq. (\ref{eq:massless}) and calculating as 
\begin{equation}
\begin{split}
  \dot{\bm s} - {\bm v}_{\rm n} &= (\dot{\bm s}_0 - {\bm v}_{\rm n}) - \alpha {\bm s}' \times (\dot{\bm s}_0 - {\bm v}_{\rm n}) + \alpha' {\bm s}' \times [ {\bm s}' \times ( \dot{\bm s}_0 - {\bm v}_{\rm n})], \\
  {\bm s}' \times (\dot{\bm s} - {\bm v}_{\rm n}) &= (1-\alpha') {\bm s}' \times (\dot{\bm s}_0 - {\bm v}_{\rm n}) - \alpha {\bm s}' \times [ {\bm s}' \times (\dot{\bm s}_0 -{\bm v}_{\rm n}) ], \\
\end{split}
\end{equation}
then we obtain
\begin{equation}
\begin{split}
  - &\left[ -\gamma_0 \alpha + (\rho_{\rm s} \kappa - \gamma'_0) \alpha' + \gamma'_0 \right] {\bm s}' \times (\dot{\bm s}_0 - {\bm v}_{\rm n}) \\
  + &\left[(-\rho_{\rm s} \kappa + \gamma'_0) \alpha - \gamma_0 \alpha' + \gamma_0 \right] {\bm s}' \times [ {\bm s}' \times (\dot{\bm s}_0 - {\bm v}_{\rm n})] = {\bm 0}.
\end{split}
\end{equation}
The linear independence of the two vectors in Eq. (152) leads to
\begin{equation}
  \alpha = \frac{\rho_{\rm s}\kappa\gamma_0}{\gamma_0^2 + (\rho_{\rm s} \kappa - \gamma_0')^2}, ~~~
  \alpha' = \frac{\gamma_0^2 - \gamma_0' (\rho_{\rm s}\kappa - \gamma_0')}{\gamma_0^2 + (\rho_{\rm s} \kappa - \gamma_0')^2}.
\end{equation}

\section*{Appendix B: Derivation of Eqs. (\ref{Numerics1})--(\ref{Numerics2})}
We follow the approach considering a circle, which has the center ${\bm m}$ and passes through ${\bm s}_{i-1}$, ${\bm s}_{i}$, and ${\bm s}_{i+1}$ \cite{schwarz85}.
The relation
\begin{equation}
  |{\bm m} - {\bm s}_{i-1}|^2 = |{\bm m} - {\bm s}_{i}|^2 = |{\bm m} - {\bm s}_{i+1}|^2,
\end{equation}
is satisfied, and is rewritten as
\begin{equation}
\begin{split}
  2({\bm m}-{\bm s}_i) \cdot {\bm l}_{-} + l_{-}^2 &= 0, \\
  -2({\bm m}-{\bm s}_i) \cdot {\bm l}_{+} + l_{+}^2 &= 0,
\end{split}
\end{equation}
with ${\bm l}_{+} = {\bm s}_{i+1} - {\bm s}_{i}$ and ${\bm l}_{-} = {\bm s}_{i} - {\bm s}_{i-1}$.
Because these four vectors indicate the positions in the same plane, we can write
\begin{equation}
  {\bm m} - {\bm s}_i = a_{+} {\bm l}_{+} - a_{-} {\bm l}_{-}.
\end{equation}
The parameters $a_{\pm}$ are obtained by calculating these equations as
\begin{equation}
  a_{\pm} = \frac{1}{2} \frac{ l_{+}^2 l_{-}^2 + l_{\mp}^2 ({\bm l}_{+} \cdot {\bm l}_{-})}{l_{+}^2 l_{-}^2 - ({\bm l}_{+} \cdot {\bm l}_{-})^2}.
\end{equation}
Because ${\bm s}_{i}'' \parallel ({\bm m} - {\bm s}_i)$ and $|{\bm s}_{i}''| = |{\bm m} - {\bm s}_i |^{-1}$, we obtain
\begin{equation}
  {\bm s}_{i}'' = c_i^{+} {\bm l}_{+} - c_i^{-} {\bm l}_{-},
\end{equation}
with
\begin{equation}
  c_i^{\pm} = \frac{a_{\pm}}{ |a_{+} {\bm l}_{+} - a_{-} {\bm l}_{-} |^2 }.
\end{equation}
Similarly, the first derivative ${\bm s}_{i}'$ can be obtained from ${\bm s}_{i}' \cdot ({\bm m}-{\bm s}_i) = 0$ and $ |{\bm s}_{i}'| = 1$.

\begin{figure}
  \centering
  \includegraphics[width=0.4\textwidth]{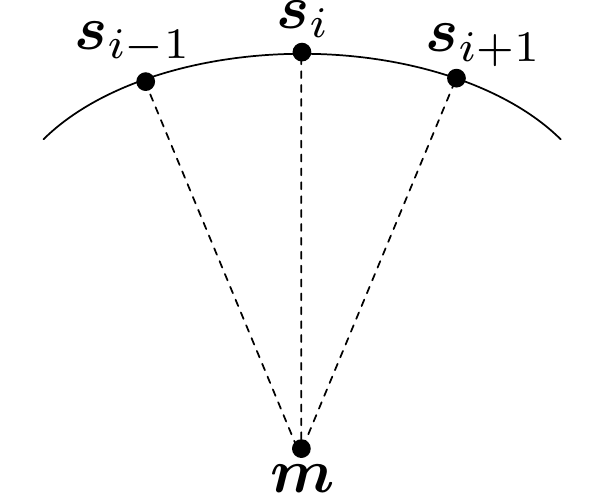}
  \caption
  {
  Circle passing through the adjacent points ${\bm s}_{i-1}$, ${\bm s}_{i}$, and ${\bm s}_{i+1}$.
  The center of the circle is pointed by ${\bm m}$.
  }
  \label{circle_adjacent.eps}
\end{figure}

\section*{Appendix C: Derivation of Eqs. (\ref{Ekin_c}) and (\ref{Ekin_i})}
To begin with, we comment on $\bm{A}_0(t)$ in Eq. (\ref{Ekin_v}). A constant field for $A(\bm{r},t)$, which is a zero-wave-number mode, satisfies both the compressible and the incompressible conditions. Then, we denote it as $\bm{A}_0(t)$ and separate it from the compressible and the incompressible fields. Hence, the Fourier components for $\bm{A}_{\rm c}(\bm{r},t)$ and $\bm{A}_{\rm i}(\bm{r},t)$ do not include the zero-wave-number mode.

Under this situation, we show the derivation of Eqs. (\ref{Ekin_c}). The divergence of Eq. (\ref{Ekin_v}) leads to 
\begin{equation}
{\rm div} {\bm A}(\bm{r},t) = {\rm div} {\bm A}_{\rm c}(\bm{r},t), \label{C1}
\end{equation}
where we use the condition ${\rm div} {\bm A}_{\rm i}(\bm{r},t)=0$. Substituting the Fourier expansions for ${\bm A}(\bm{r},t)$ and ${\bm A}_{\rm c}(\bm{r},t)$ to Eq. (\ref{C1}), we obtain
\begin{equation}
{\bm k} \cdot {\tilde{\bm A}}(\bm{k},t) = {\bm k} \cdot {\tilde{\bm A}}_{\rm c}(\bm{k},t). \label{C2}
\end{equation}
The condition ${\rm rot}{\bm A}_{\rm c} =0$ leads to 
\begin{equation}
 {\bm k} \times {\tilde{\bm A}}_{\rm c}(\bm{k},t) = 0. \label{C3}
\end{equation}
Then, we obtain ${\tilde{\bm A}}_{\rm c}(\bm{k},t) =  \mathcal{C} {\bm k }$ with the constant $\mathcal{C}$. 
We substitute it to Eq. (\ref{C2}), obtaining 
\begin{equation}
\mathcal{C} = \frac{\bm{k}\cdot  {\tilde{\bm A}}(\bm{k},t)  }{k^2}. \label{C4}
\end{equation}
Thus, we can derive Eq. (\ref{Ekin_c}). 

The derivation of Eq. (\ref{Ekin_i}) is done by the Fourier transformation of Eq. (\ref{Ekin_v}), which leads to 
\begin{equation}
{\tilde{\bm A}}(\bm{k},t)  = {\tilde{\bm A}}_{\rm c}(\bm{k},t)  + {\tilde{\bm A}}_{\rm i}(\bm{k},t), \label{C5}
\end{equation}
where the wave number $\bm k$ is not a zero vector. Equations (\ref{C4}) and (\ref{C5}) give
\begin{equation}
{\tilde{\bm A}}_{\rm i} (\bm{k},t)  = {\tilde{\bm A}}(\bm{k},t) - \frac{\bm{k}\cdot  {\tilde{\bm A}}(\bm{k},t)  }{k^2} \bm{k}. \label{C6}
\end{equation}
Then, Eq. (\ref{Ekin_i}) is derived.



\end{document}